\documentclass{elsart1p}

\usepackage{graphicx}

\usepackage{amssymb,longtable,amsmath,graphicx}

\begin{document}

\begin{frontmatter}



\title{Charged-Particle Thermonuclear Reaction Rates:\\ IV. Comparison to Previous Work}


\author{C. Iliadis}, \author{R. Longland}, \author{A. E. Champagne}
\address{Department of Physics and Astronomy, University of North Carolina, Chapel Hill, NC 27599-3255, USA; Triangle Universities Nuclear Laboratory, Durham, NC 27708-0308, USA}
\author{A. Coc}
\address{Centre de Spectrom\'etrie Nucl\'eaire et de Spectrom\'etrie de Masse (CSNSM), UMR 8609, CNRS/IN2P3 and Universit\'e
        Paris Sud 11, B\^atiment 104, 91405 Orsay Campus, France}

\begin{abstract}
We compare our Monte Carlo reaction rates (see Paper II of this series) to previous results that were obtained by using the classical method of computing thermonuclear reaction rates. For each reaction, the comparison is presented using two types of graphs: the first shows the change in reaction rate uncertainties, while the second displays our new results normalized to the previously recommended reaction rate. We find that the rates have changed significantly for almost all reactions considered here. The changes are caused by (i) our new Monte Carlo method of computing reaction rates (see Paper I of this series), and (ii) newly available nuclear physics information (see Paper III of this series). 
\end{abstract}


\end{frontmatter}

\section{Introduction}\label{intro}
In the final paper of this series, referred to as Paper IV, we compare our Monte Carlo reaction rates to previously derived results. In Paper I we point out that previously reported (or ``classical") reaction rates have a different meaning from our Monte Carlo rates. The latter are presented in Paper II and are subject to a straightforward statistical interpretation, while the former are statistically meaningless. Consequently, one has to be very careful when comparing such qualitatively and quantitatively different results. Since we did not want to complicate the discussion in Paper II by comparing ``apples with oranges", we avoided so far a comparison between present and previous reaction rates. 

On the other hand, we realize that for practical purposes many readers are precisely interested in such a comparison: by how much did a particular reaction rate change? How accurate was a previous reaction rate? Does a given rate used in a particular stellar model need updating or not? How significant is this new Monte Carlo method compared to classical techniques of computing thermonuclear reaction rates? In order to answer these questions, we present a comparison of our new rates with previous results.

Section 2 provides a few useful comments. A brief summary is presented in Sec. 3. Results are presented in graphical form in the Appendix, together with information on how to interpret the figures.

\section{Discussion}
The comparison of Monte Carlo with classical reaction rates is presented in graphical form in the Appendix. For each reaction considered, two types of graphs are presented. The first shows the change in reaction rate uncertainties, while the second graph displays the new results normalized to the previously recommended reaction rate. Detailed information on how to interpret the graphs is also given in the Appendix. 

Generally, the previous results used for the comparison are adopted from the latest reference that lists recommended reaction rates and, if available, associated uncertainties in {\it tabular form}. This reference is provided after each rate comparison figure. Most of the previous reaction rates are adopted from Angulo et al. \cite{Ang99} or Iliadis et al. \cite{Ili01}. For some reactions, newer rates have been presented in the literature, but in graphical form only. In such cases we requested (and obtained) from the original authors the numerical reaction rates used to produce the literature plots.

Visual inspection of the graphs reveals that for almost all reactions our new thermonuclear rates differ substantially from previous results. The changes are caused both by our new Monte Carlo method of computing reaction rates (see Paper I of this series), and by newly available nuclear physics information (see Paper III of this series). We did not analyze in detail for each reaction the sources of a rate change at a given temperature. Such a comprehensive analysis would require considerably more efforts in computing time and manpower. Nevertheless, we feel that many readers may find the graphs presented here useful for a variety of purposes.

\section{Summary}
Our new Monte Carlo reaction rates (see Paper II) are compared with previous results. For each reaction the comparison is presented in graphical form. It is found that the rates of almost all reactions considered here have changed significantly. The changes are caused by our new Monte Carlo method of computing reaction rates (see Paper I) and by newly available nuclear physics information (see Paper III). 

\section{Acknowledgement}
The authors would like to thank Dan Bardayan, Shawn Bishop, Barry Davids and Hendrik Schatz for providing us with numerical results for their previous reaction rates. This work was supported in part by the U.S. Department of Energy under Contract No. DE-FG02-97ER41041. 

\clearpage
\appendix*

\section{}
For each reaction, the comparison of present (Monte Carlo) with previous (classical) thermonuclear rates is presented in two graphs. The x-axis shows the entire temperature range of T=0.01-10.0 GK, while the y-axis displays a reaction rate ratio. The meaning of the curves is given below.
\\
\begin{tabbing}
AAAAAAAAAAAAAAAA\= \kill
$N_A\left<\sigma v\right>_{high}^{MC}$	\> Present high Monte Carlo reaction rate\\
$N_A\left<\sigma v\right>_{med}^{MC}$	\> Present median Monte Carlo reaction rate\\
$N_A\left<\sigma v\right>_{low}^{MC}$	\> Present low Monte Carlo reaction rate\\
$N_A\left<\sigma v\right>_{high}^{clas}$	\> Previous high classical reaction rate\\
$N_A\left<\sigma v\right>_{ad}^{clas}$	\> Previous adopted classical reaction rate\\
$N_A\left<\sigma v\right>_{low}^{clas}$	\> Previous low classical reaction rate\\
\\
{\it Top graph:}	\\  
\\
upper solid line	    \> $N_A\left<\sigma v\right>_{high}^{MC}/N_A\left<\sigma v\right>_{med}^{MC}$  \\
lower solid line	    \> $N_A\left<\sigma v\right>_{low}^{MC}/N_A\left<\sigma v\right>_{med}^{MC}$  \\
upper dashed line	\> $N_A\left<\sigma v\right>_{high}^{clas}/N_A\left<\sigma v\right>_{ad}^{clas}$  \\
lower dashed line    \> $N_A\left<\sigma v\right>_{low}^{clas}/N_A\left<\sigma v\right>_{ad}^{clas}$  \\
\\
{\it Bottom graph:}	\\  
\\
upper thin solid line	\> $N_A\left<\sigma v\right>_{high}^{MC}/N_A\left<\sigma v\right>_{ad}^{clas}$  \\
thick solid line   	\> $N_A\left<\sigma v\right>_{med}^{MC}/N_A\left<\sigma v\right>_{ad}^{clas}$  \\
lower thin solid line	\> $N_A\left<\sigma v\right>_{low}^{MC}/N_A\left<\sigma v\right>_{ad}^{clas}$  \\
\end{tabbing}
\clearpage
%
%
\begin{figure}[b]
\includegraphics[height=17.05cm]{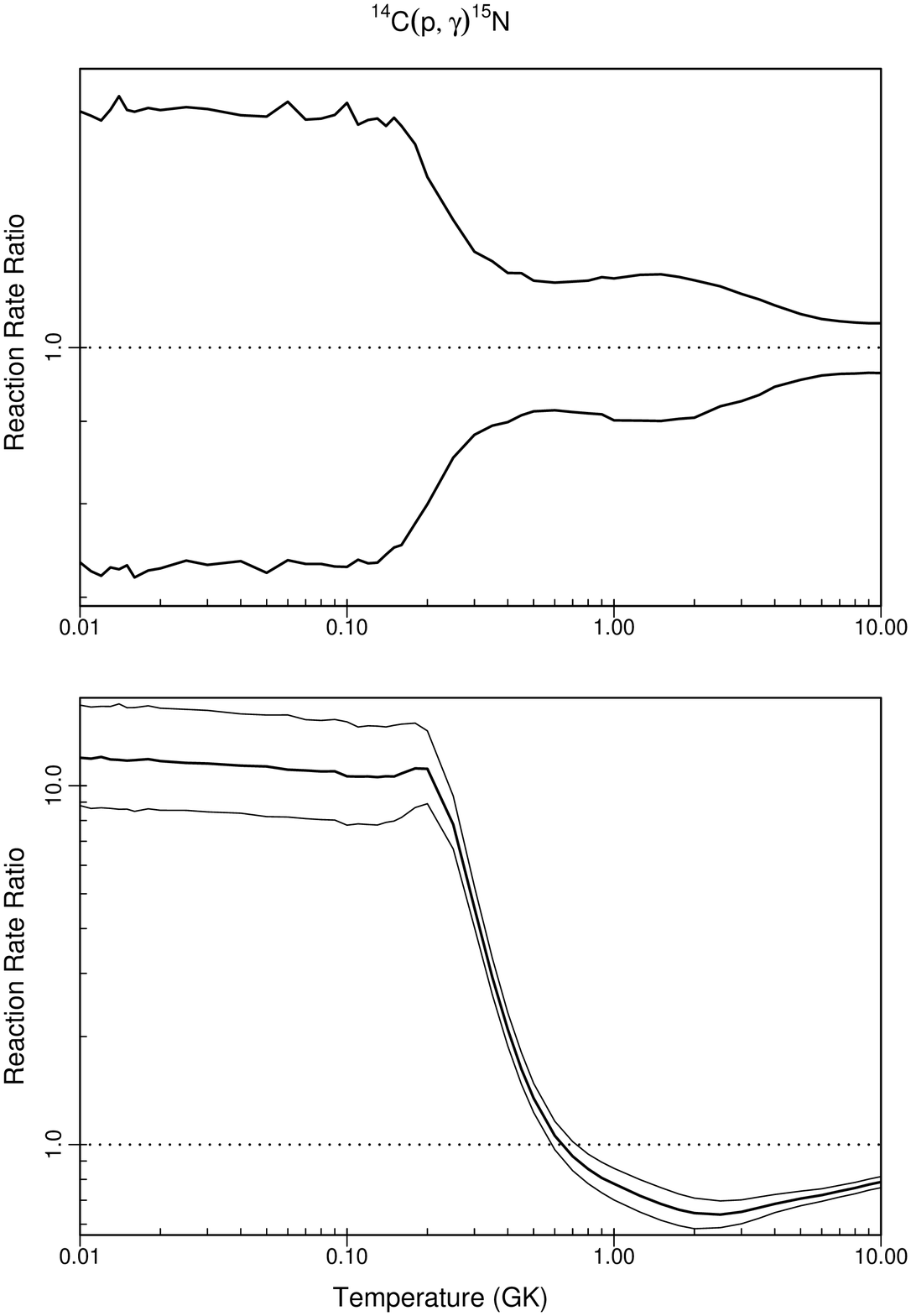}
\caption{\label{} 
Previous reaction rates: Ref. \cite{CF88}. No numerical reaction rates are presented in more recent work of Ref. \cite{GO90}.}
\end{figure}
\clearpage
\begin{figure}[b]
\includegraphics[height=17.05cm]{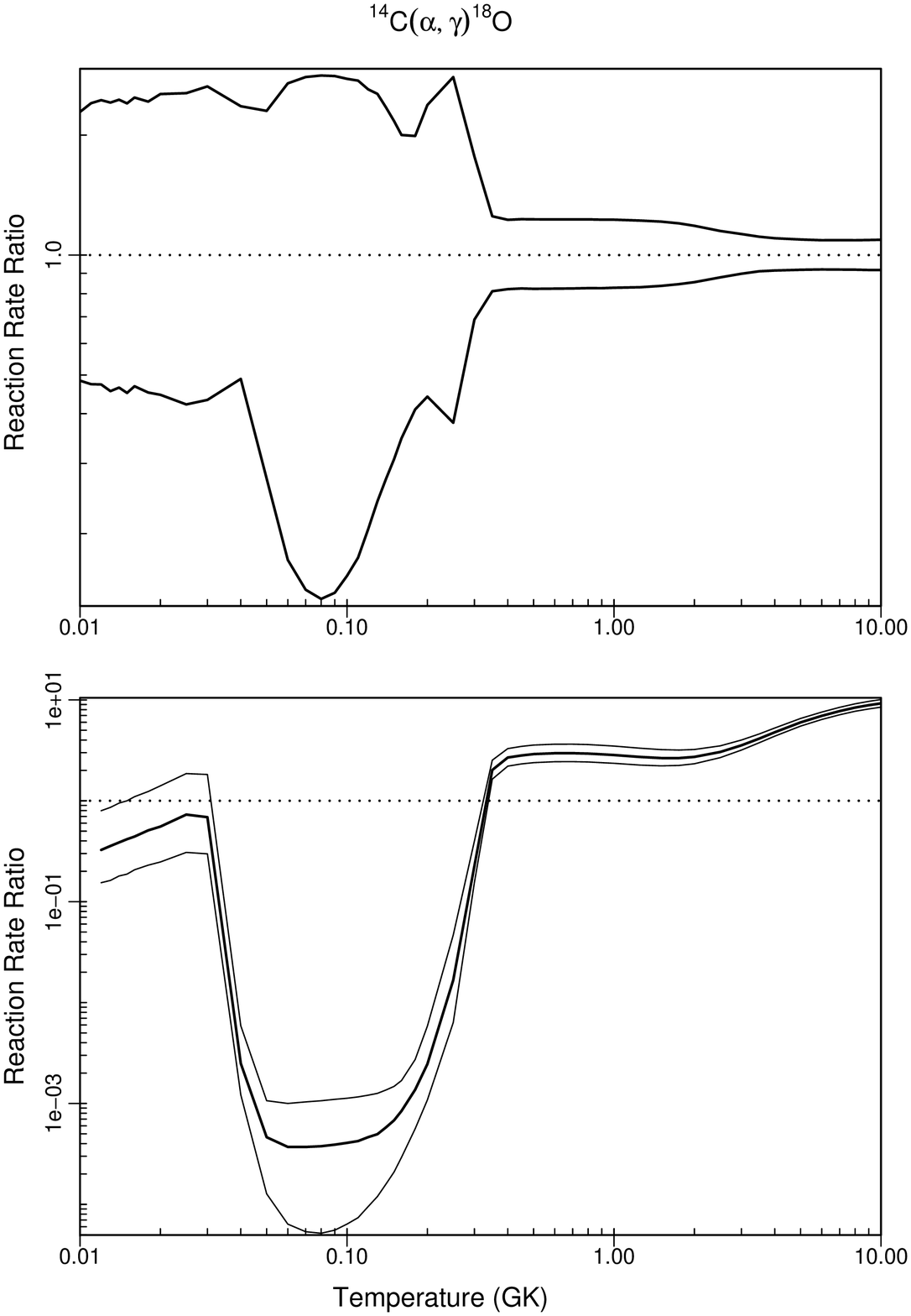}
\caption{\label{} 
Previous reaction rates: Ref. \cite{CF88}. No numerical reaction rates are presented in more recent works of Refs. \cite{BU07,GA87,GO92,LU04}.}
\end{figure}
\clearpage
\begin{figure}[b]
\includegraphics[height=17.05cm]{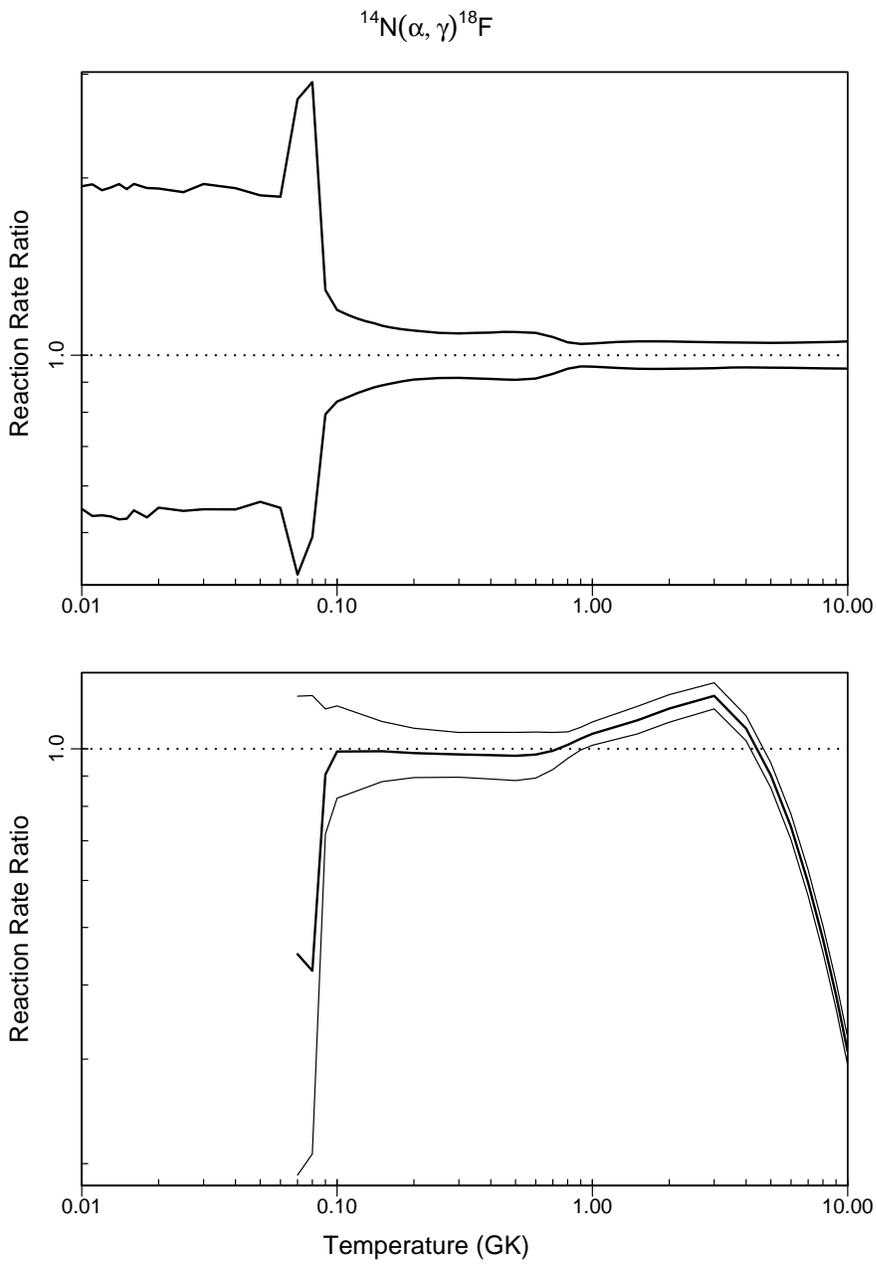}
\caption{\label{} 
Previous reaction rates: Ref. \cite{Goe00}. Rate uncertainties have not been determined in their work.}
\end{figure}
\clearpage
\begin{figure}[b]
\includegraphics[height=17.05cm]{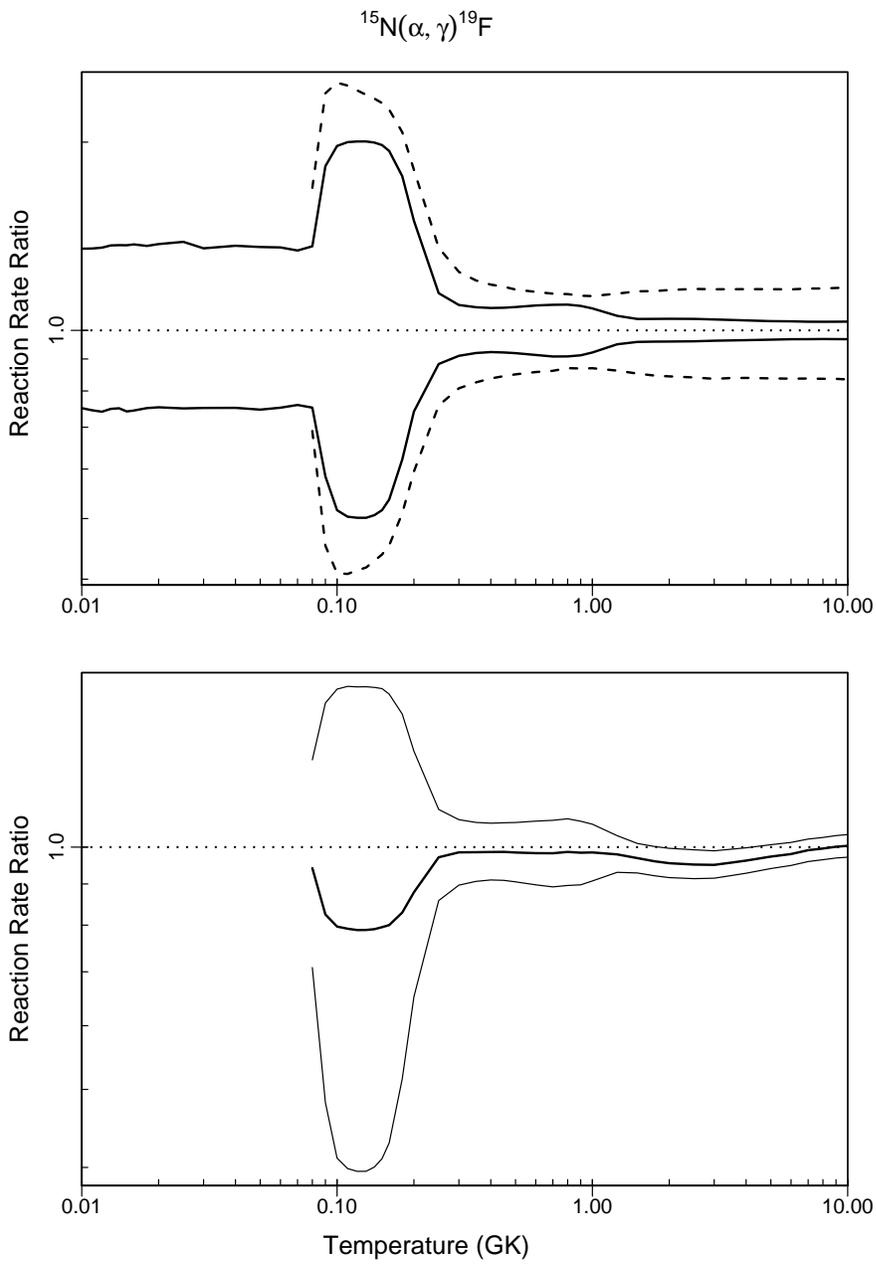}
\caption{\label{} 
Previous reaction rates: Ref. \cite{Ang99}.}
\end{figure}
\clearpage
\begin{figure}[b]
\includegraphics[height=17.05cm]{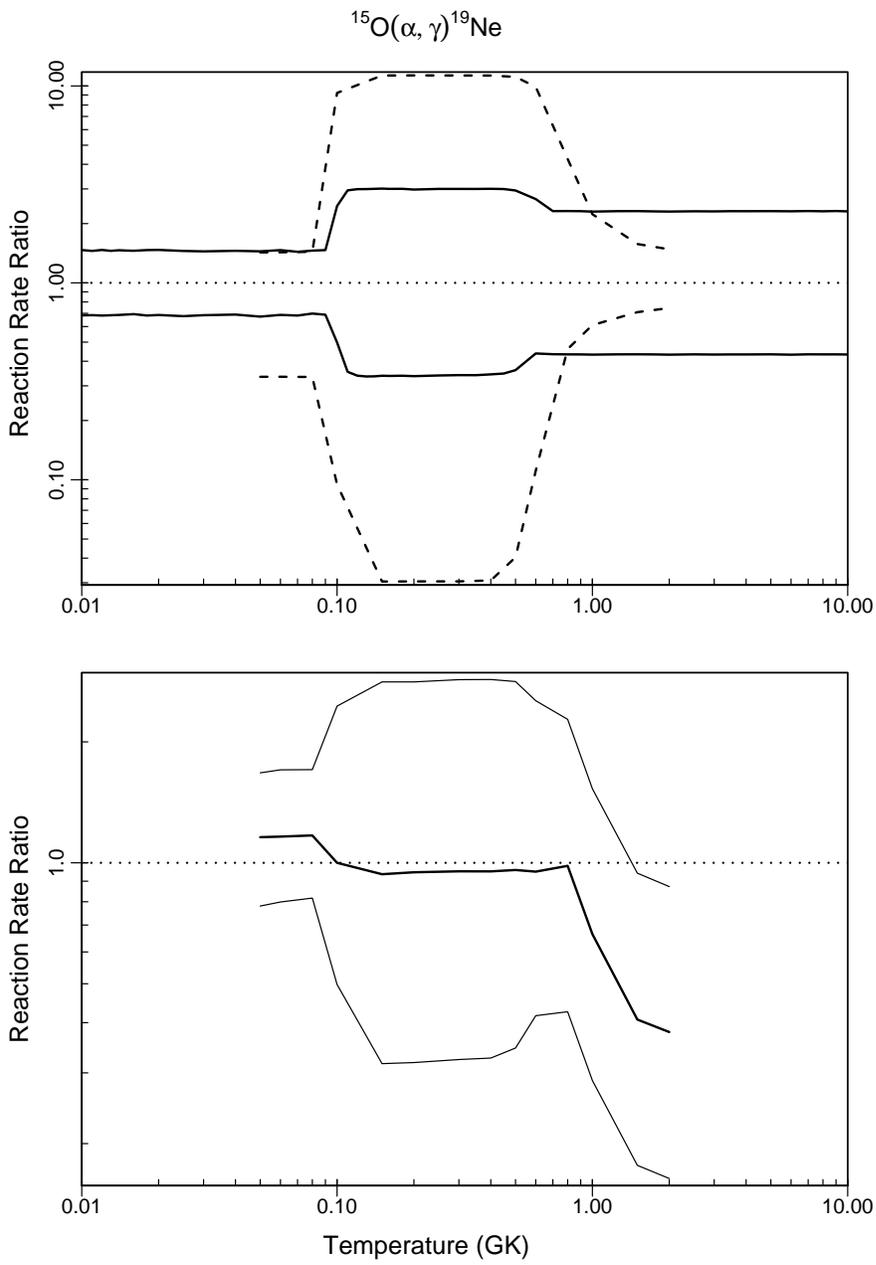}
\caption{\label{} 
Previous reaction rates: Ref. \cite{Fis06}.}
\end{figure}
\clearpage
\begin{figure}[b]
\includegraphics[height=17.05cm]{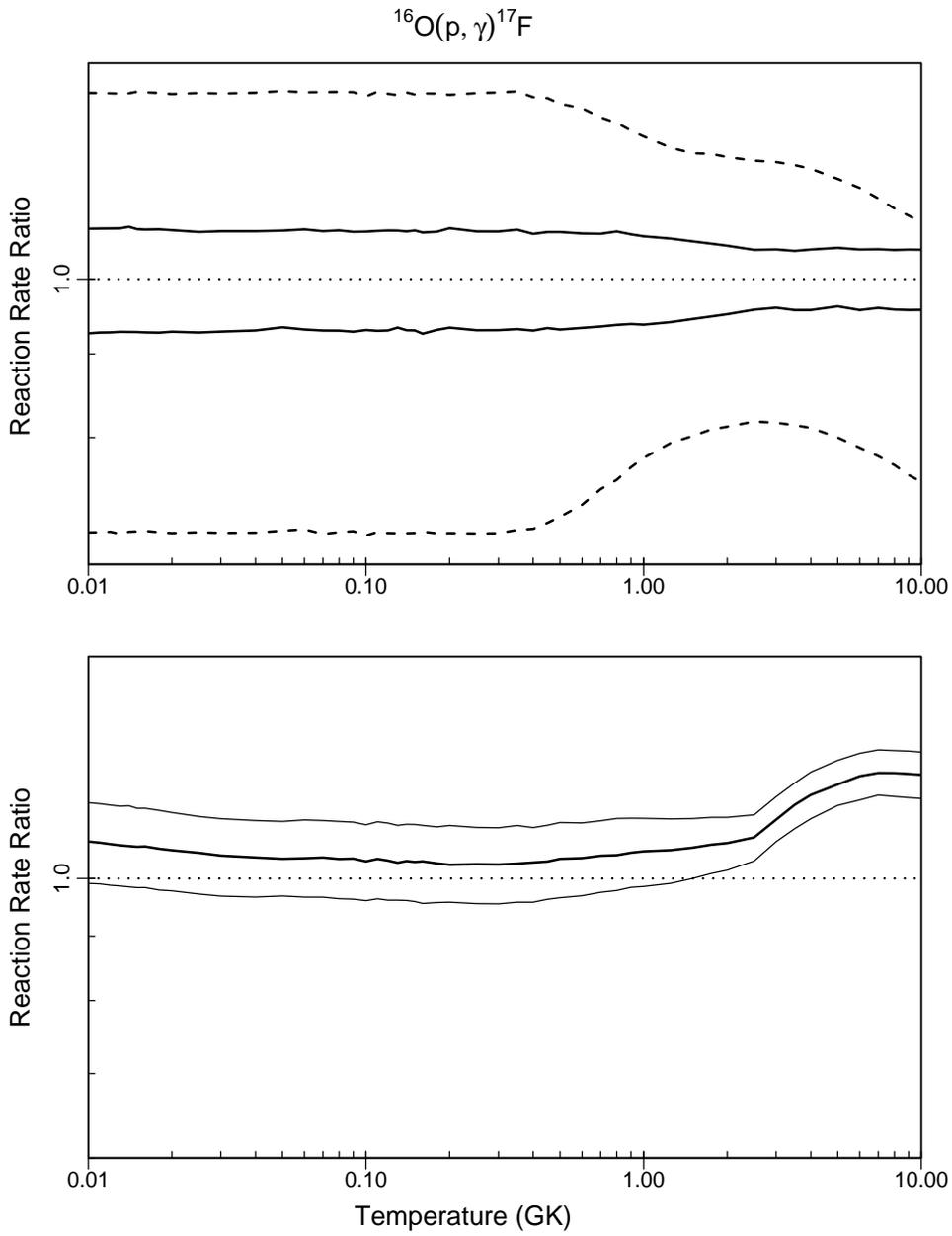}
\caption{\label{} 
Previous reaction rates: Ref. \cite{Ang99}.}
\end{figure}
\clearpage
\begin{figure}[b]
\includegraphics[height=17.05cm]{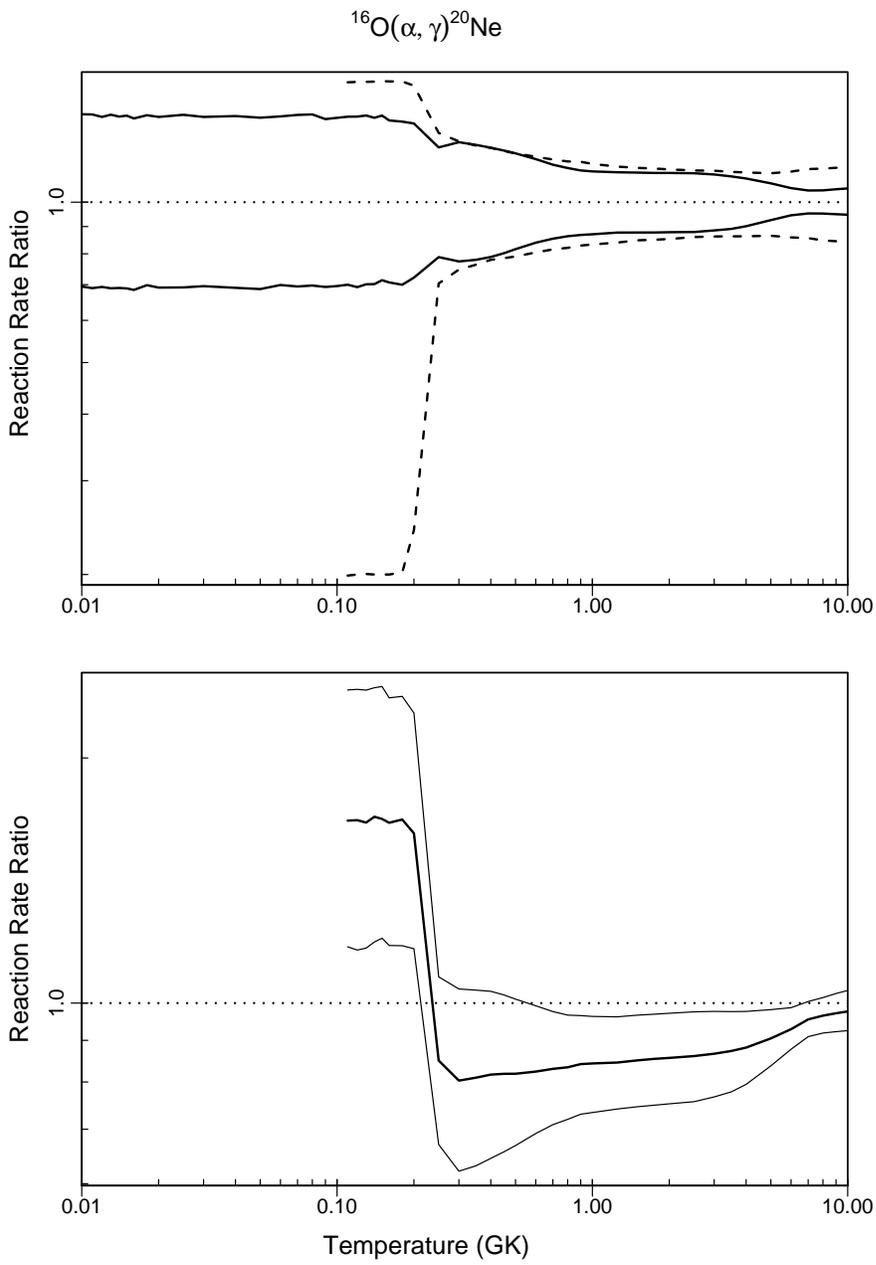}
\caption{\label{} 
Previous reaction rates: Ref. \cite{Ang99}.}
\end{figure}
\clearpage
\begin{figure}[b]
\includegraphics[height=17.05cm]{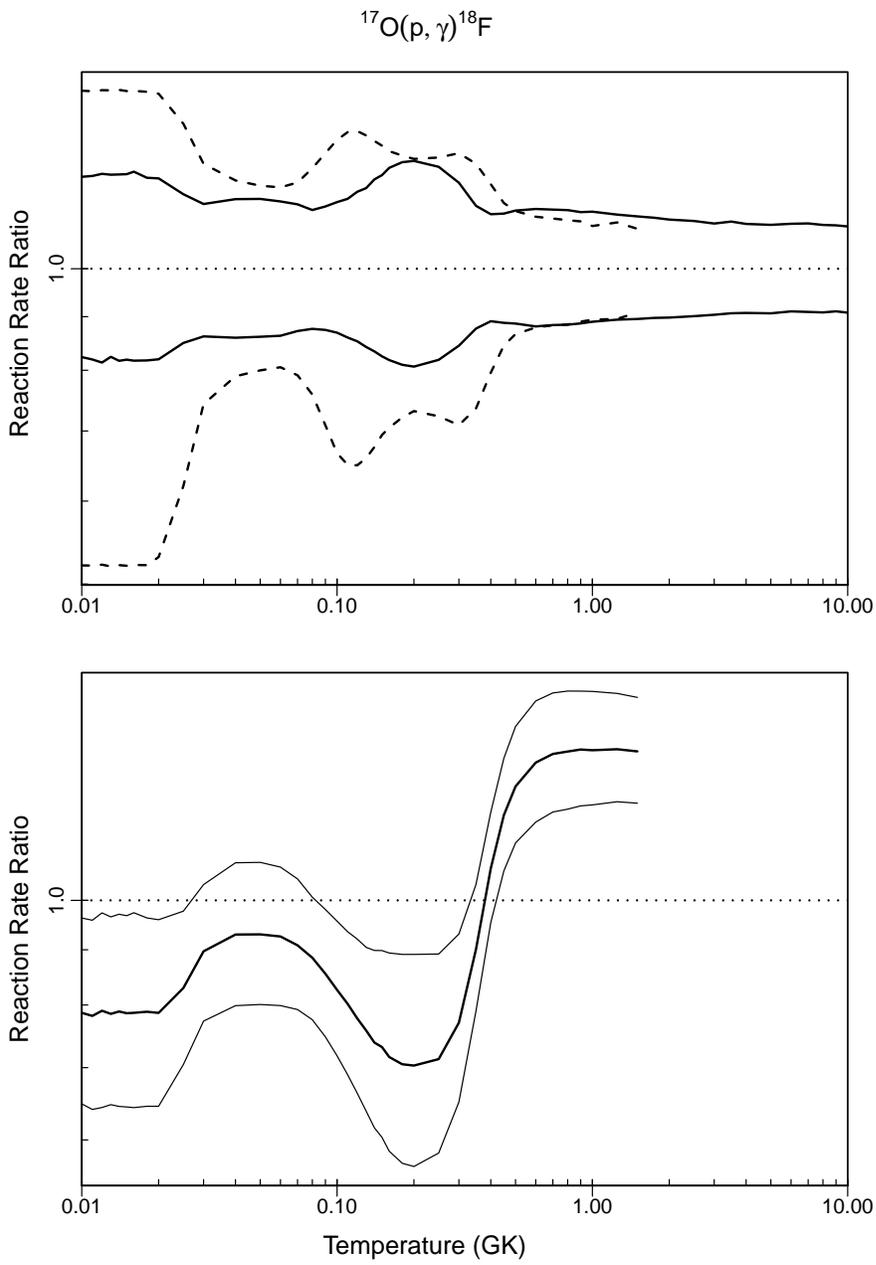}
\caption{\label{} 
Previous reaction rates: Ref. \cite{Cha07}.}
\end{figure}
\clearpage
\begin{figure}[b]
\includegraphics[height=17.05cm]{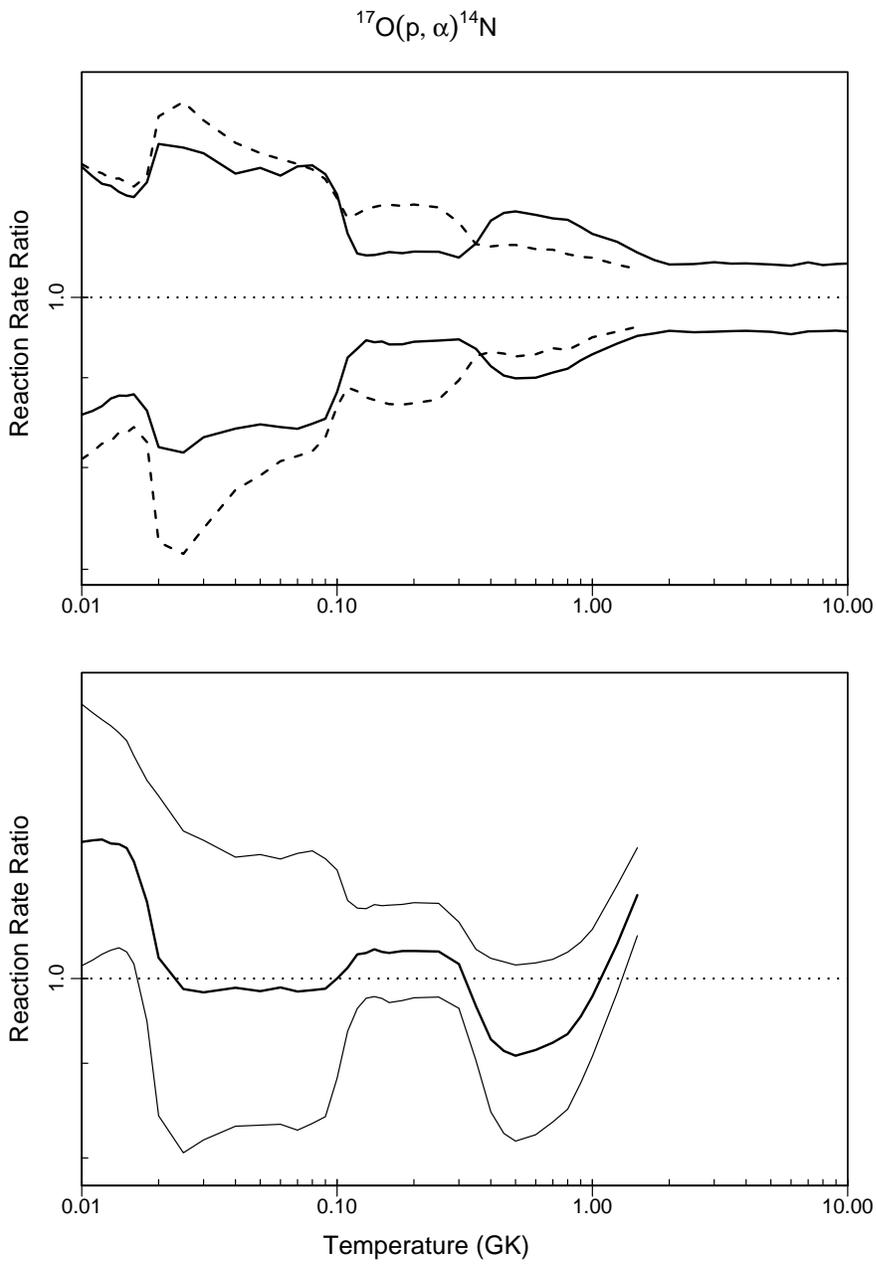}
\caption{\label{} 
Previous reaction rates: Ref. \cite{Cha07}.}
\end{figure}
\clearpage
\begin{figure}[b]
\includegraphics[height=17.05cm]{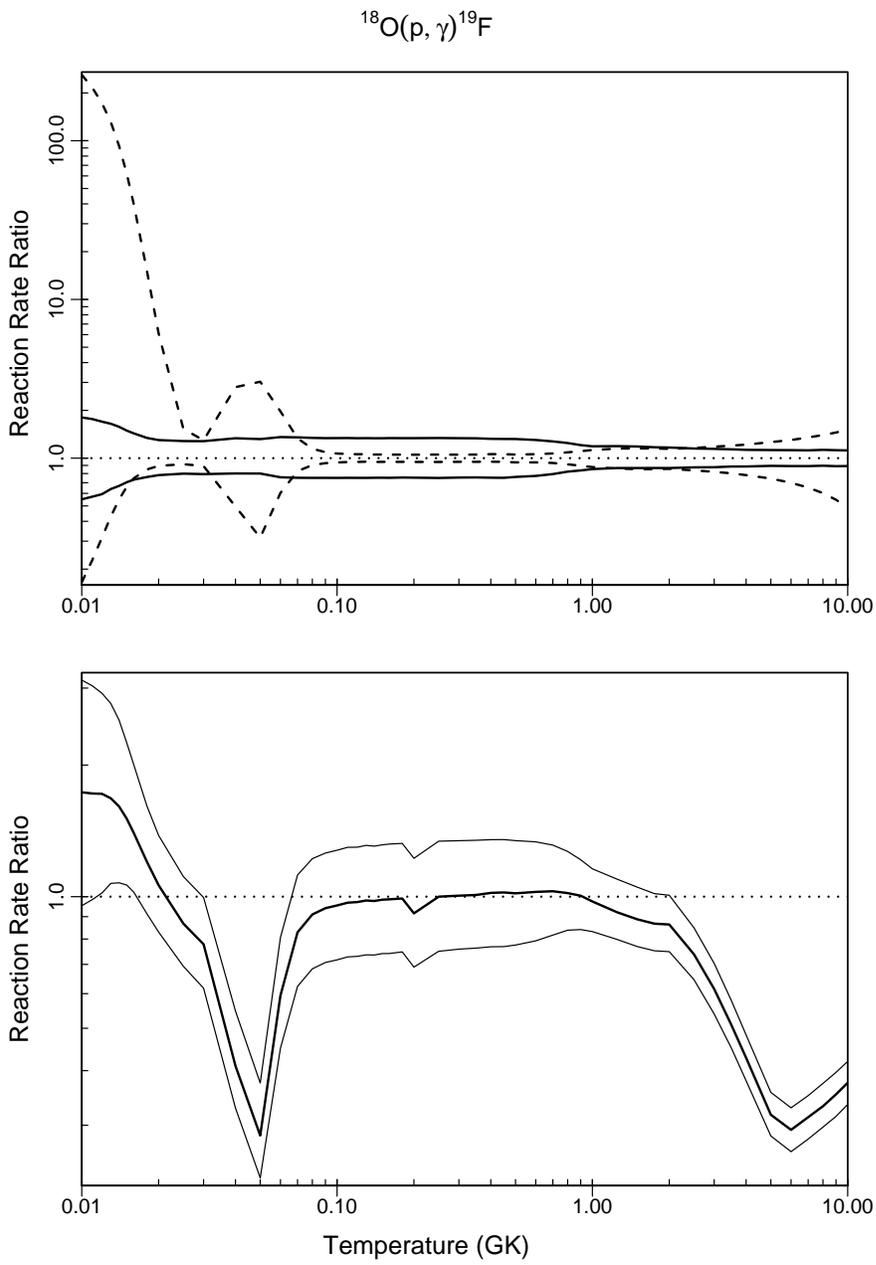}
\caption{\label{} 
Previous reaction rates: Ref. \cite{Ang99}.}
\end{figure}
\clearpage
\begin{figure}[b]
\includegraphics[height=17.05cm]{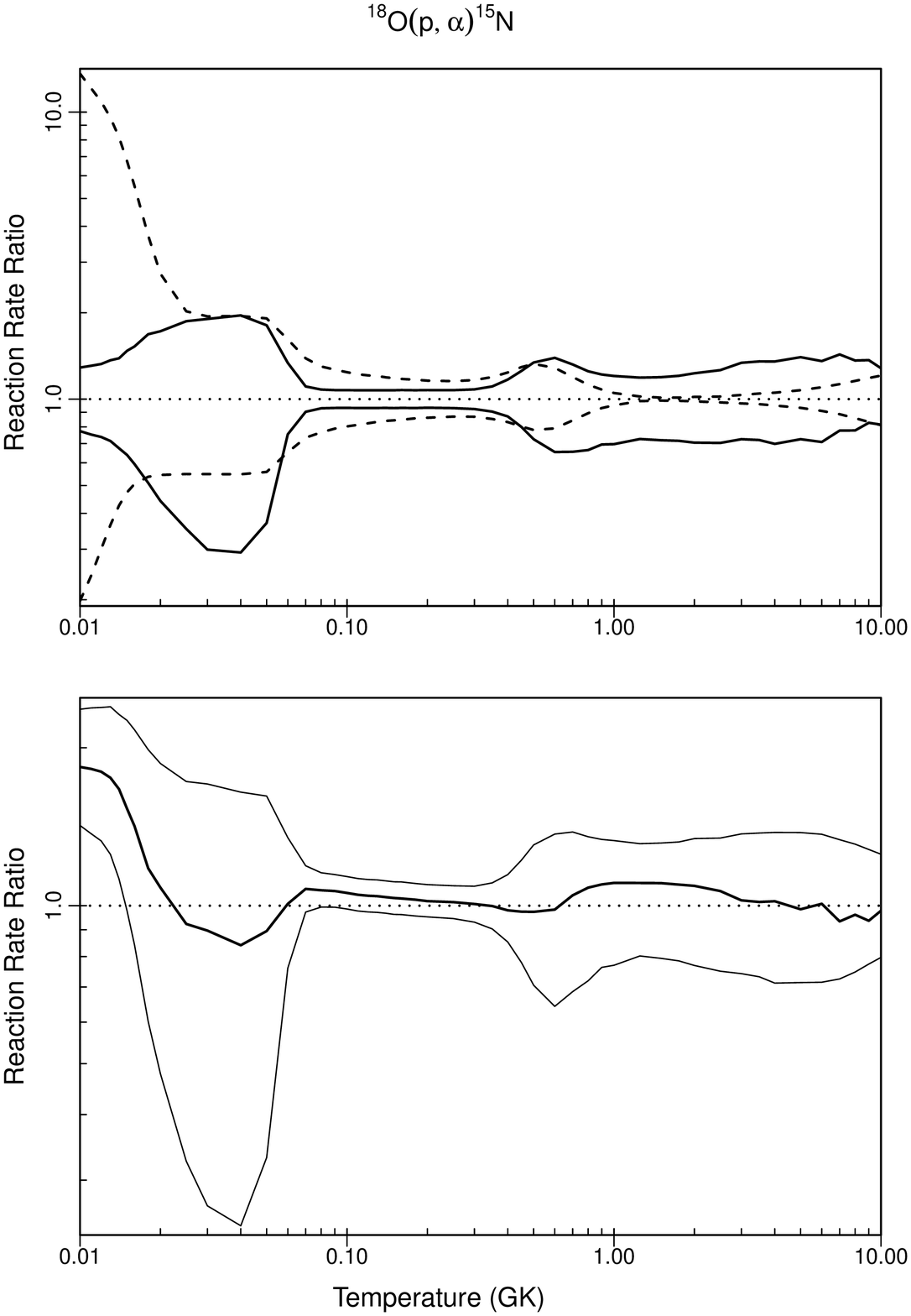}
\caption{\label{} 
Previous reaction rates: Ref. \cite{Ang99}. A more recent rate is displayed in Fig. 3 of Ref. \cite{La08}, but numerical rate values are not available.}
\end{figure}
\clearpage
\begin{figure}[b]
\includegraphics[height=17.05cm]{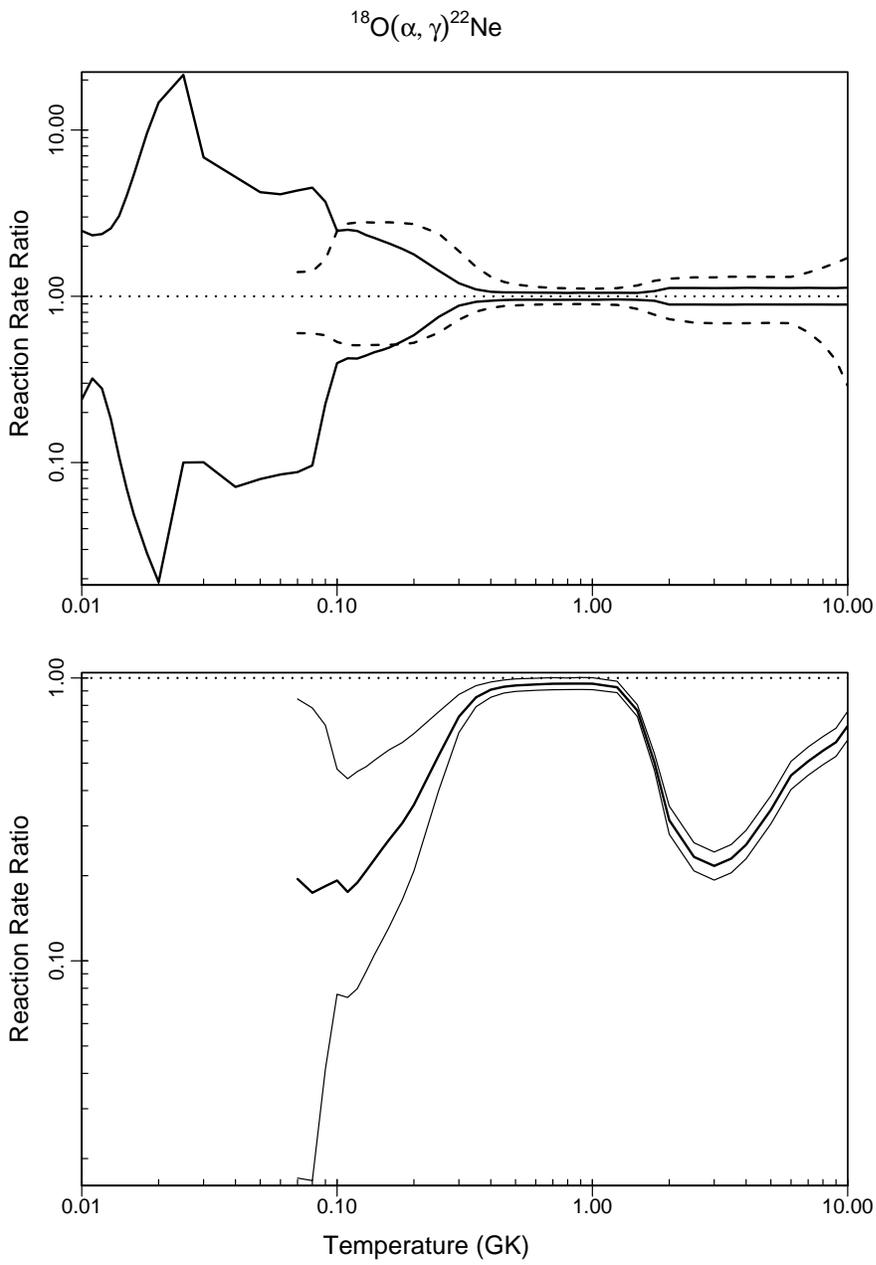}
\caption{\label{} 
Previous reaction rates: Ref. \cite{Ang99}.}
\end{figure}
\clearpage
\begin{figure}[b]
\includegraphics[height=17.05cm]{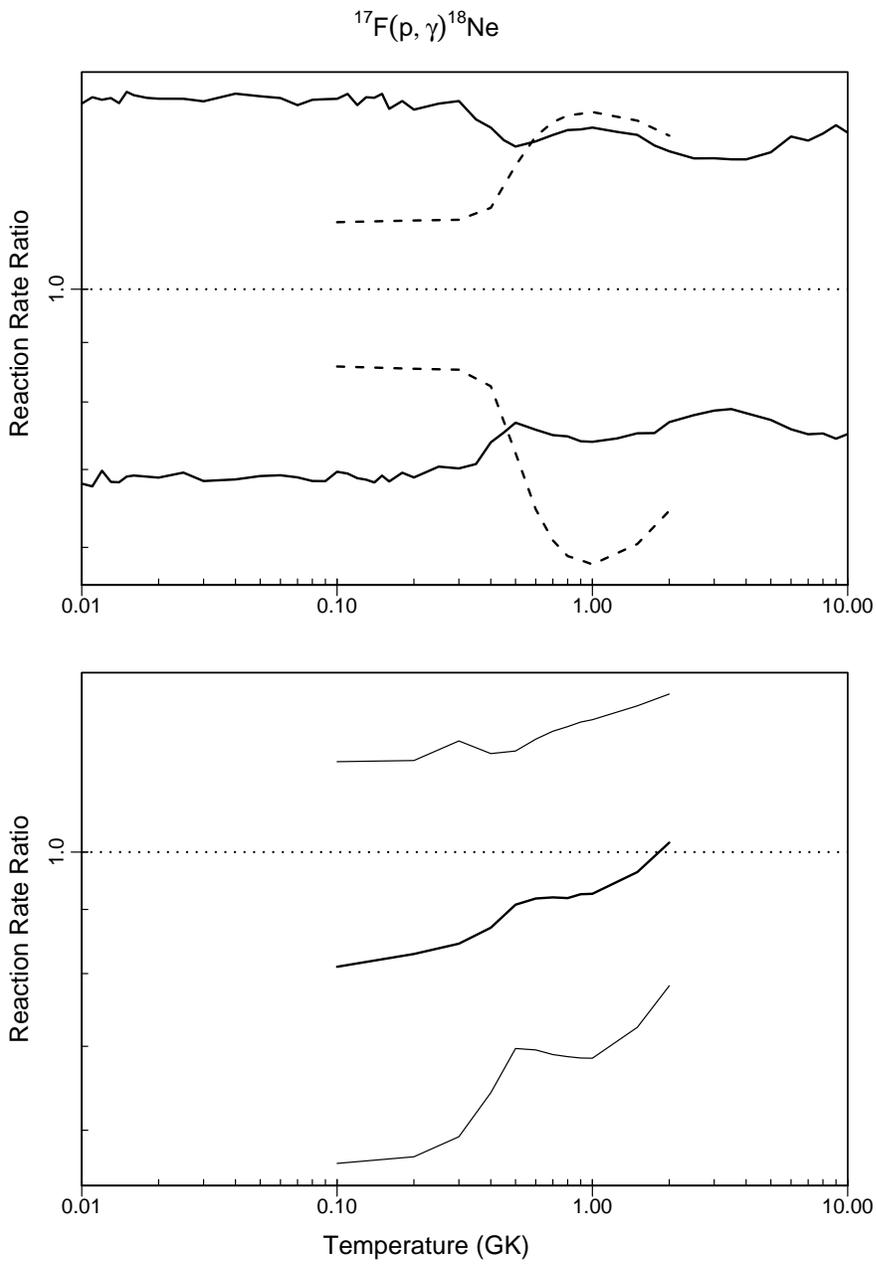}
\caption{\label{} 
Previous reaction rates: Ref. \cite{Bar00}.}
\end{figure}
\clearpage
\begin{figure}[b]
\includegraphics[height=17.05cm]{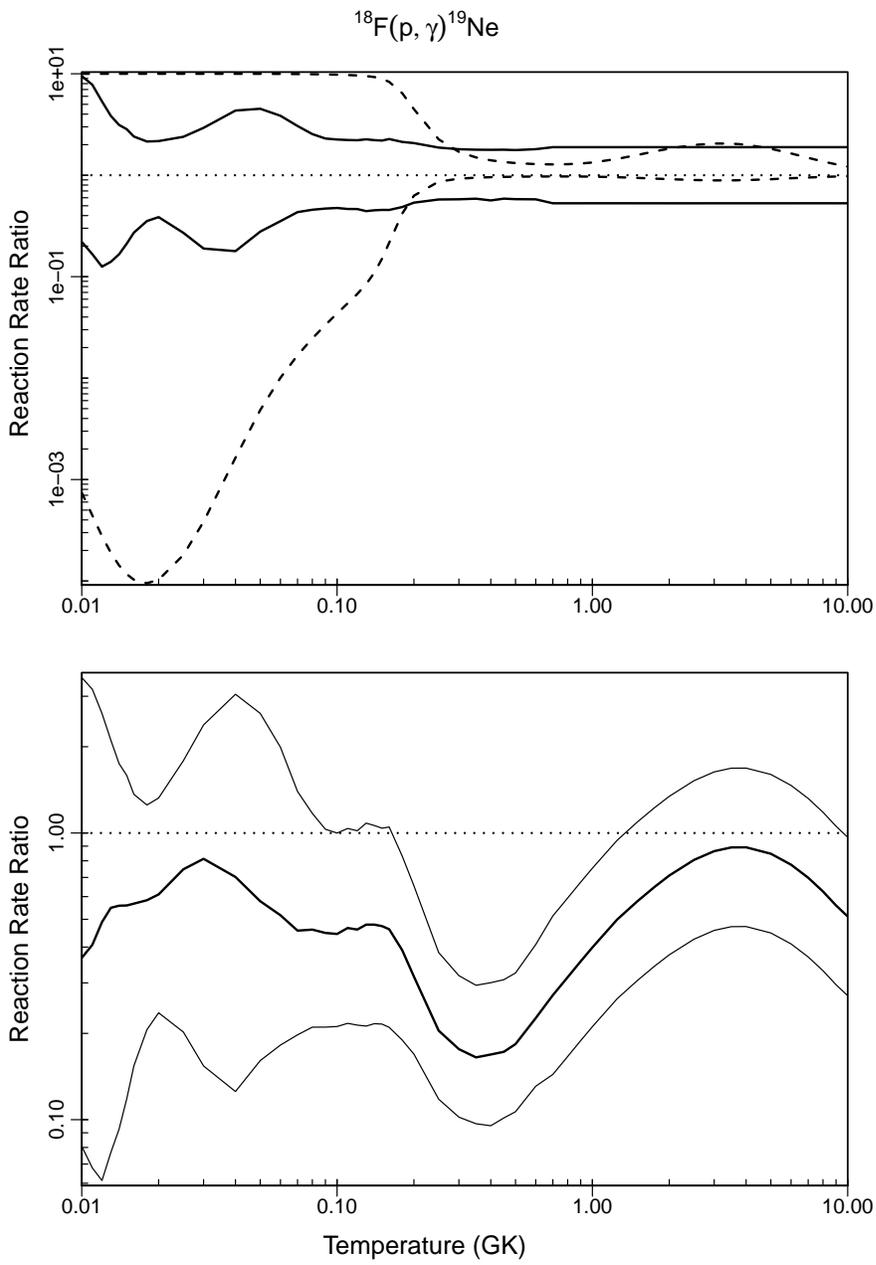}
\caption{\label{} 
Previous reaction rates: Ref. \cite{CO00}.}
\end{figure}
\clearpage
\begin{figure}[b]
\includegraphics[height=17.05cm]{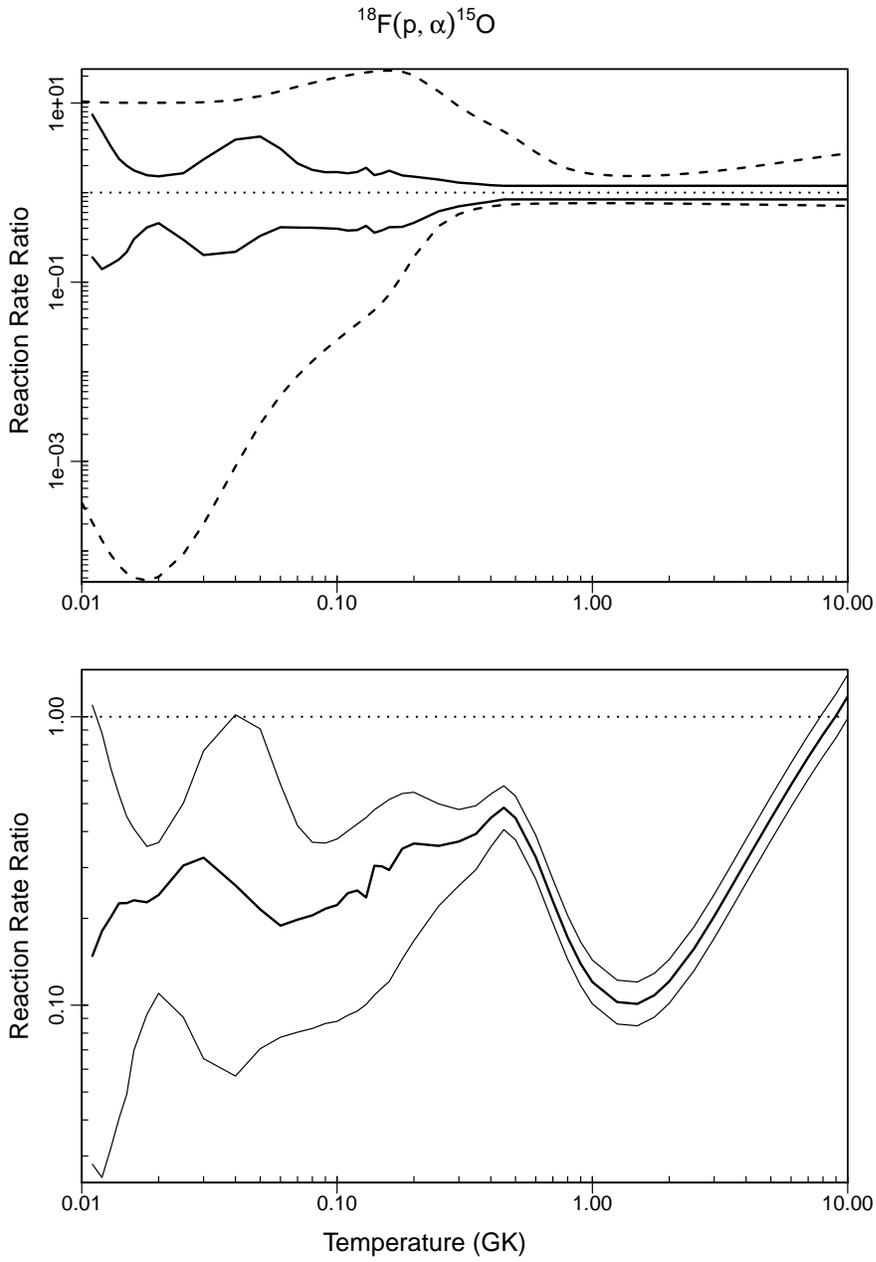}
\caption{\label{} 
Previous reaction rates: Ref. \cite{CO00}.}
\end{figure}
\clearpage
\begin{figure}[b]
\includegraphics[height=17.05cm]{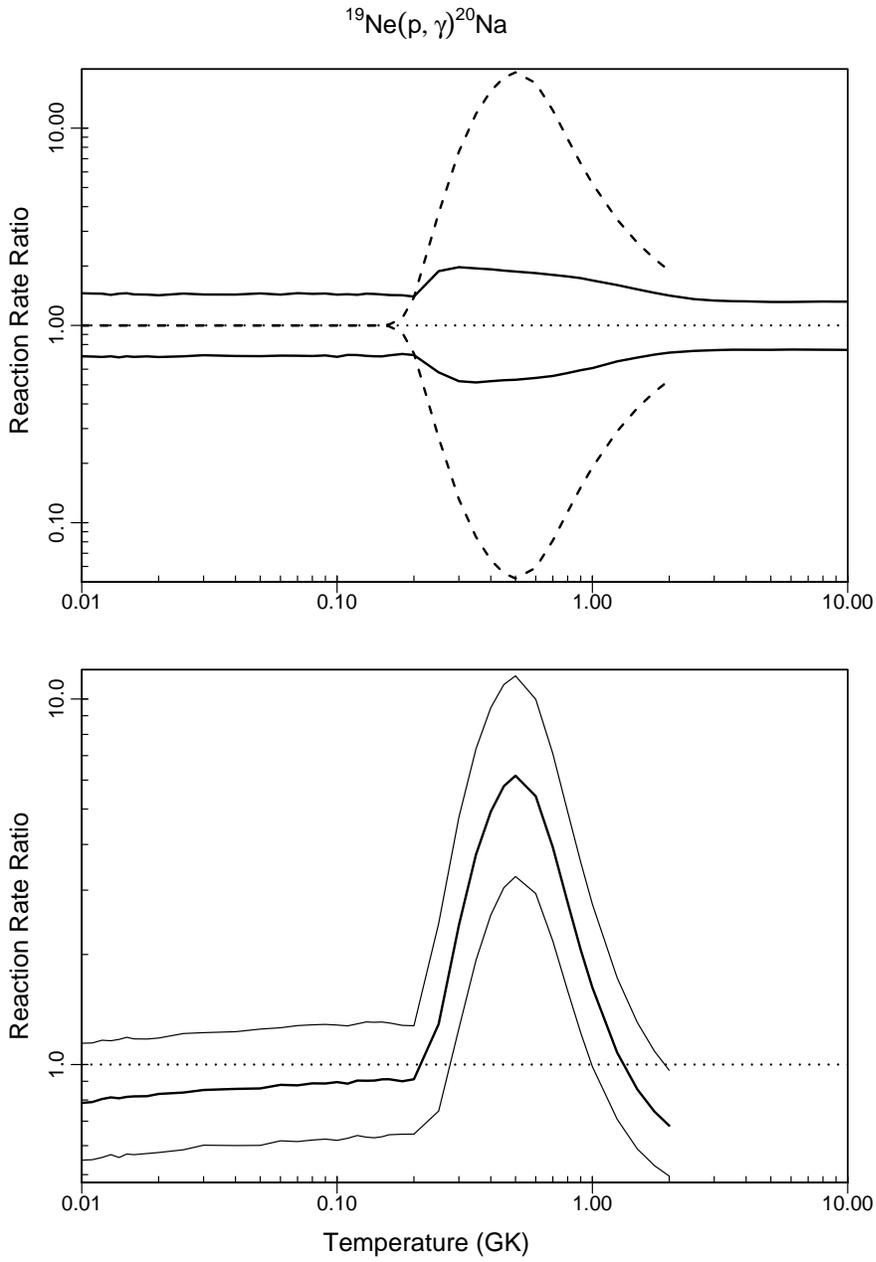}
\caption{\label{} 
Previous reaction rates: Ref. \cite{VA98}. Only lower and upper rate limits are shown in Fig. 7 of Ref. \cite{VA98}; we estimated the previous adopted rate by using the geometric mean of the rate limit values.}
\end{figure}
\clearpage
\begin{figure}[b]
\includegraphics[height=17.05cm]{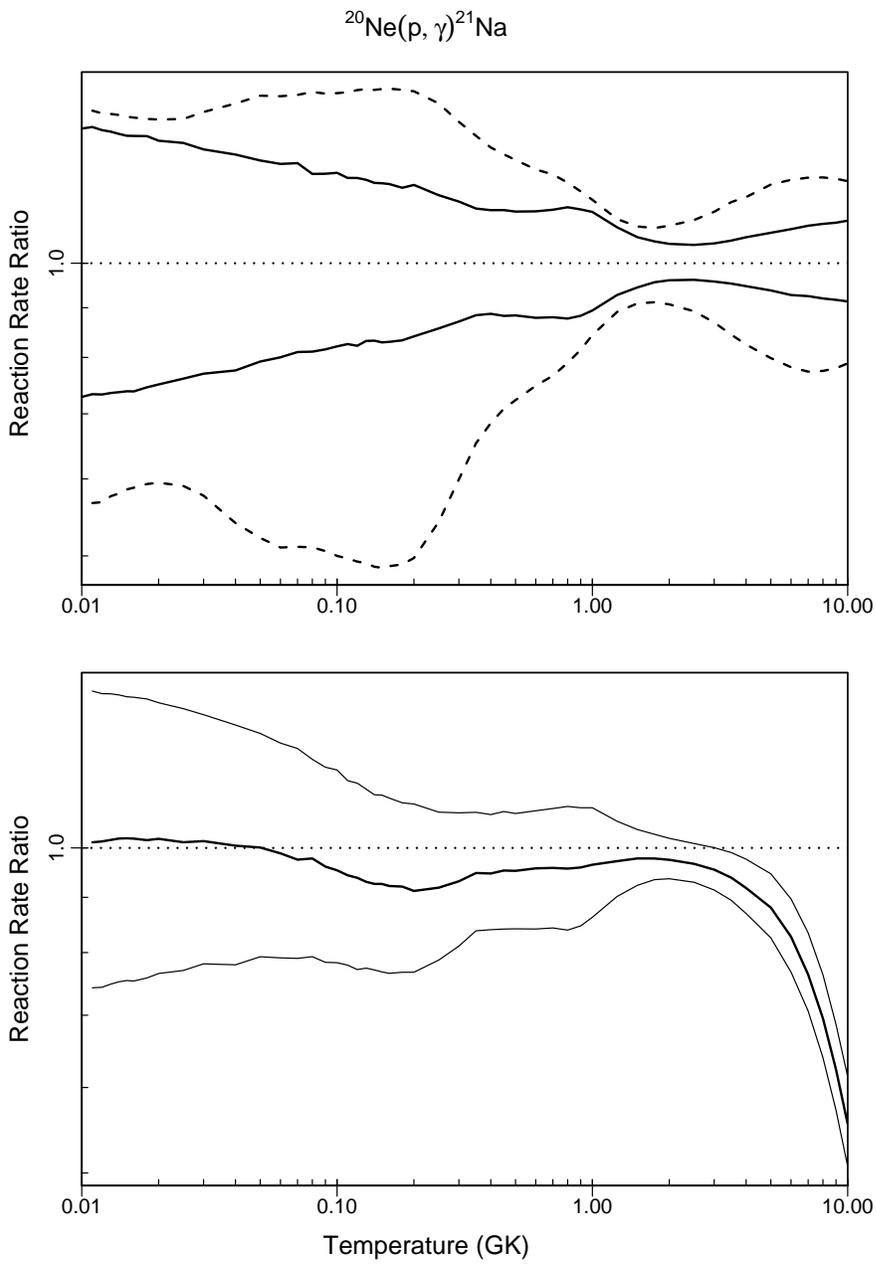}
\caption{\label{} 
Previous reaction rates: Ref. \cite{Ang99}.}
\end{figure}
\clearpage
\begin{figure}[b]
\includegraphics[height=17.05cm]{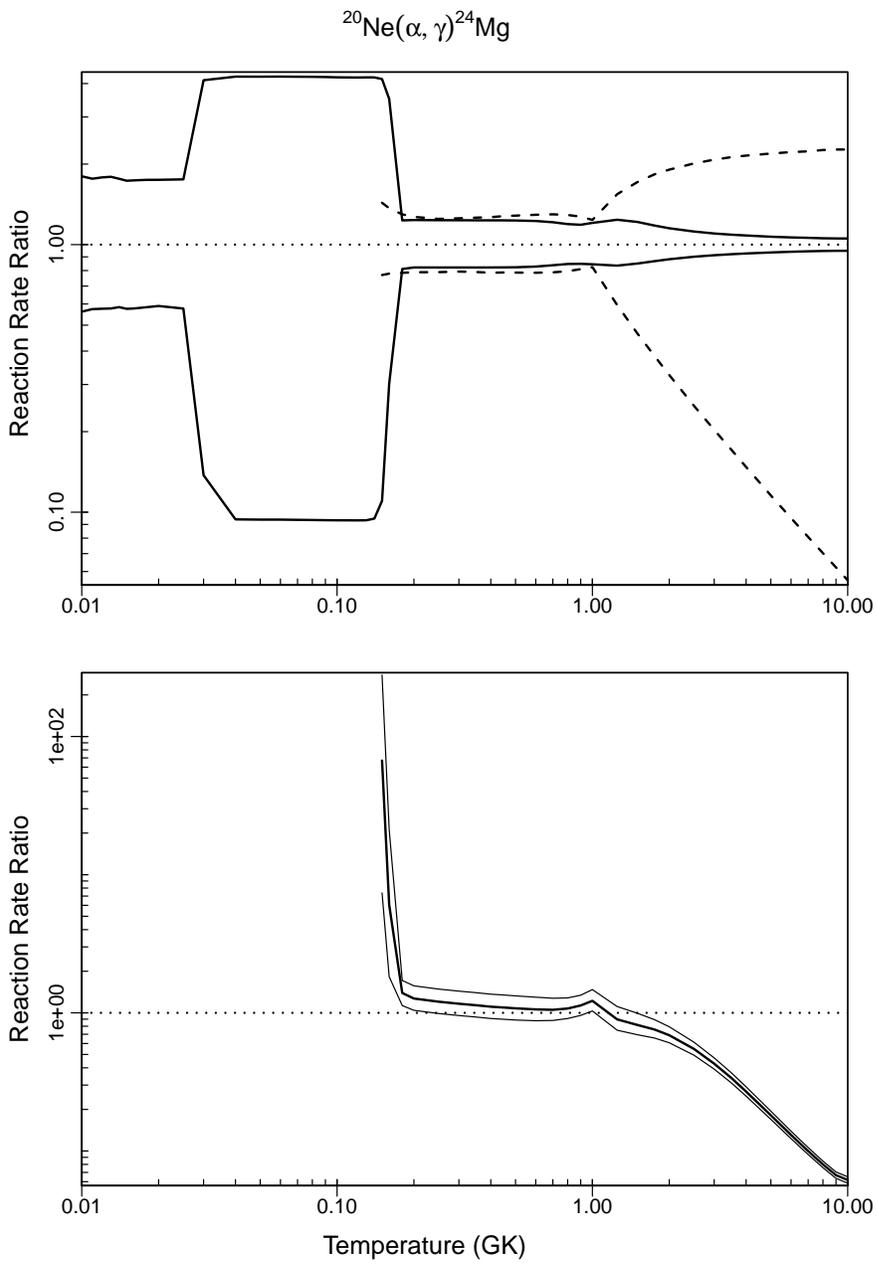}
\caption{\label{} 
Previous reaction rates: Ref. \cite{Ang99}.}
\end{figure}
\clearpage
\begin{figure}[b]
\includegraphics[height=17.05cm]{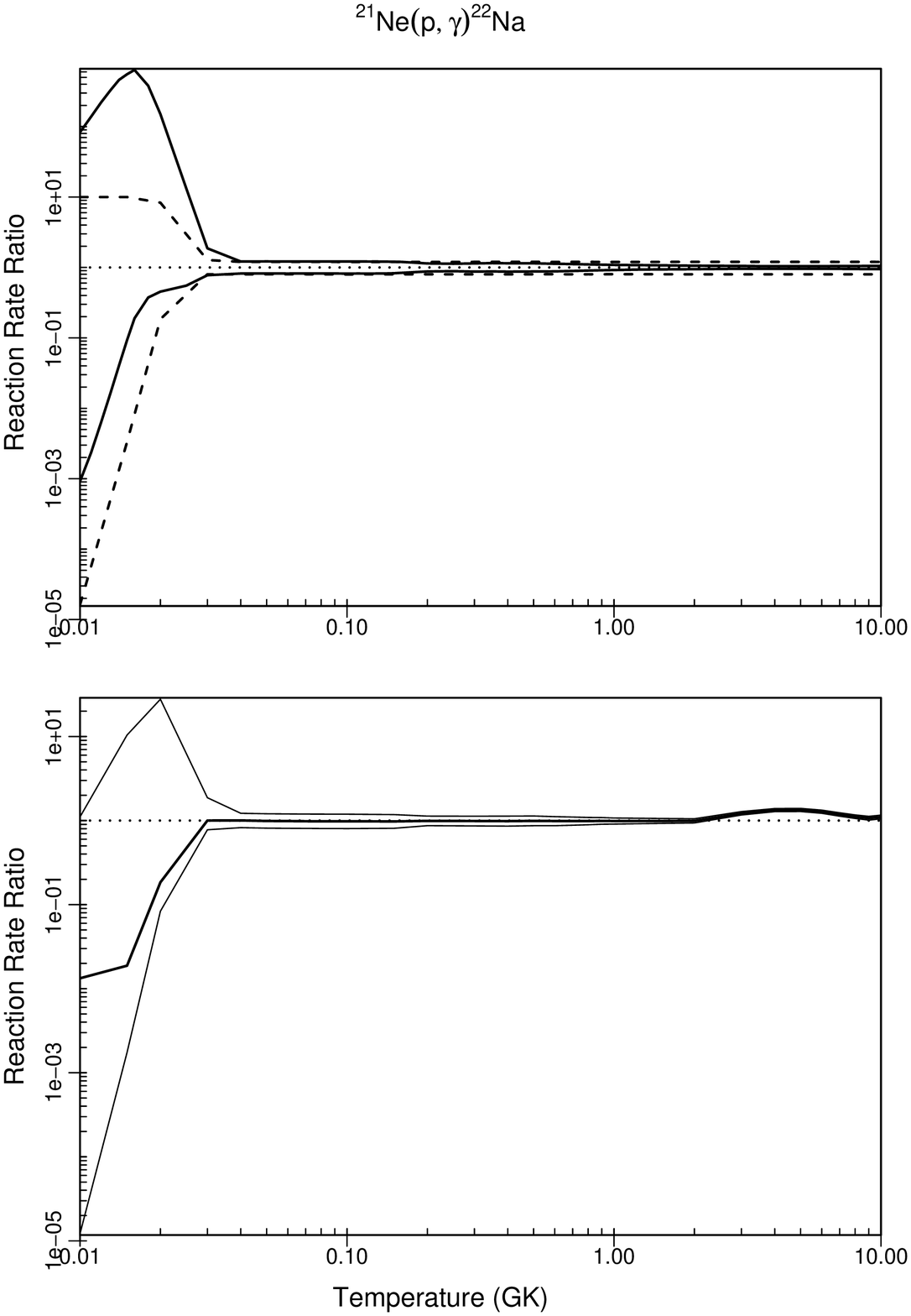}
\caption{\label{} 
Previous reaction rates: Ref. \cite{Ili01}.}
\end{figure}
\clearpage
\begin{figure}[b]
\includegraphics[height=17.05cm]{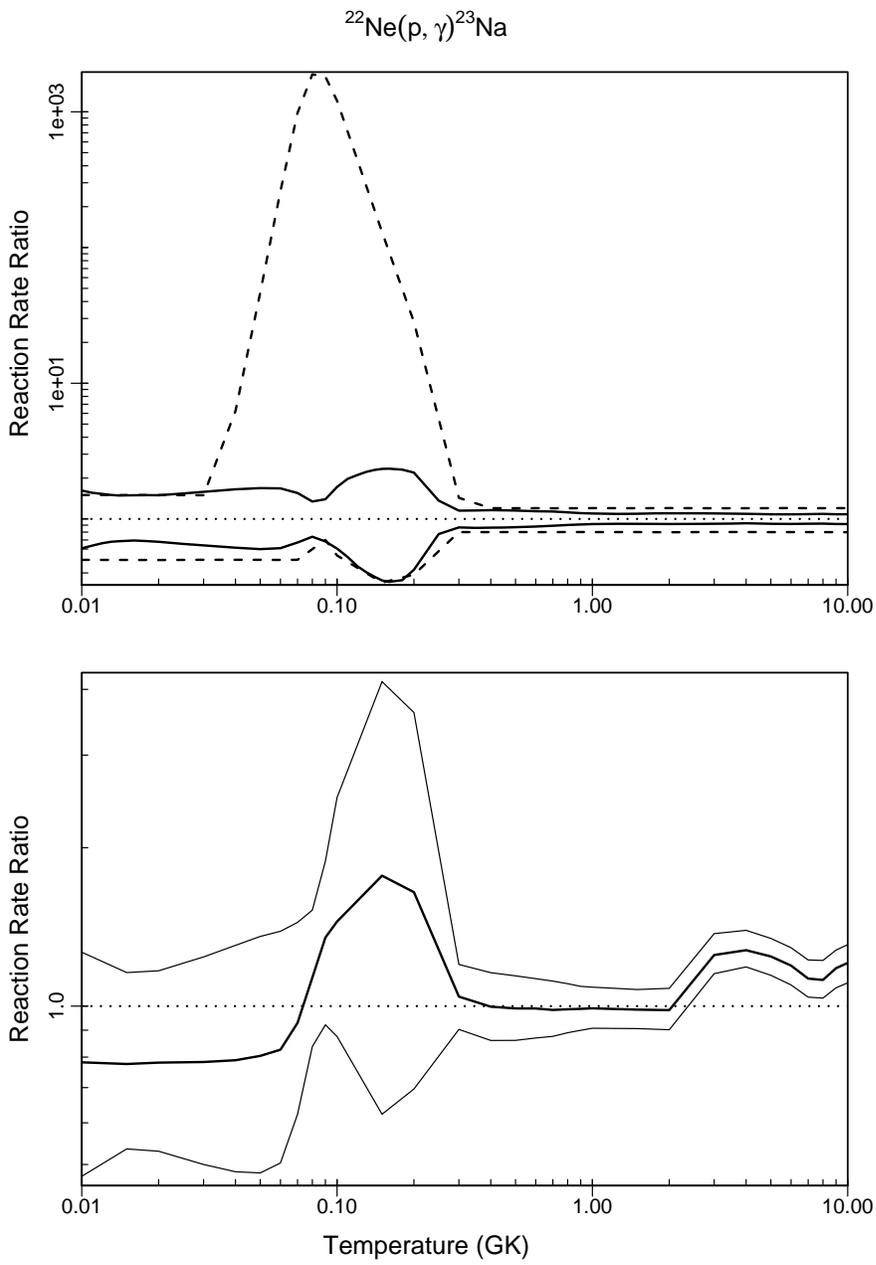}
\caption{\label{} 
Previous reaction rates: Ref. \cite{Ili01}.}
\end{figure}
\clearpage
\begin{figure}[b]
\includegraphics[height=17.05cm]{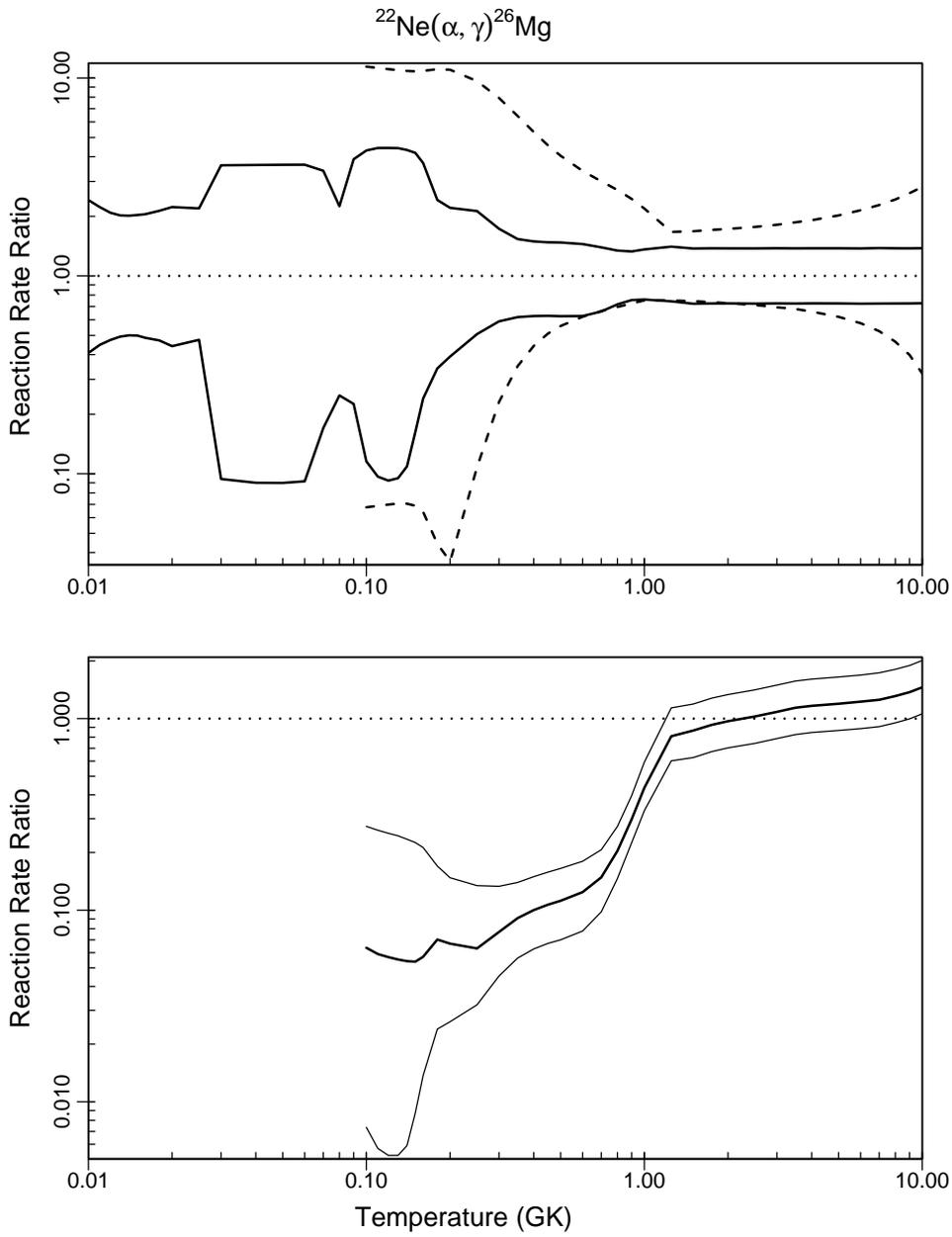}
\caption{\label{} 
Previous reaction rates: Ref. \cite{Ang99}.}
\end{figure}
\clearpage
\begin{figure}[b]
\includegraphics[height=17.05cm]{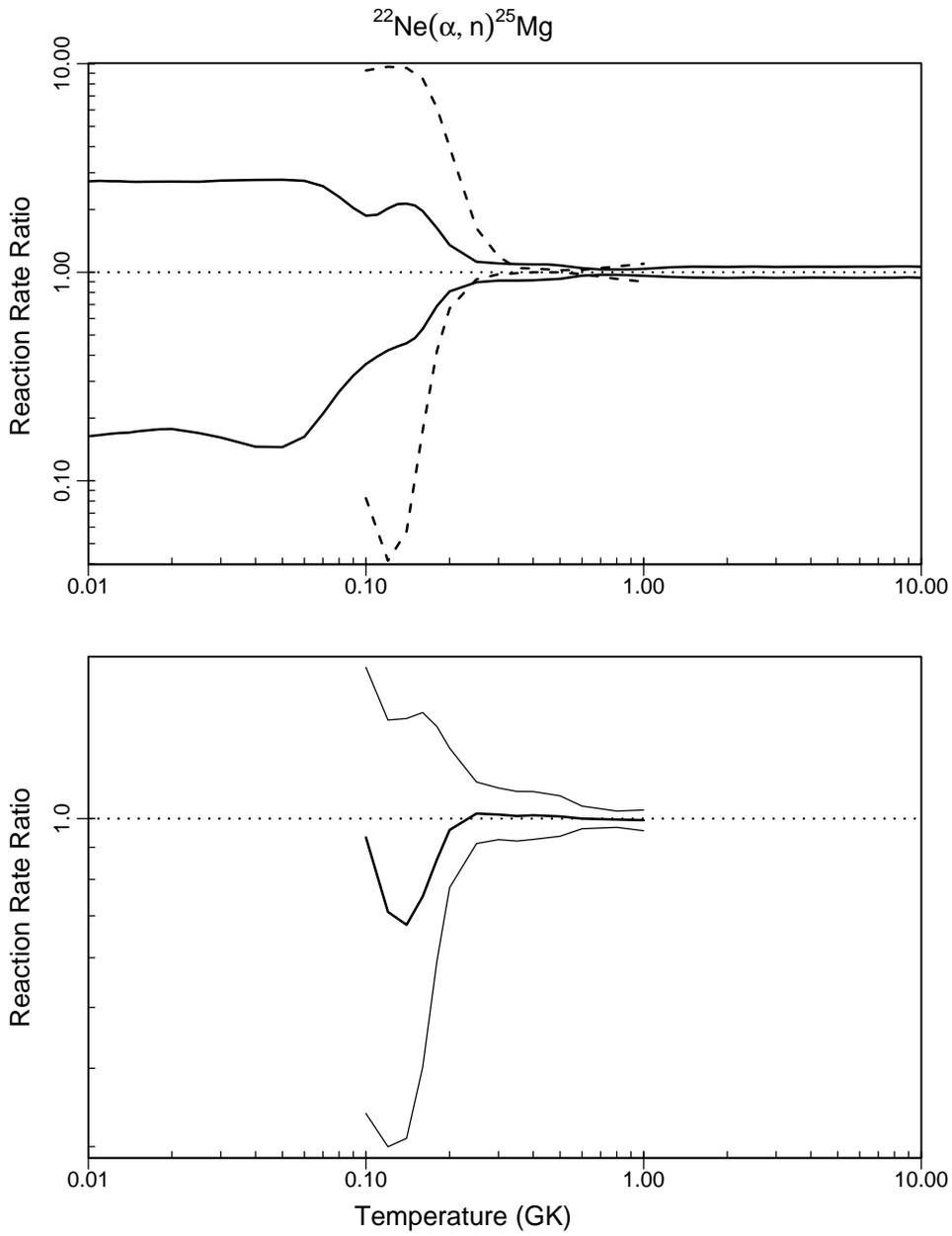}
\caption{\label{} 
Previous reaction rates: Ref. \cite{Jae01}.}
\end{figure}
\clearpage
\begin{figure}[b]
\includegraphics[height=17.05cm]{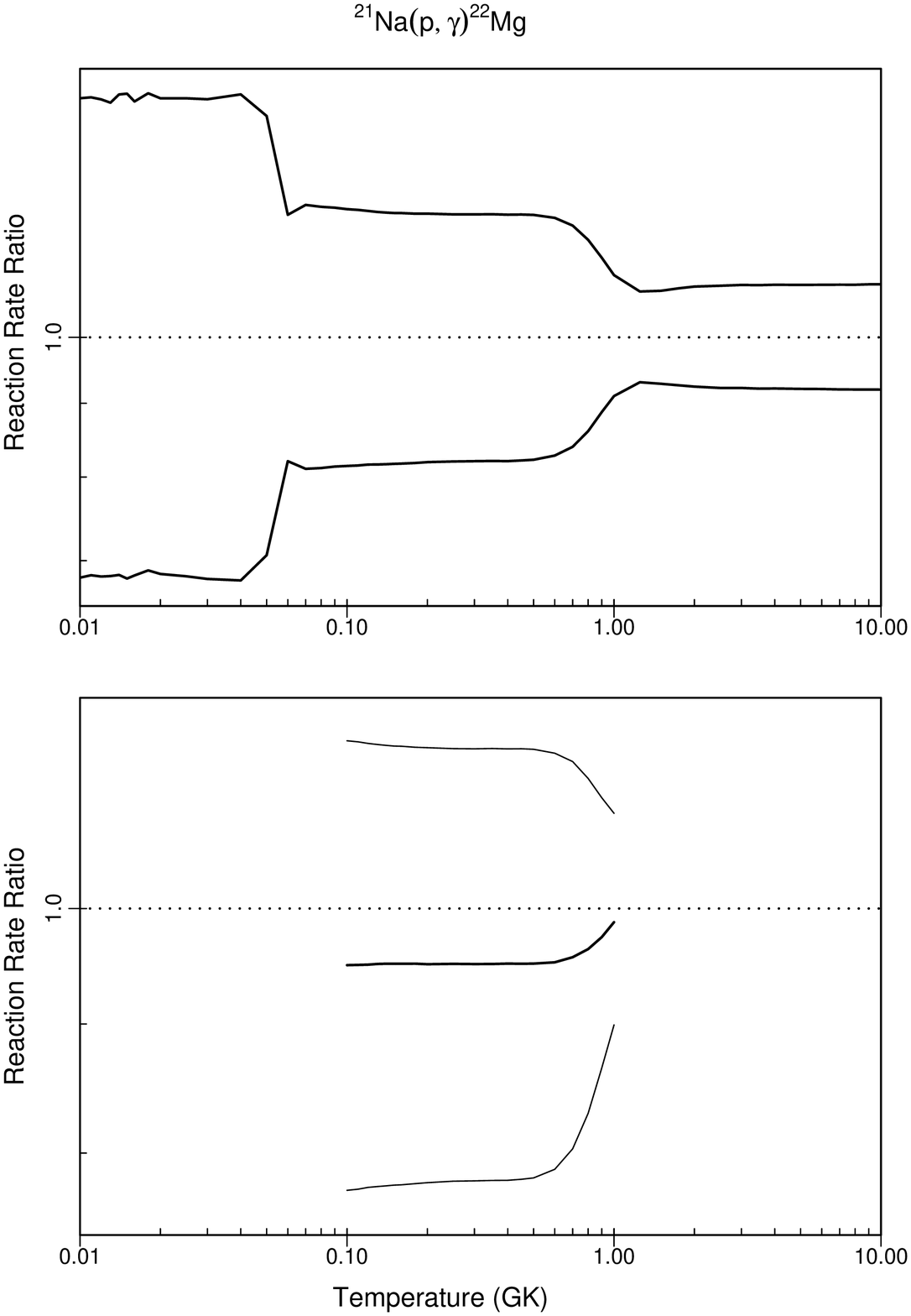}
\caption{\label{} 
Previous reaction rates: Ref. \cite{DAu04}. Rate uncertainties have not been determined previously.}
\end{figure}
\clearpage
\begin{figure}[b]
\includegraphics[height=17.05cm]{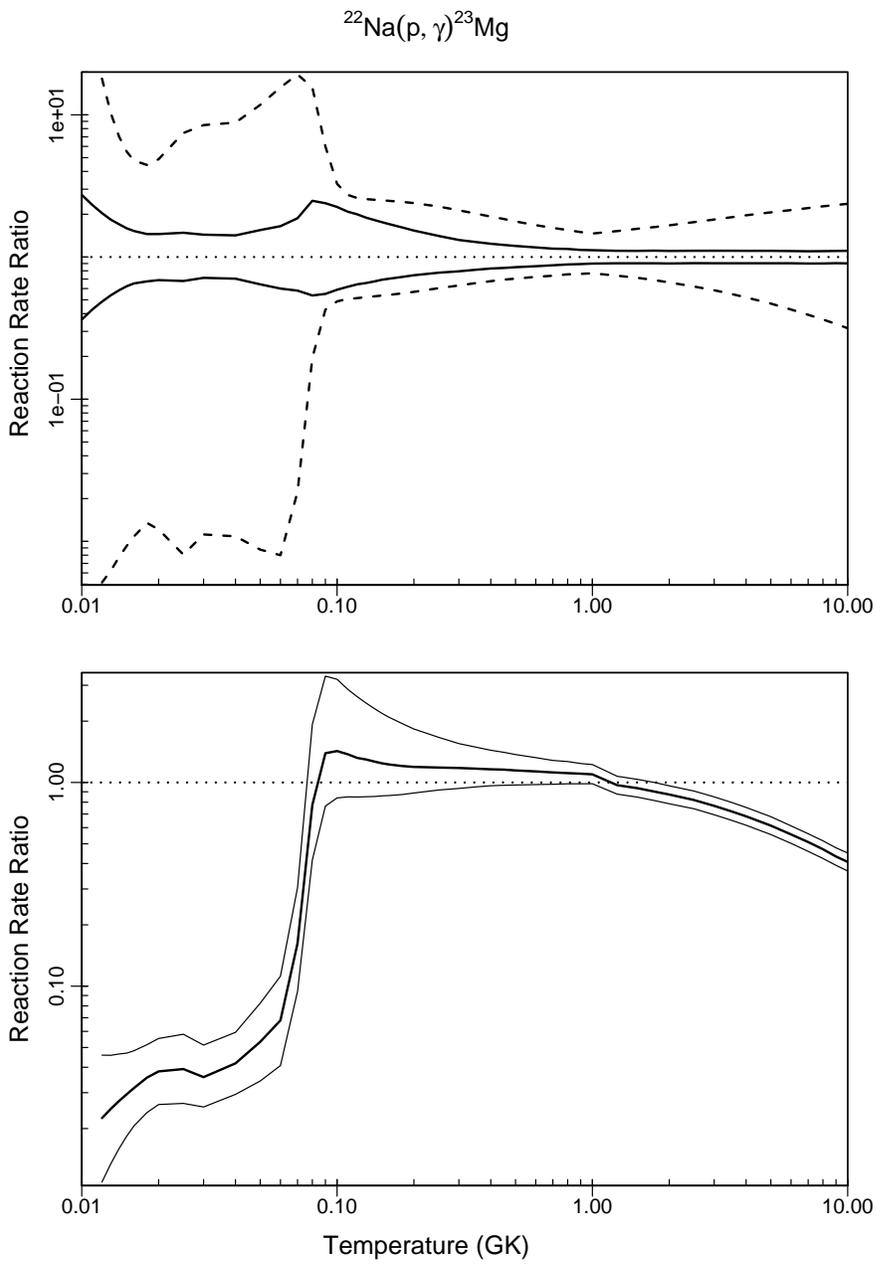}
\caption{\label{} 
Previous reaction rates: Ref. \cite{Ang99}.}
\end{figure}
\clearpage
\begin{figure}[b]
\includegraphics[height=17.05cm]{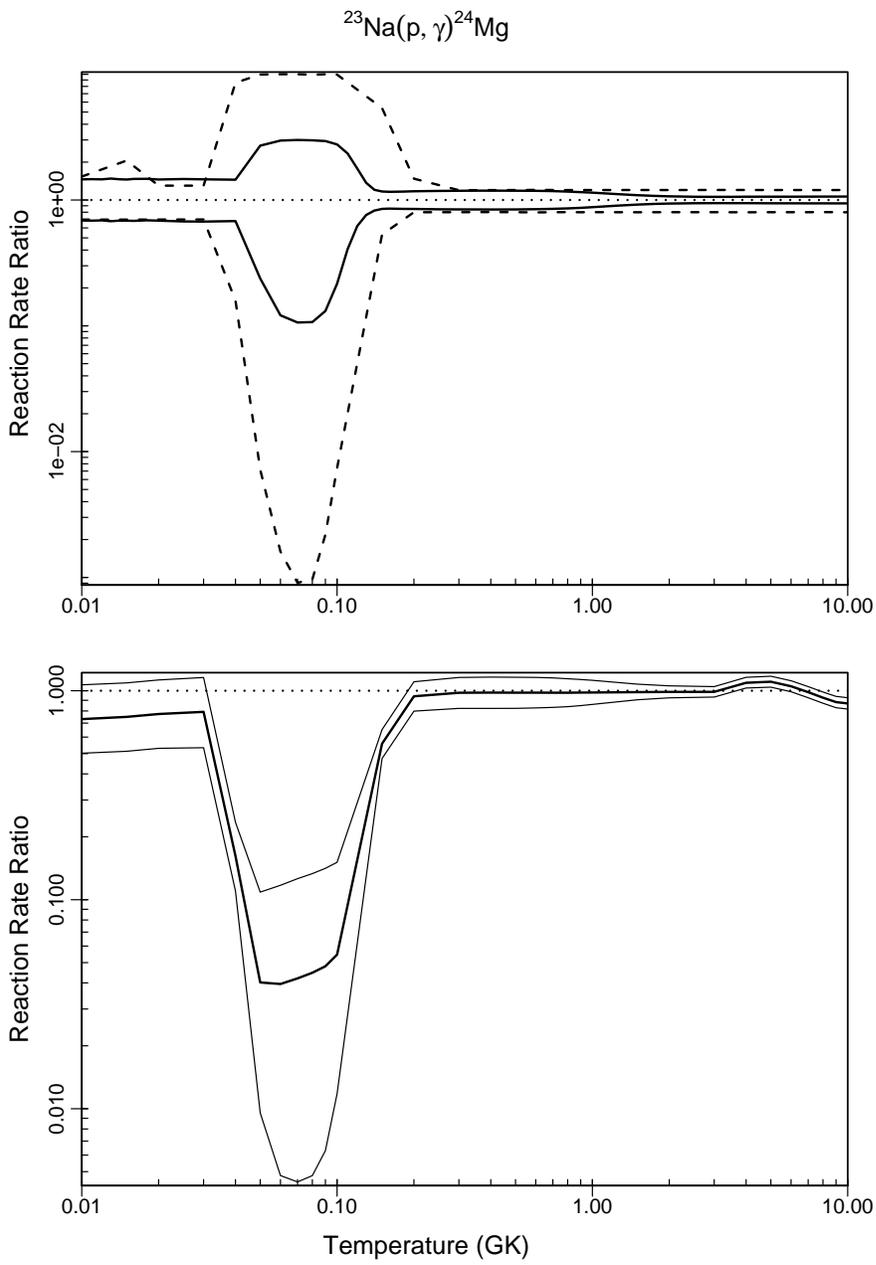}
\caption{\label{} 
Previous reaction rates: Ref. \cite{Ili01}.}
\end{figure}
\clearpage
\begin{figure}[b]
\includegraphics[height=17.05cm]{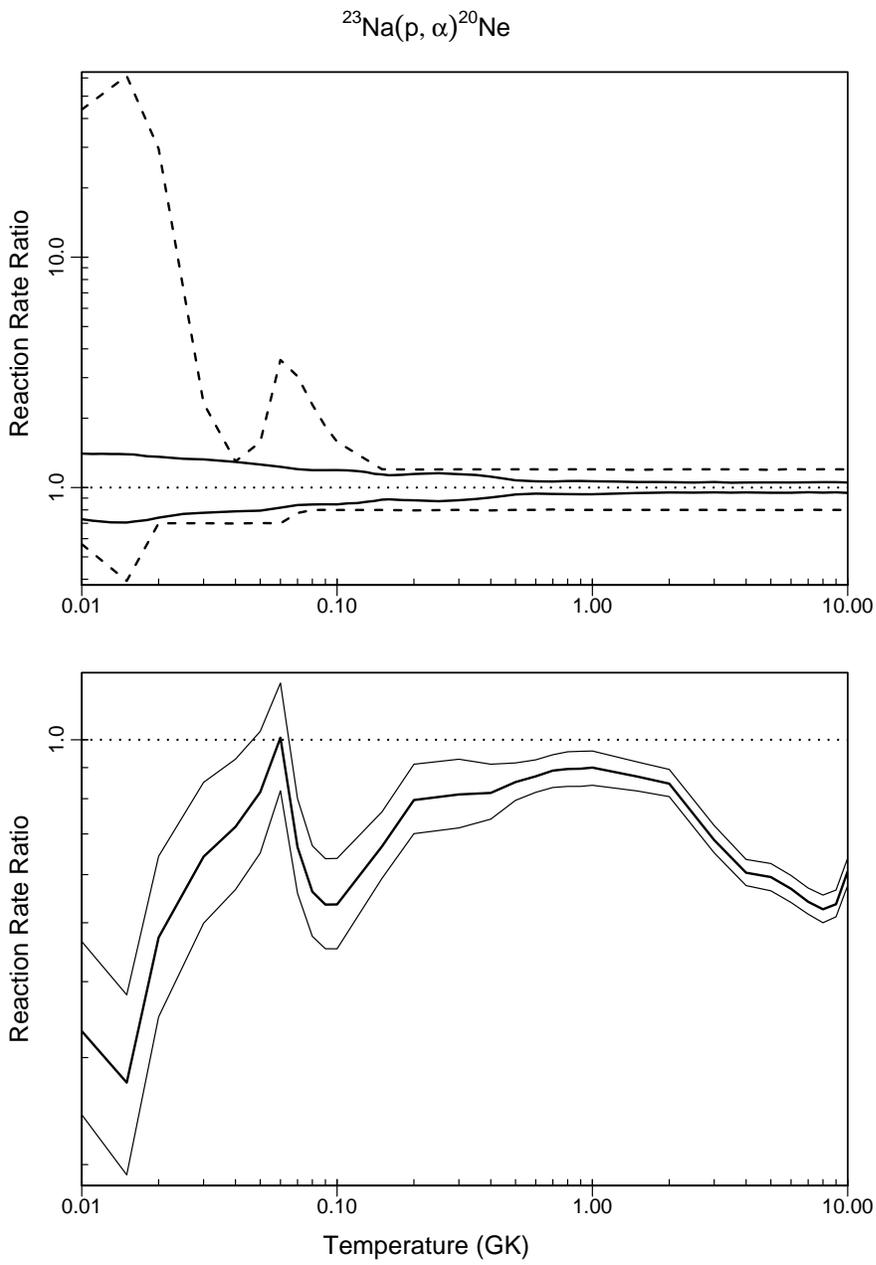}
\caption{\label{} 
Previous reaction rates: Ref. \cite{Ili01}.}
\end{figure}
\clearpage
\begin{figure}[b]
\includegraphics[height=17.05cm]{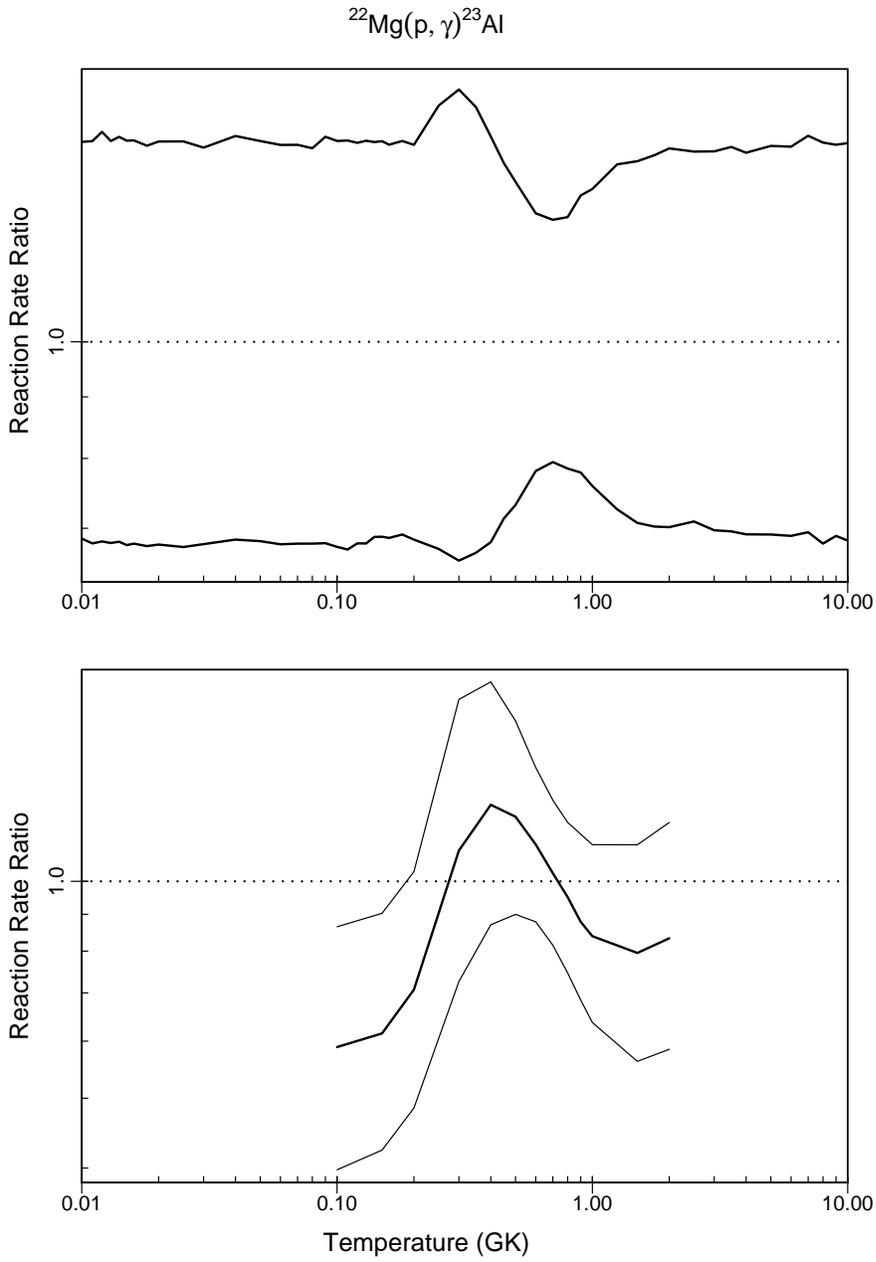}
\caption{\label{} 
Previous reaction rates: Ref. \cite{Cag01}. A more recent rate is reported in Tab. IV of Ref. \cite{He07}, but the differences compared to Ref. \cite{Cag01} are relatively small.}
\end{figure}
\clearpage
\begin{figure}[b]
\includegraphics[height=17.05cm]{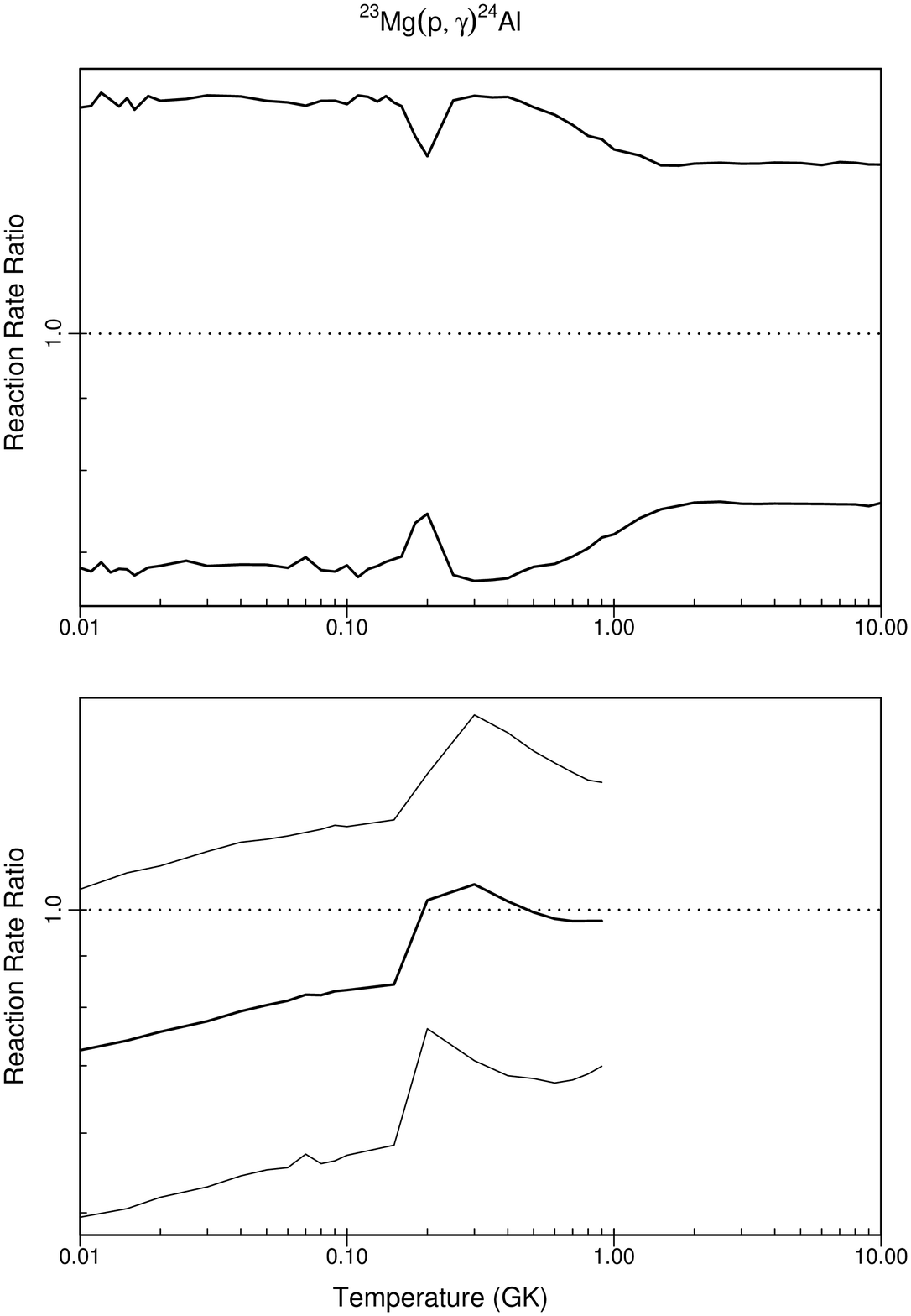}
\caption{\label{} 
Previous reaction rates: Ref. \cite{Her98}. A more recent rate is displayed in Fig. 2 of Ref. \cite{Vis07}, but numerical rate values are not available.}
\end{figure}
\clearpage
\begin{figure}[b]
\includegraphics[height=17.05cm]{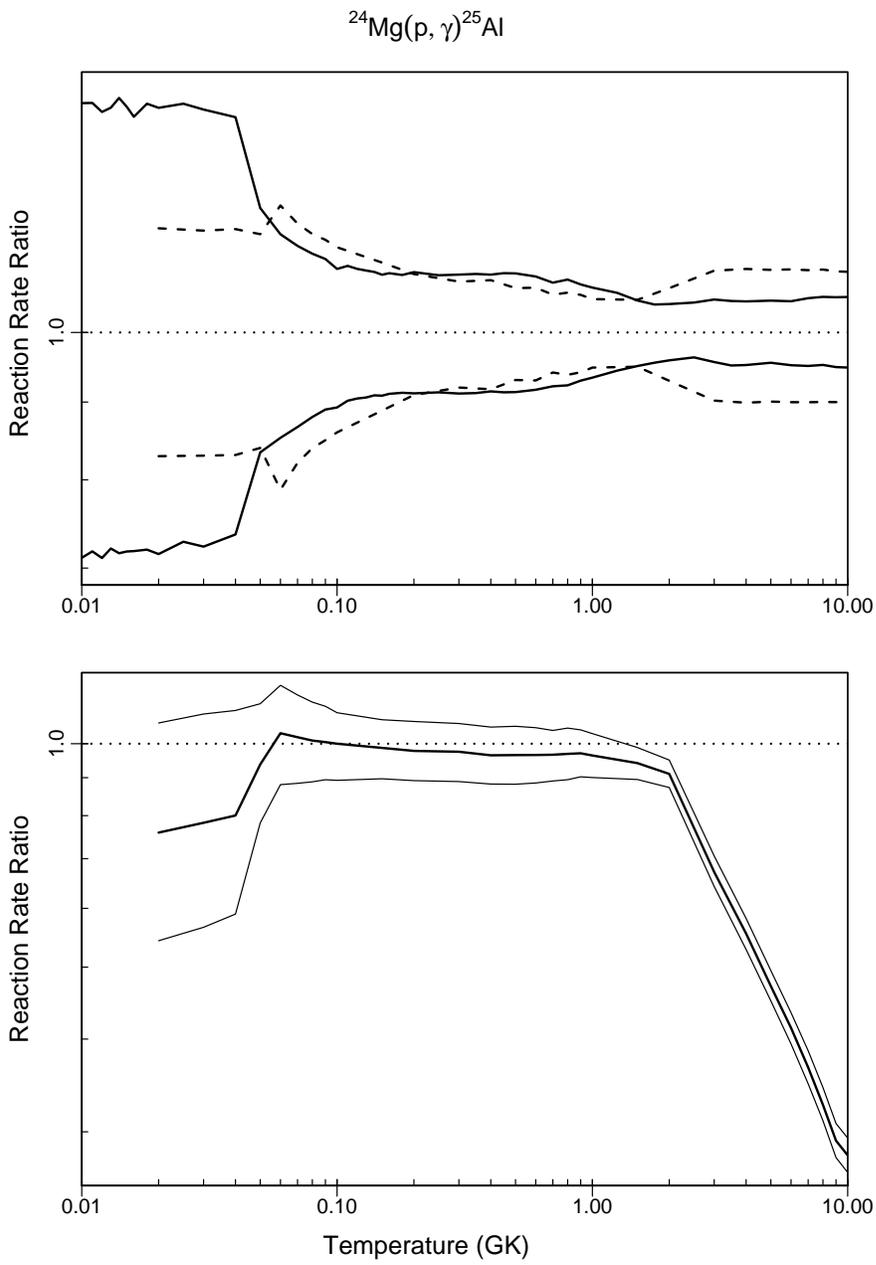}
\caption{\label{} 
Previous reaction rates: Ref. \cite{Ili01}.}
\end{figure}
\clearpage
\begin{figure}[b]
\includegraphics[height=17.05cm]{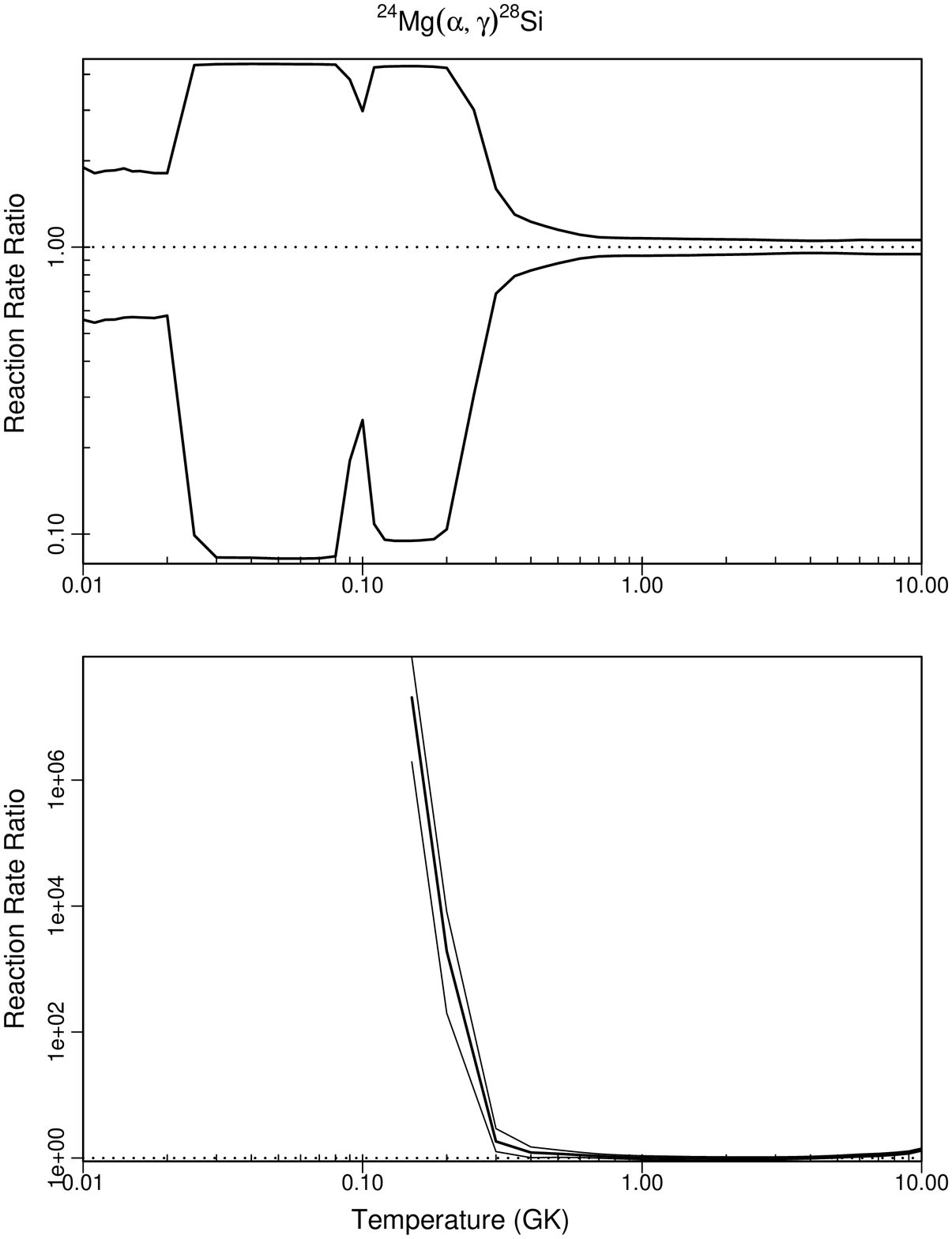}
\caption{\label{} 
Previous reaction rates: Ref. \cite{str08}. Rate uncertainties have not been determined previously.}
\end{figure}
\clearpage
\begin{figure}[b]
\includegraphics[height=17.05cm]{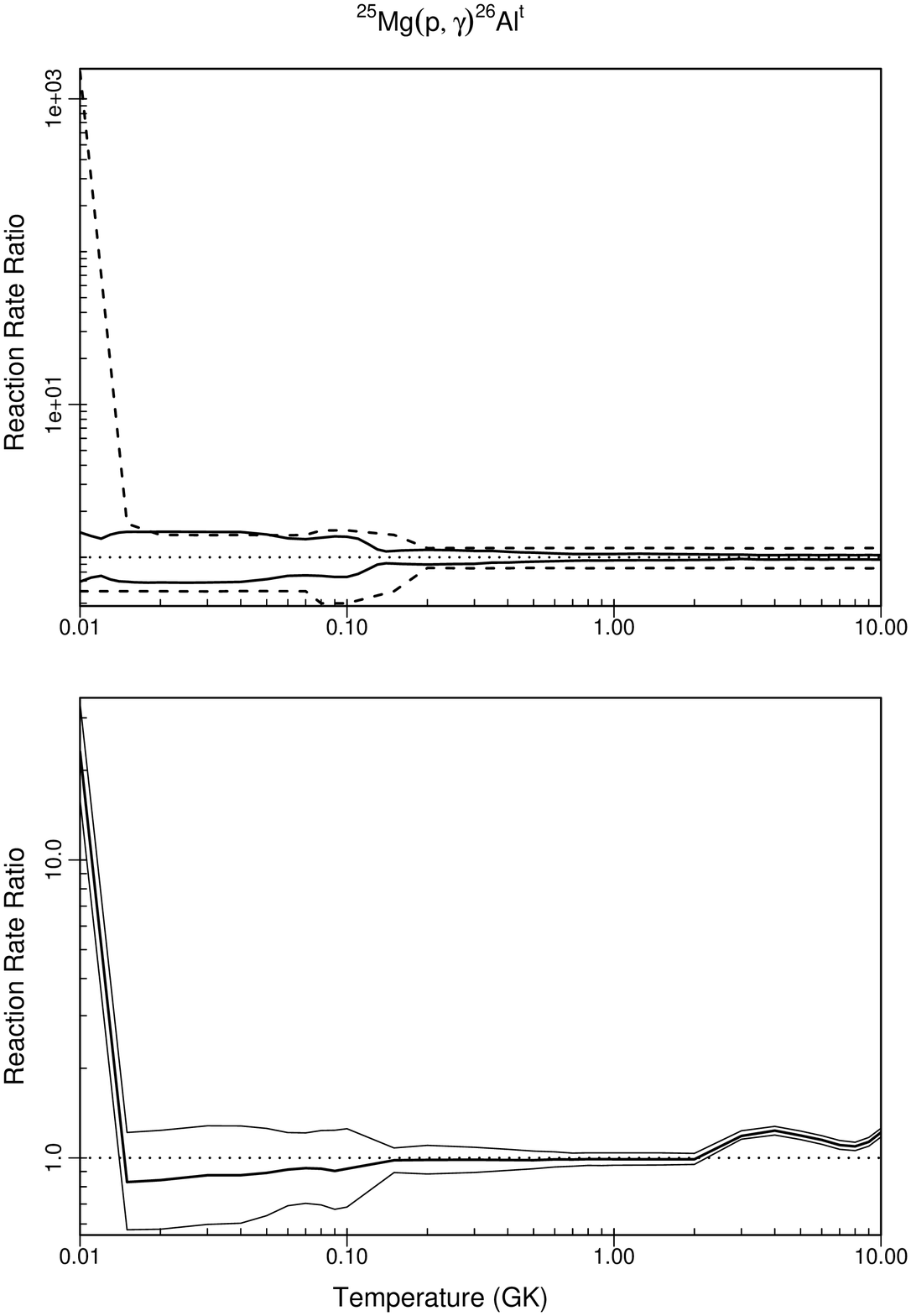}
\caption{\label{} 
Previous reaction rates: Ref. \cite{Ili01}.}
\end{figure}
\clearpage
\begin{figure}[b]
\includegraphics[height=17.05cm]{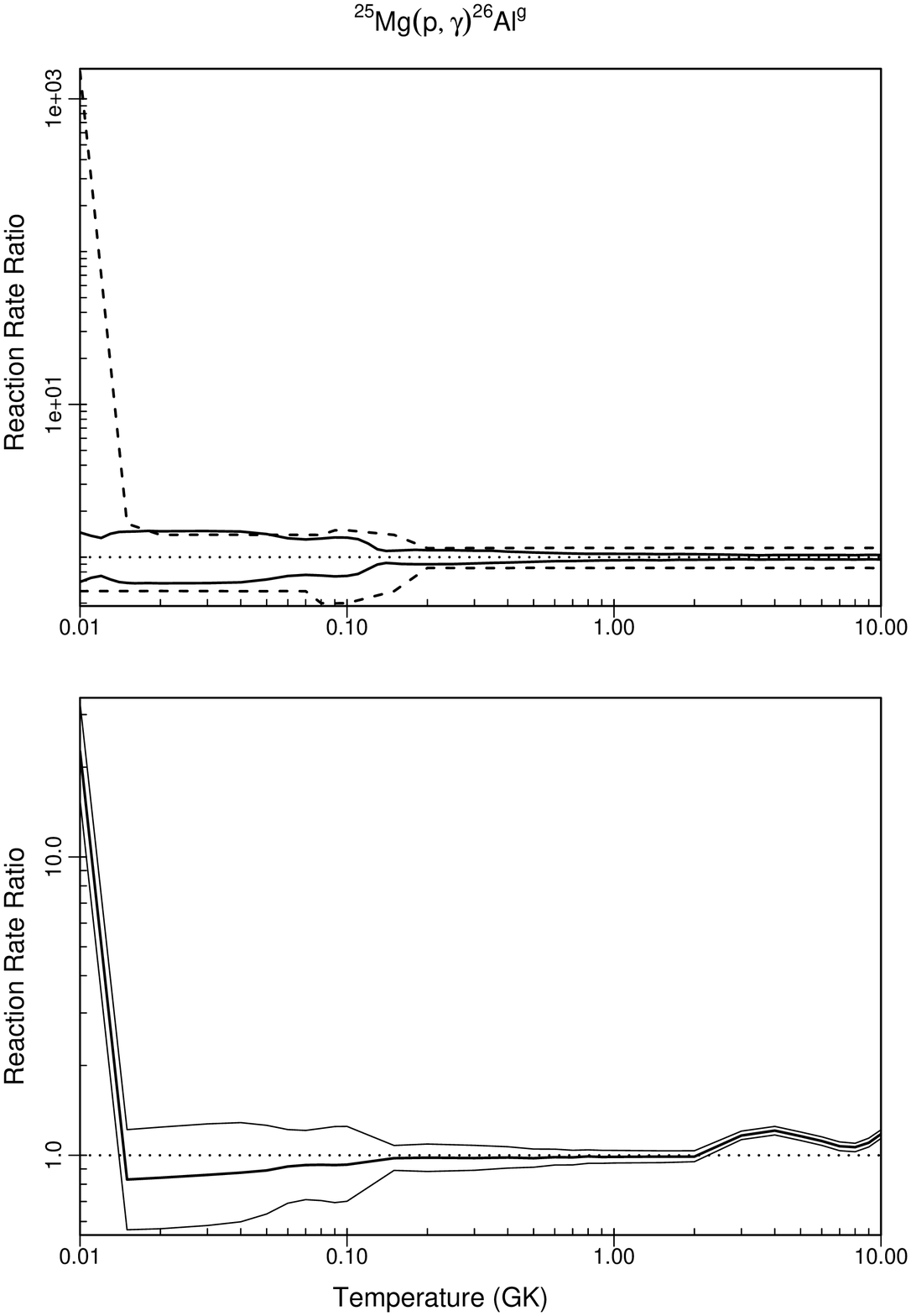}
\caption{\label{} 
Previous reaction rates: Ref. \cite{Ili01}.}
\end{figure}
\clearpage
\begin{figure}[b]
\includegraphics[height=17.05cm]{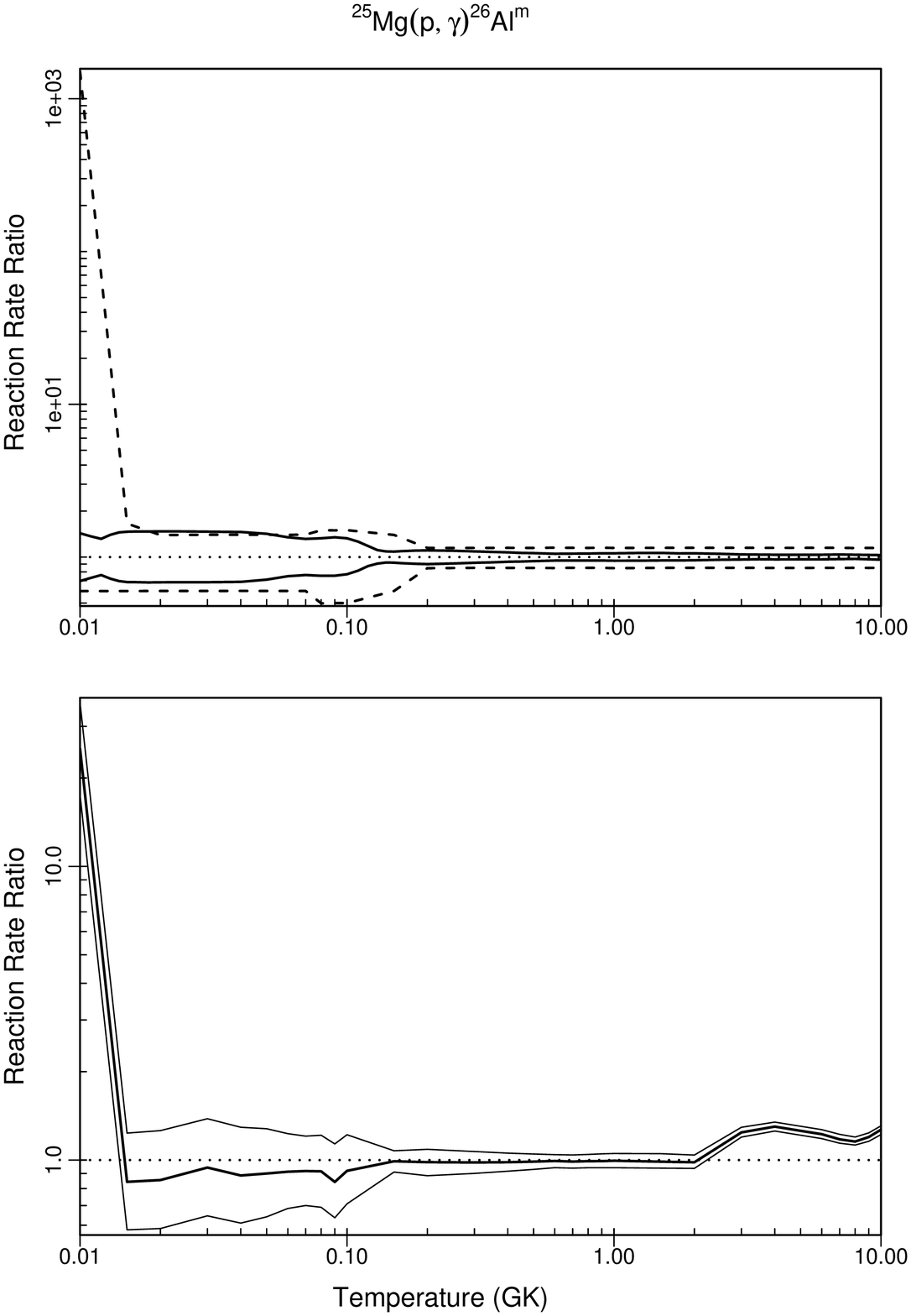}
\caption{\label{} 
Previous reaction rates: Ref. \cite{Ili01}.}
\end{figure}
\clearpage
\begin{figure}[b]
\includegraphics[height=17.05cm]{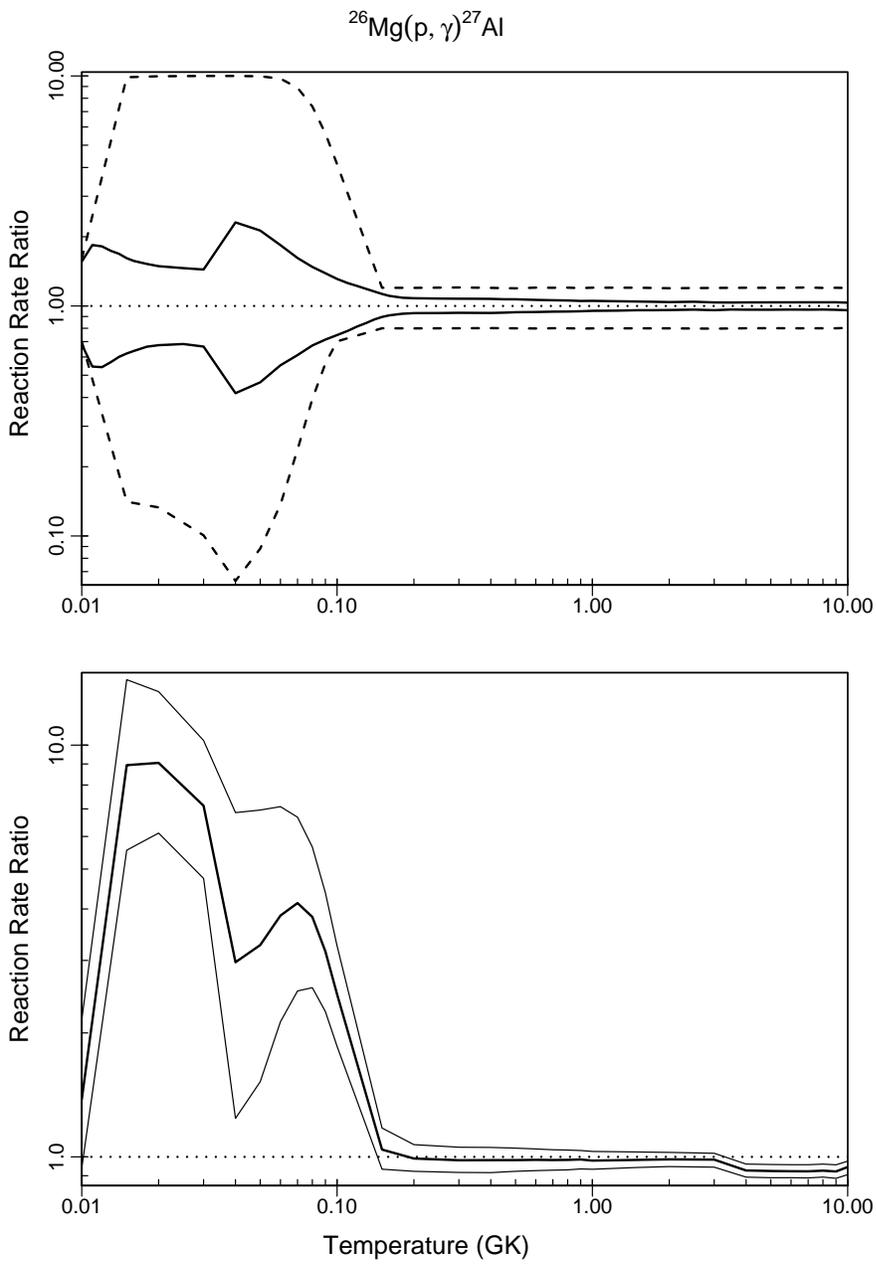}
\caption{\label{} 
Previous reaction rates: Ref. \cite{Ili01}.}
\end{figure}
\clearpage
\begin{figure}[b]
\includegraphics[height=17.05cm]{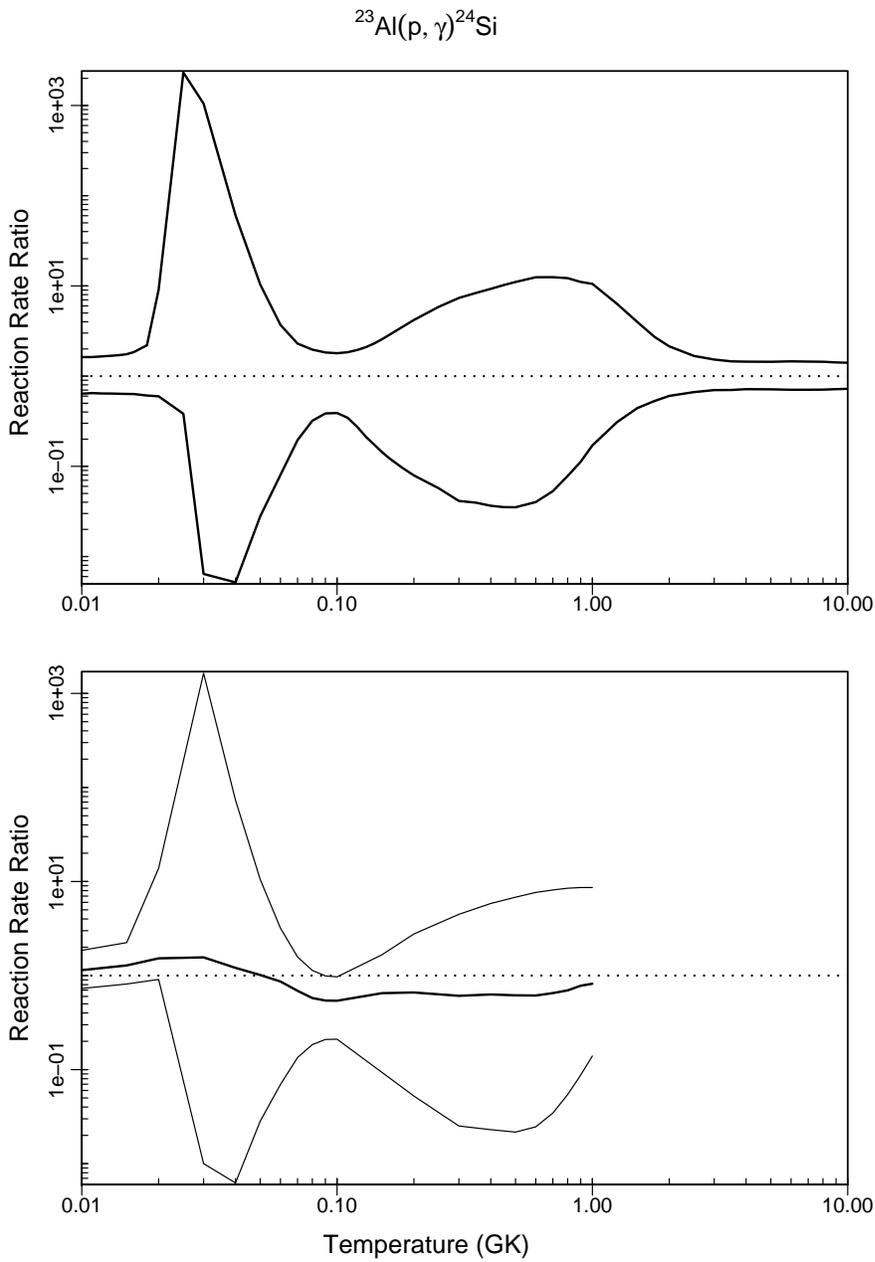}
\caption{\label{} 
Previous reaction rates: Ref. \cite{Sch97}. Previous rate uncertainties are displayed in Fig. 2 of Ref. \cite{Sch97}, but numerical values for the upper and lower rate limits are not available.}
\end{figure}
\clearpage
\begin{figure}[b]
\includegraphics[height=17.05cm]{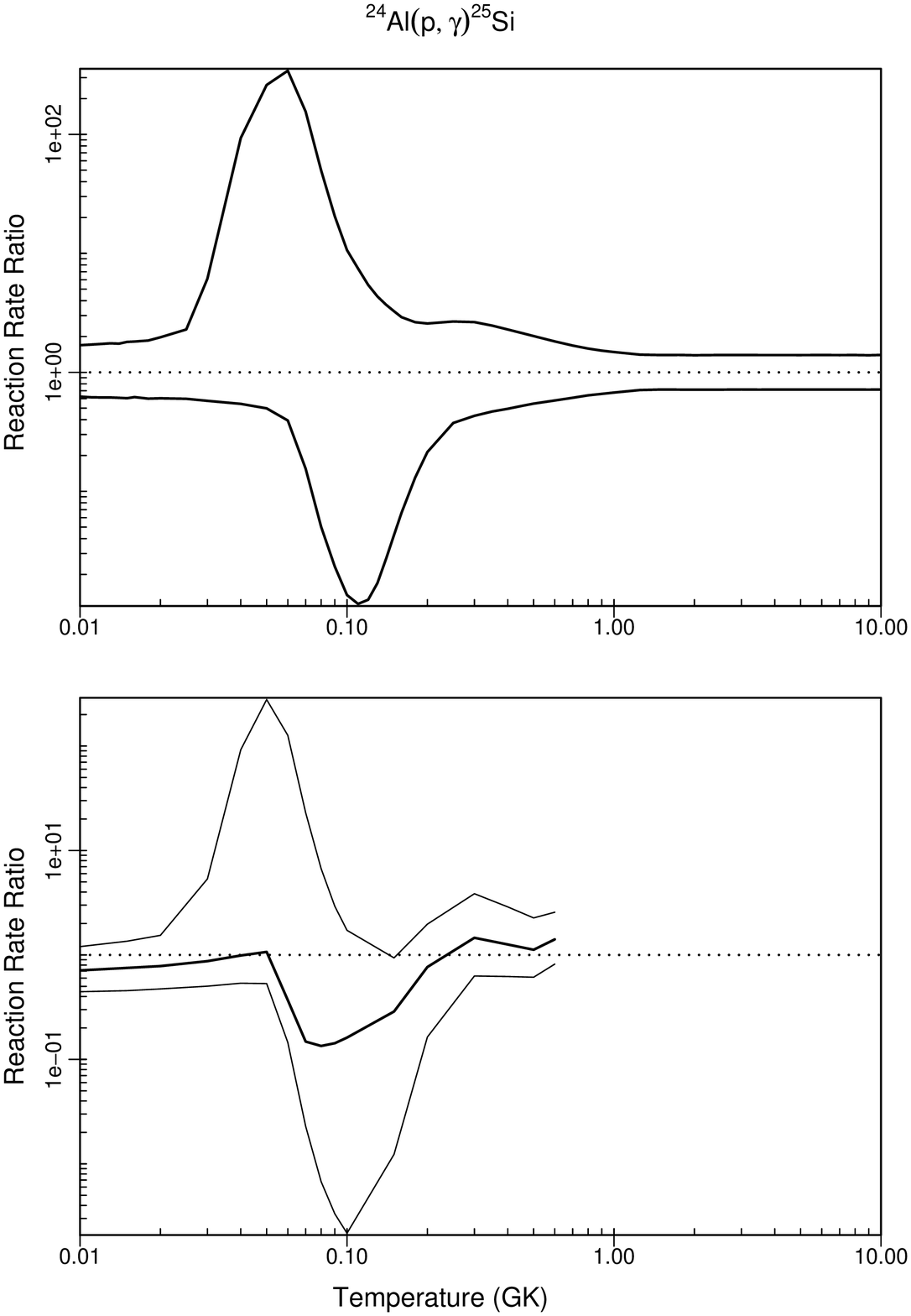}
\caption{\label{} 
Previous reaction rates: Ref. \cite{Her95}. Rate uncertainties have not been determined previously.}
\end{figure}
\clearpage
\begin{figure}[b]
\includegraphics[height=17.05cm]{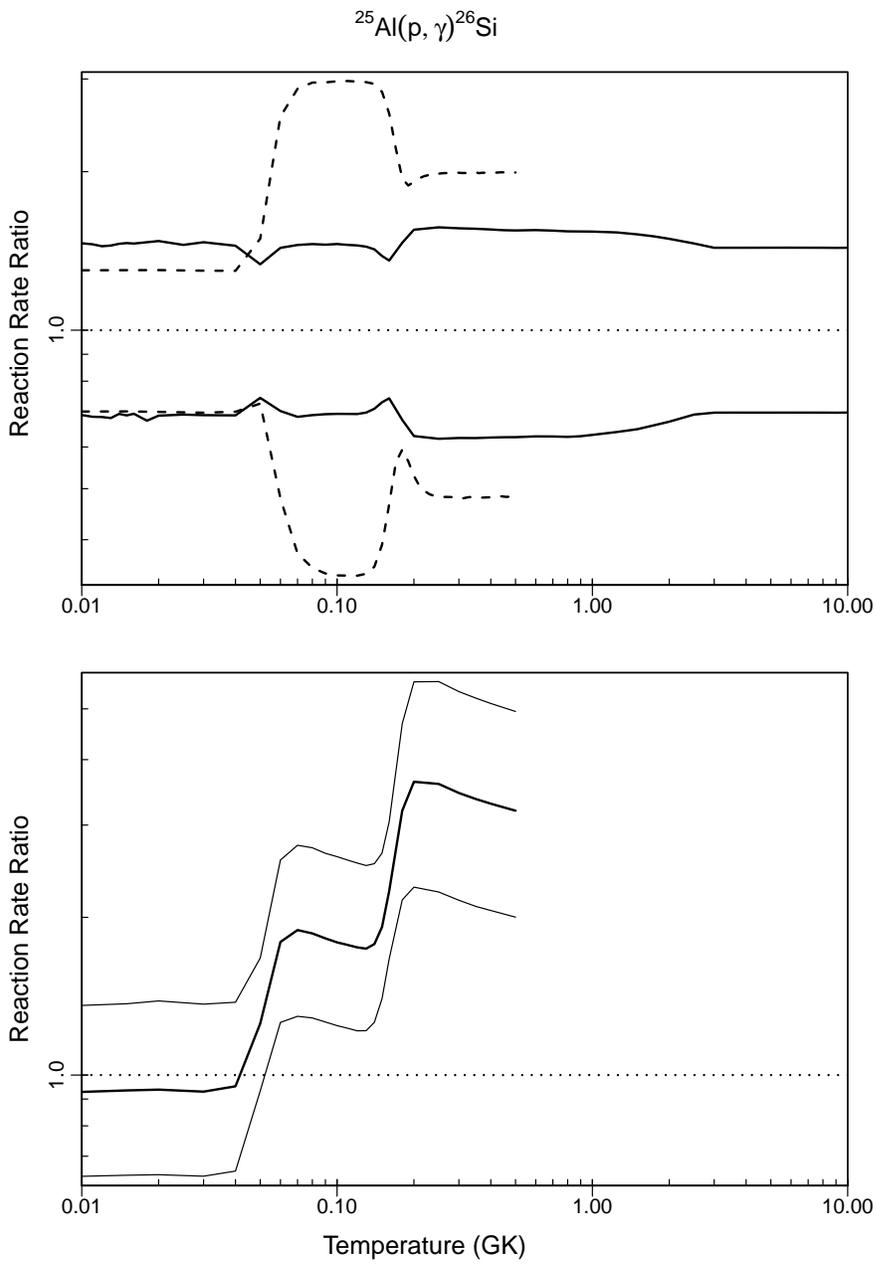}
\caption{\label{} 
Previous reaction rates: Ref. \cite{Wre09b}.}
\end{figure}
\clearpage
\begin{figure}[b]
\includegraphics[height=17.05cm]{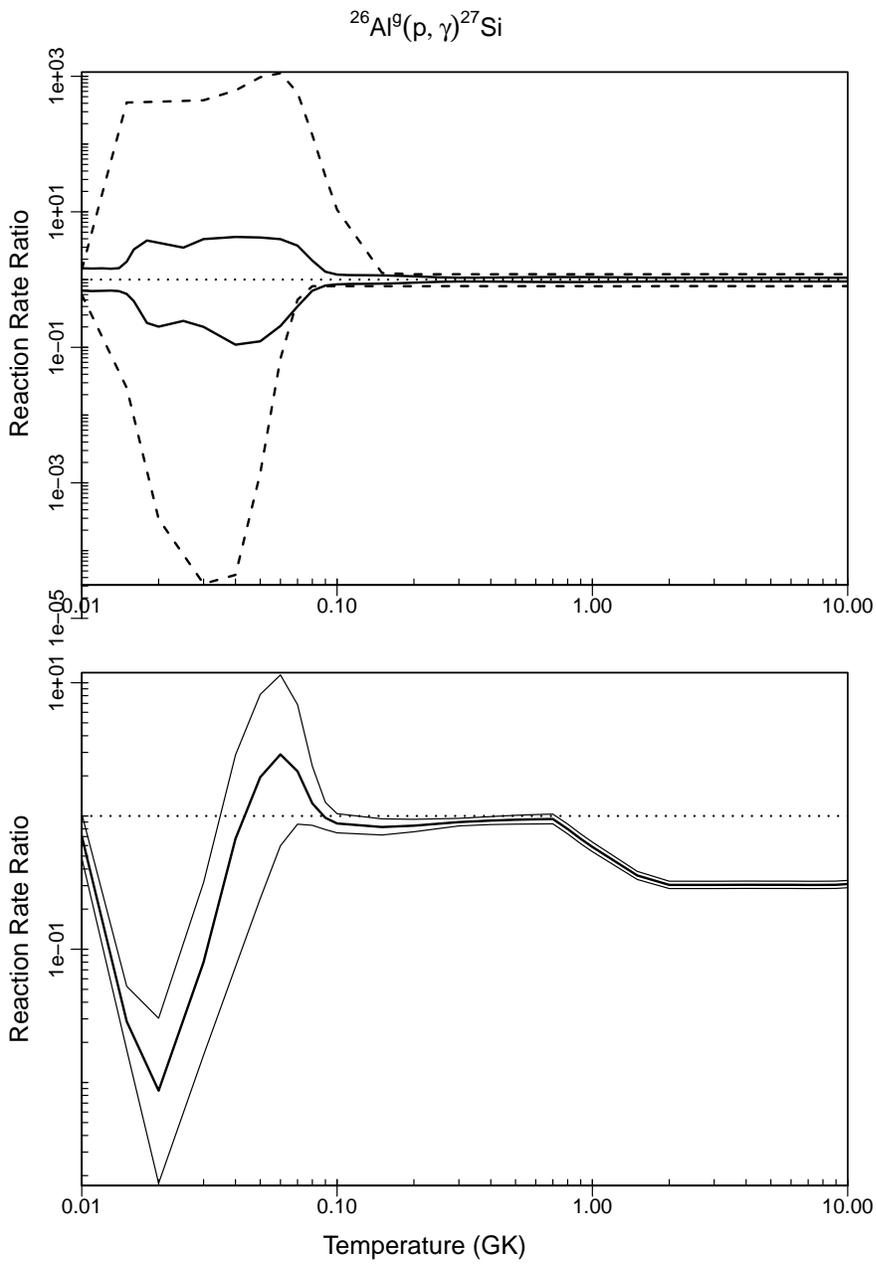}
\caption{\label{} 
Previous reaction rates: Ref. \cite{Ili01}.}
\end{figure}
\clearpage
\begin{figure}[b]
\includegraphics[height=17.05cm]{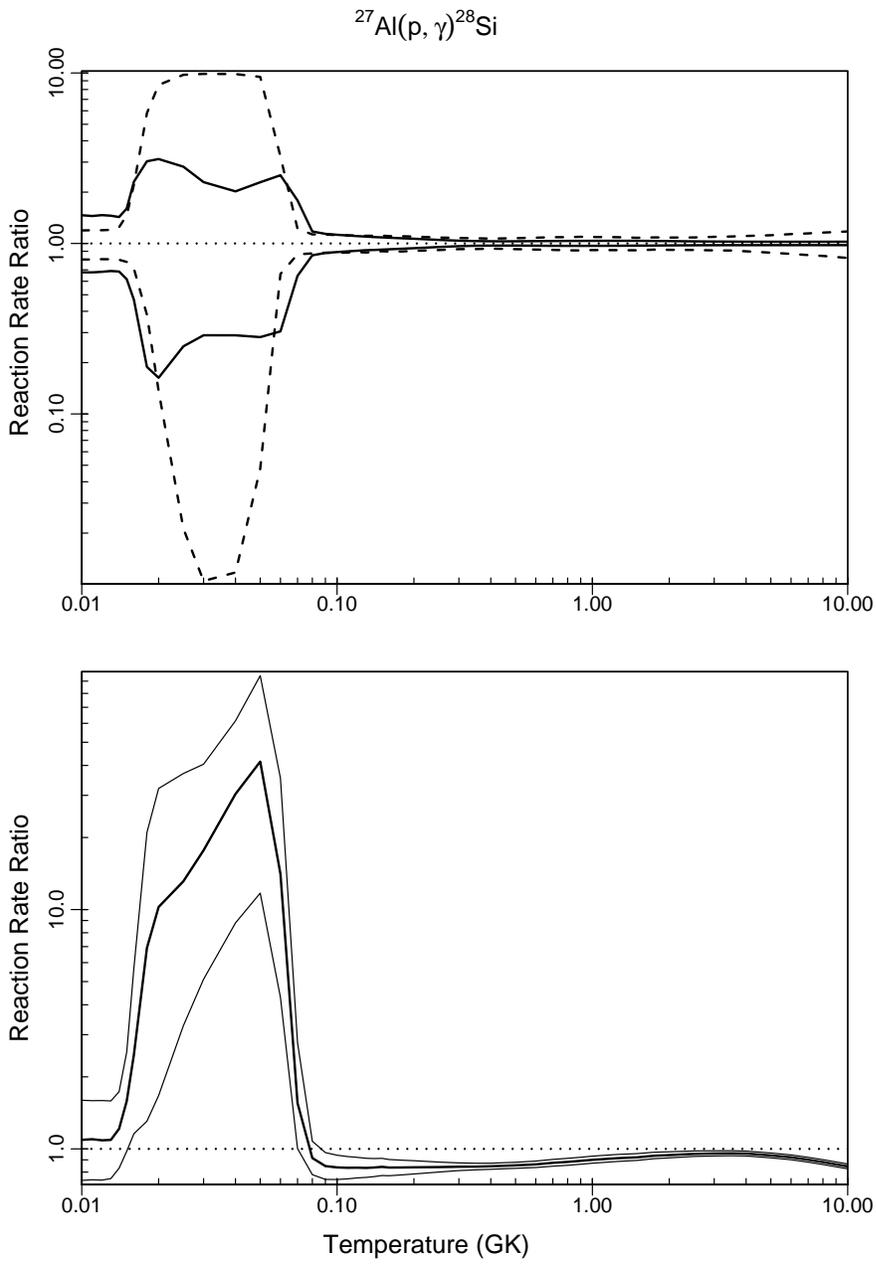}
\caption{\label{} 
Previous reaction rates: Ref. \cite{Har00}. The entry for the recommended rate at T=0.5 GK in Tab. 5 of Ref. \cite{Har00} is presumably a misprint.}
\end{figure}
\clearpage
\begin{figure}[b]
\includegraphics[height=17.05cm]{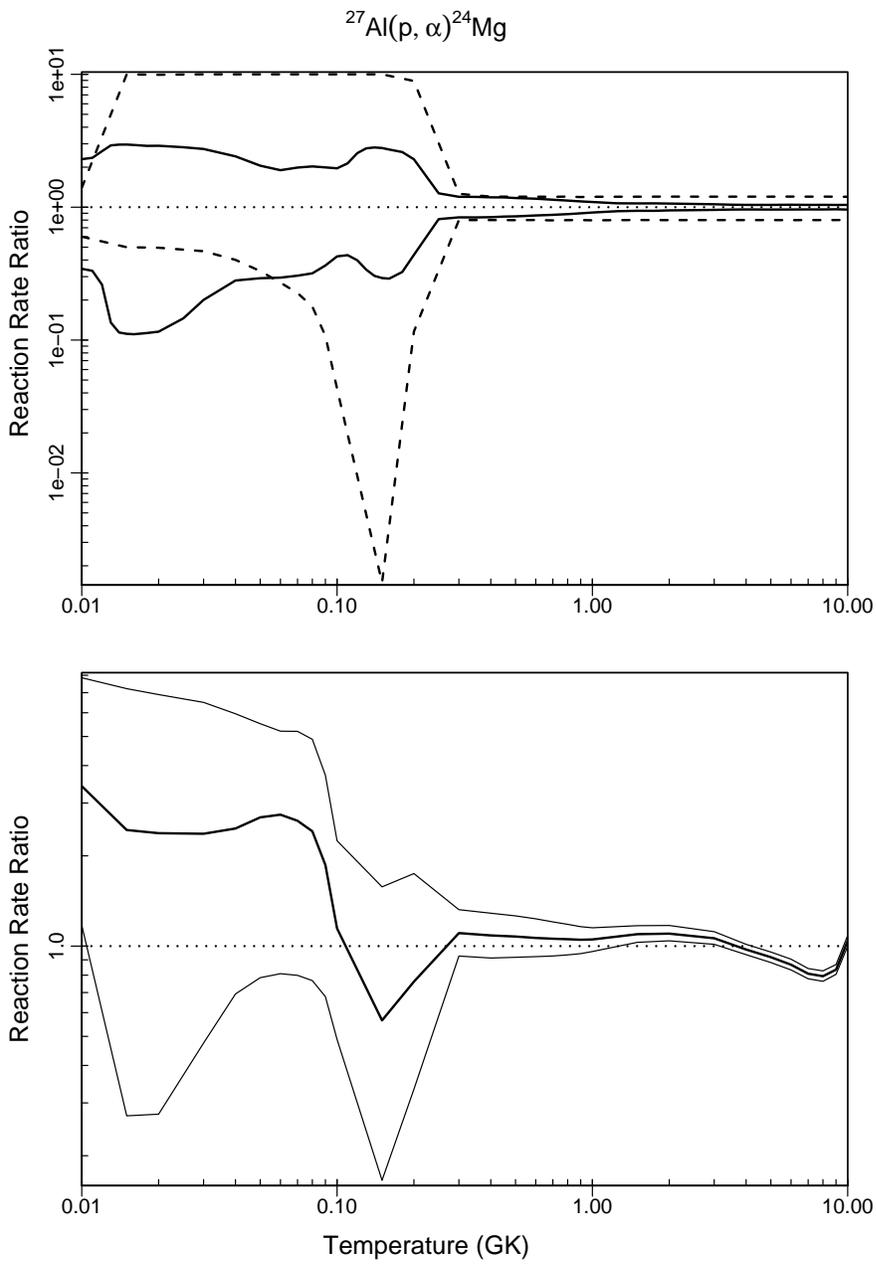}
\caption{\label{} 
Previous reaction rates: Ref. \cite{Ili01}.}
\end{figure}
\clearpage
\begin{figure}[b]
\includegraphics[height=17.05cm]{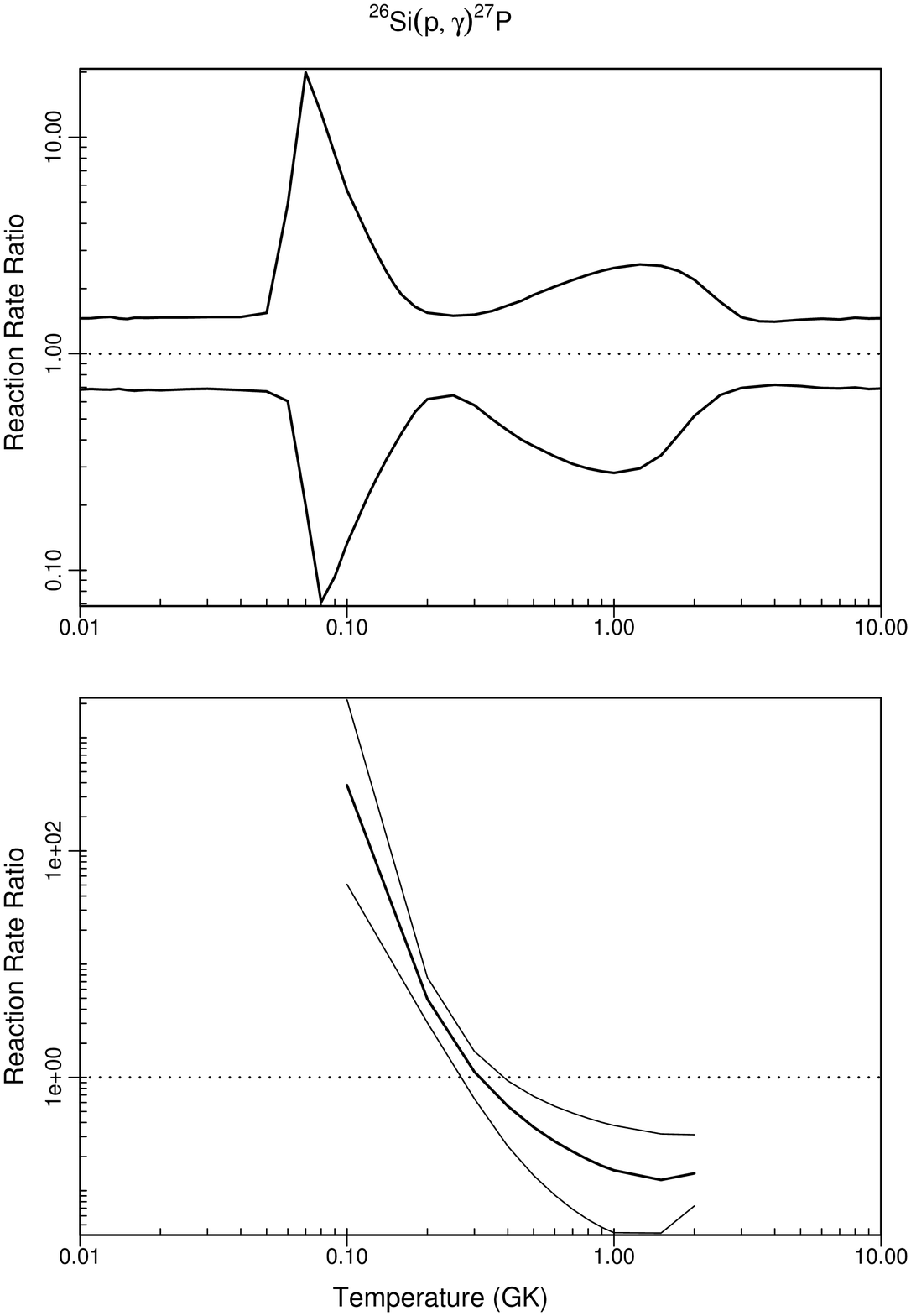}
\caption{\label{} 
Previous reaction rates: Ref. \cite{Guo06}. Rate uncertainties have not been determined previously.}
\end{figure}
\clearpage
\begin{figure}[b]
\includegraphics[height=17.05cm]{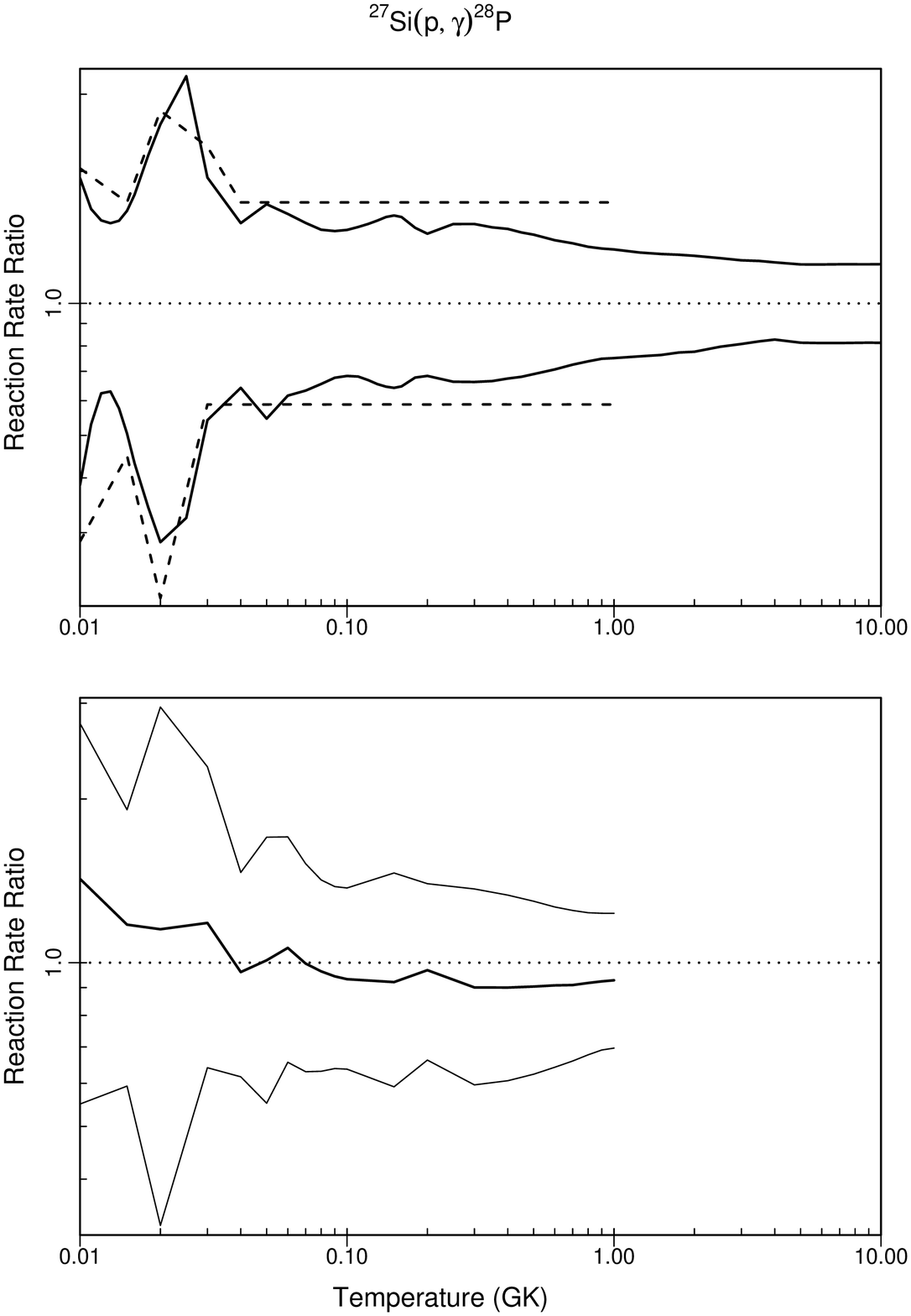}
\caption{\label{} 
Previous reaction rates: Ref. \cite{Ili99}.}
\end{figure}
\clearpage
\begin{figure}[b]
\includegraphics[height=17.05cm]{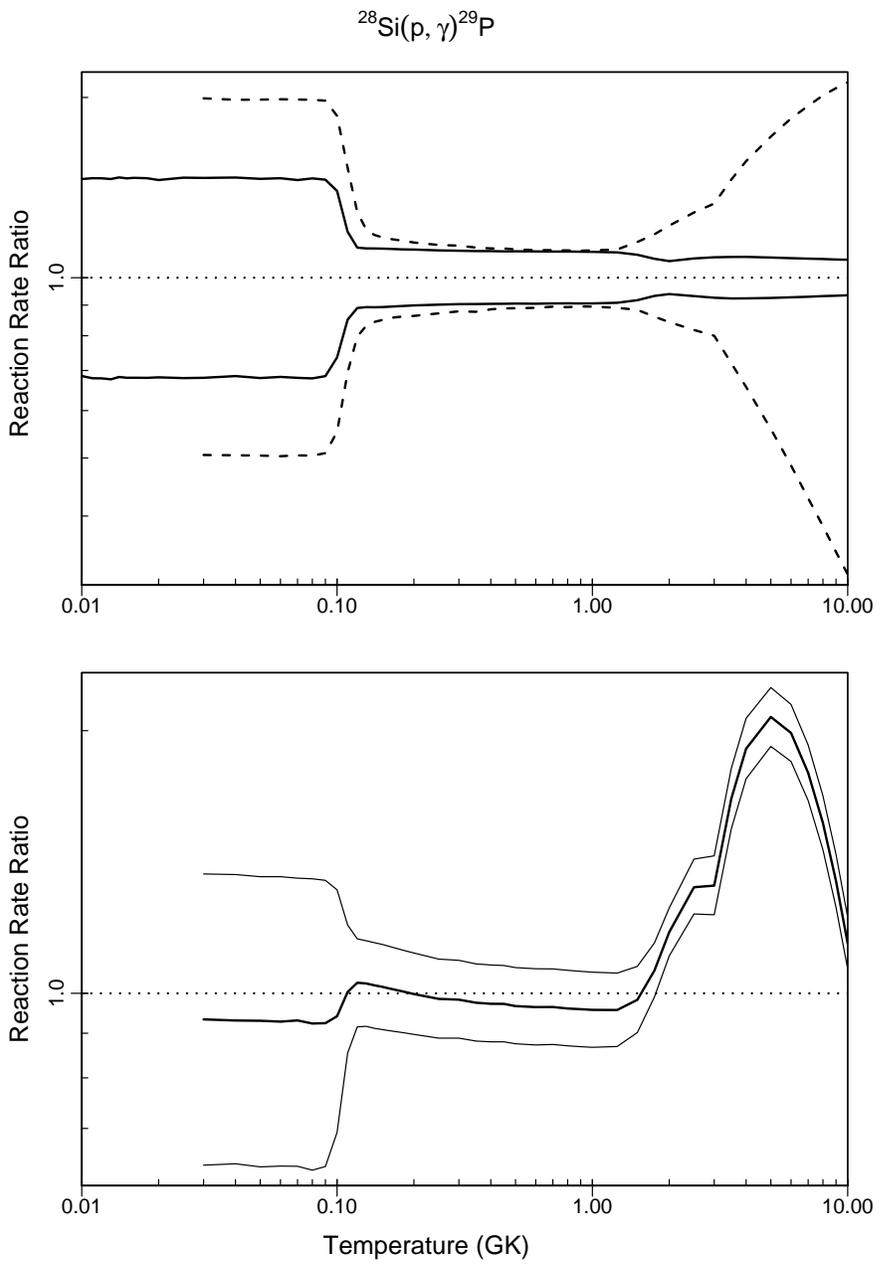}
\caption{\label{} 
Previous reaction rates: Ref. \cite{Ang99}.}
\end{figure}
\clearpage
\begin{figure}[b]
\includegraphics[height=17.05cm]{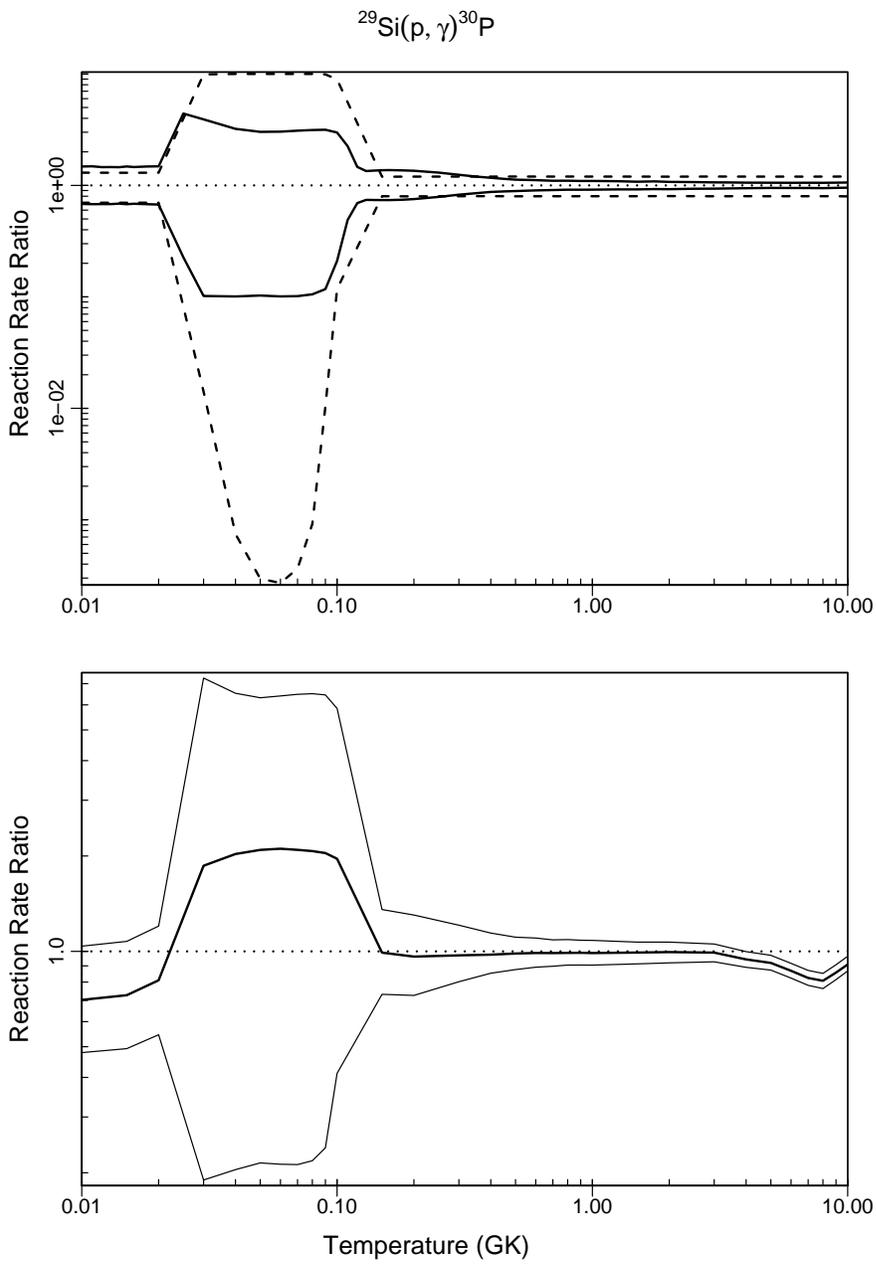}
\caption{\label{} 
Previous reaction rates: Ref. \cite{Ili01}.}
\end{figure}
\clearpage
\begin{figure}[b]
\includegraphics[height=17.05cm]{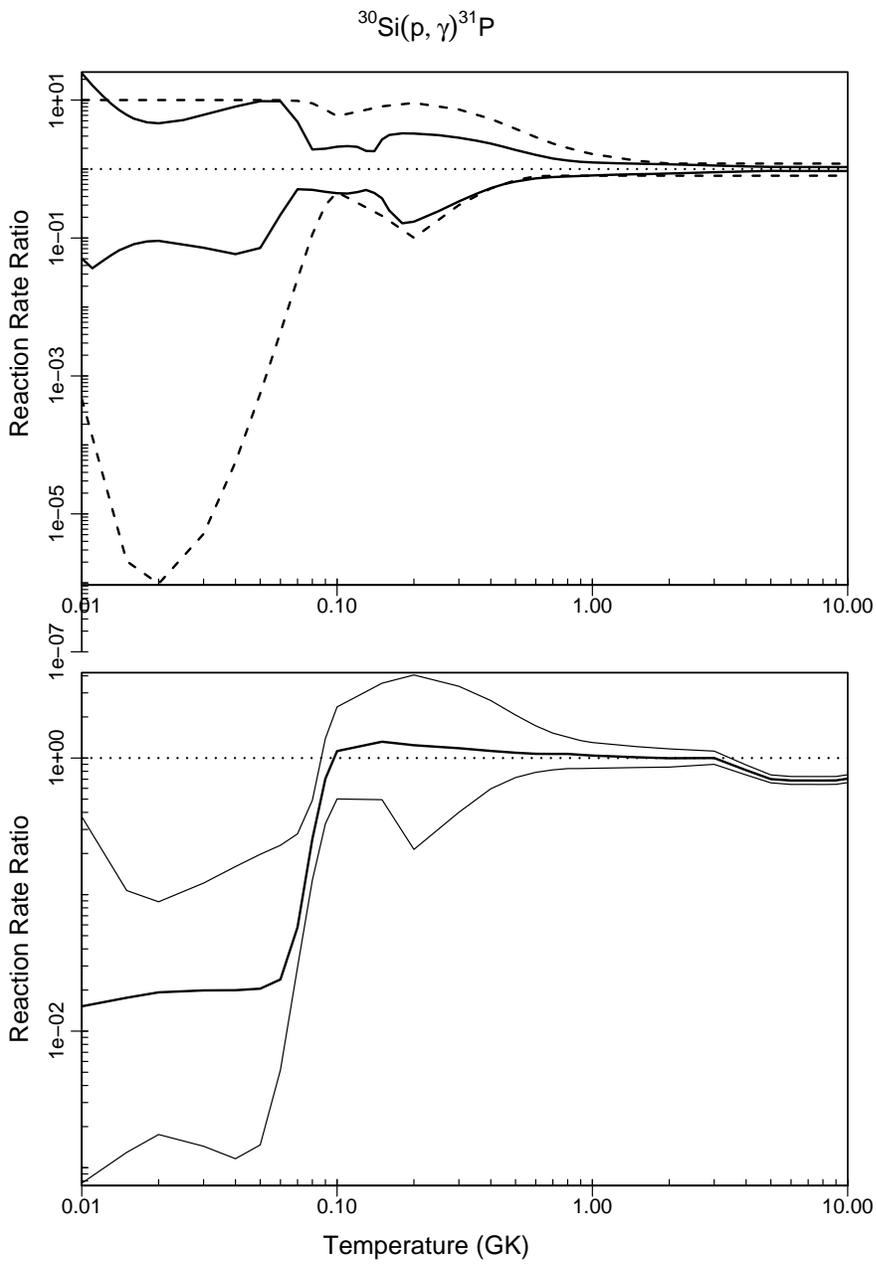}
\caption{\label{} 
Previous reaction rates: Ref. \cite{Ili01}.}
\end{figure}
\clearpage
\begin{figure}[b]
\includegraphics[height=17.05cm]{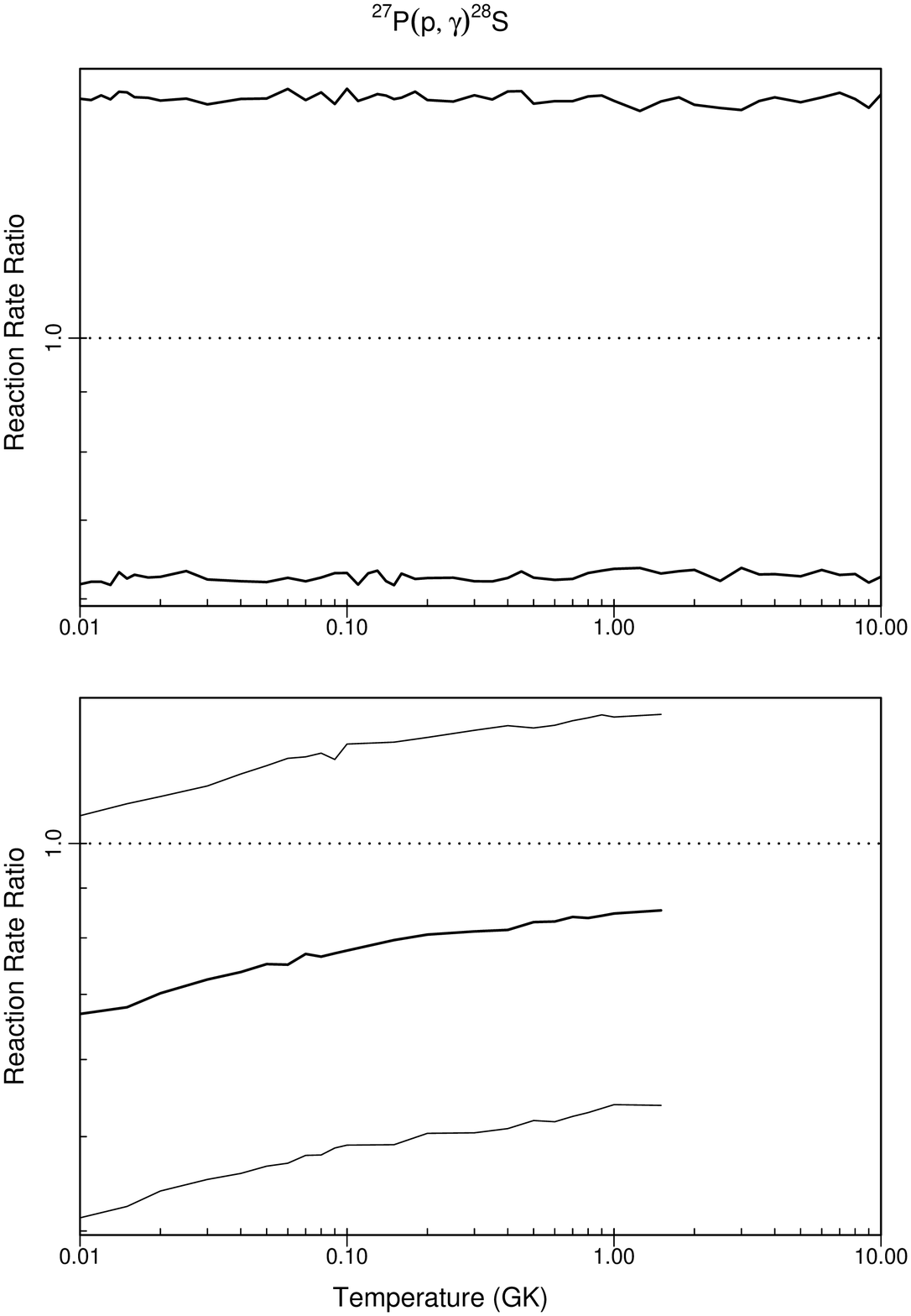}
\caption{\label{} 
Previous reaction rates: Ref. \cite{Her95}. Rate uncertainties have not been determined previously.}
\end{figure}
\clearpage
\begin{figure}[b]
\includegraphics[height=17.05cm]{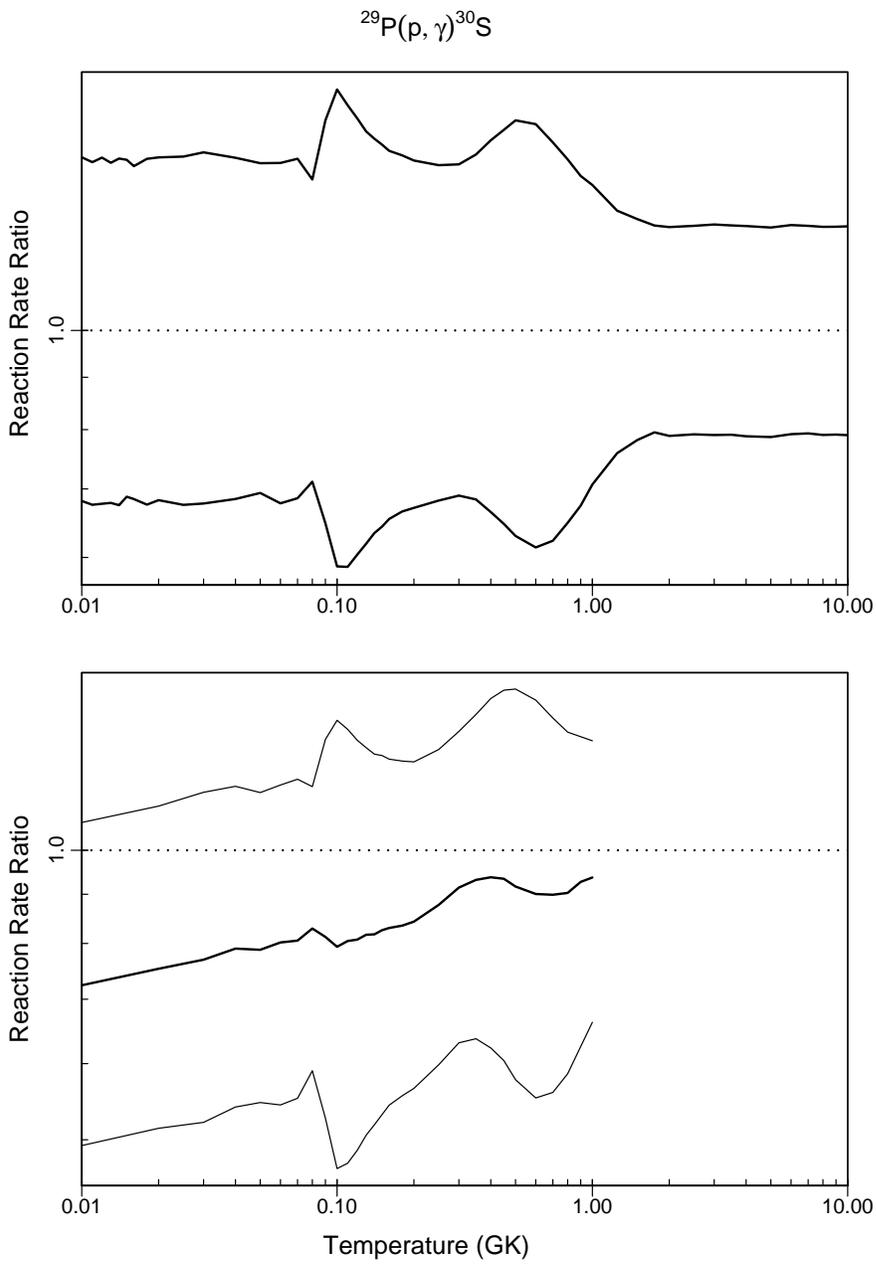}
\caption{\label{} 
Previous reaction rates: Ref. \cite{Bar07}.}
\end{figure}
\clearpage
\begin{figure}[b]
\includegraphics[height=17.05cm]{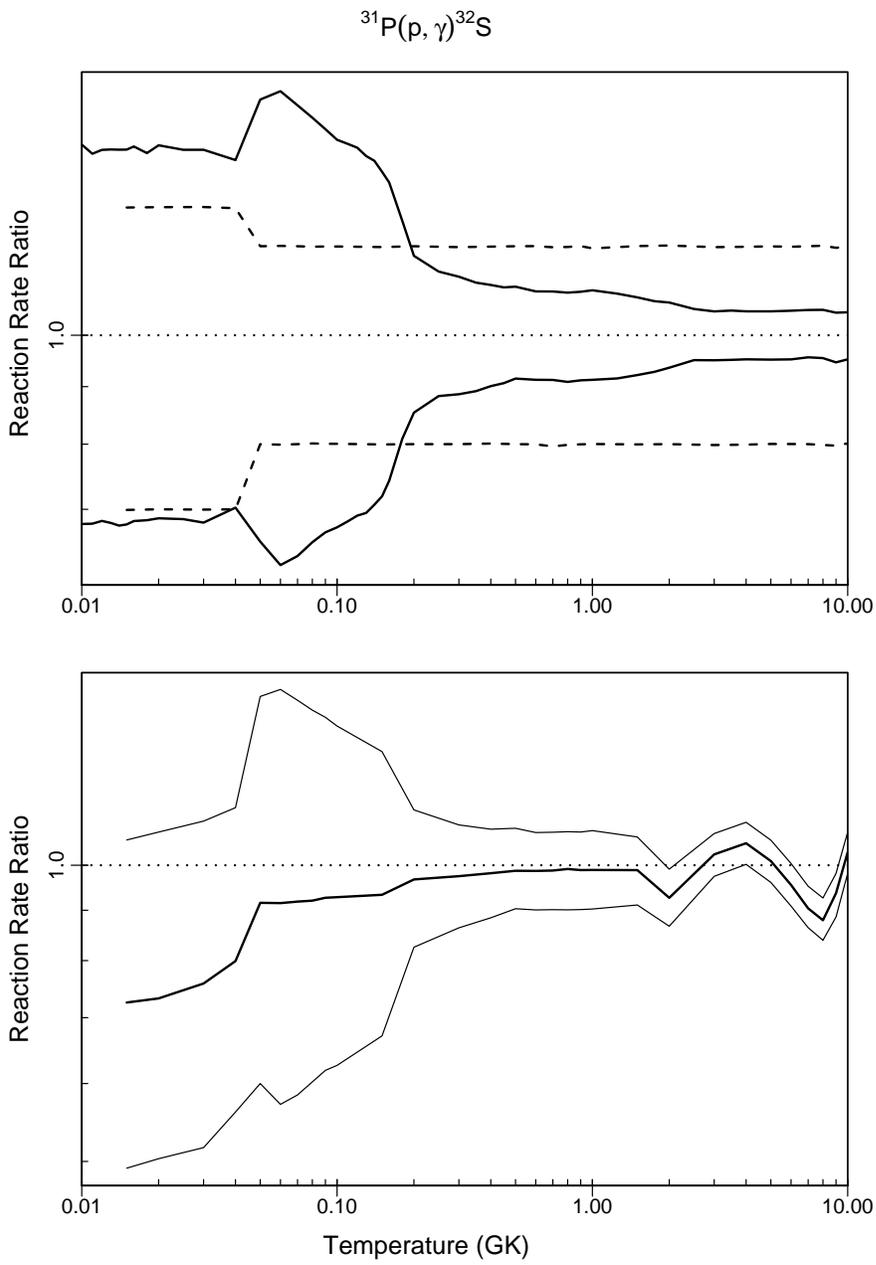}
\caption{\label{} 
Previous reaction rates: Ref. \cite{Ili01}.}
\end{figure}
\clearpage
\begin{figure}[b]
\includegraphics[height=17.05cm]{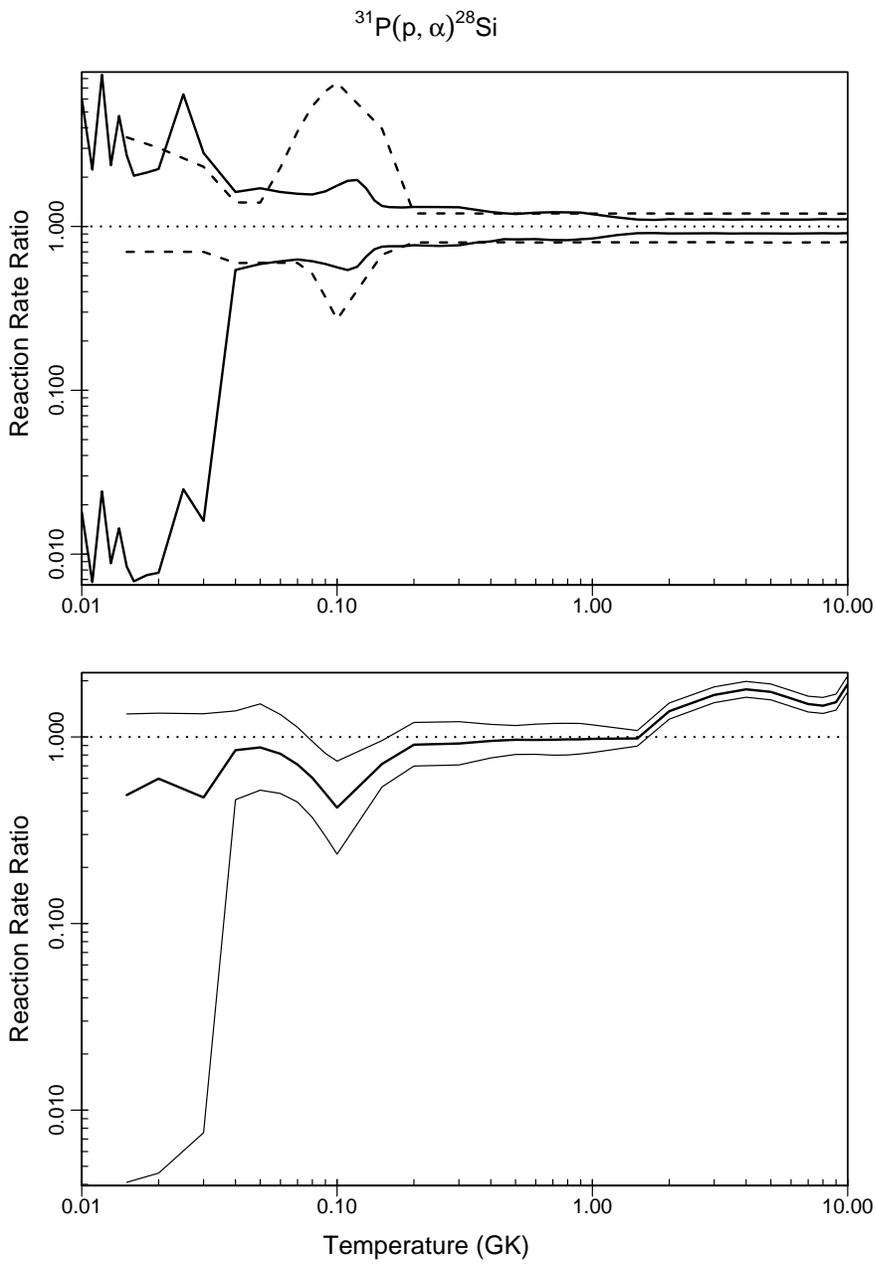}
\caption{\label{} 
Previous reaction rates: Ref. \cite{Ili01}.}
\end{figure}
\clearpage
\begin{figure}[b]
\includegraphics[height=17.05cm]{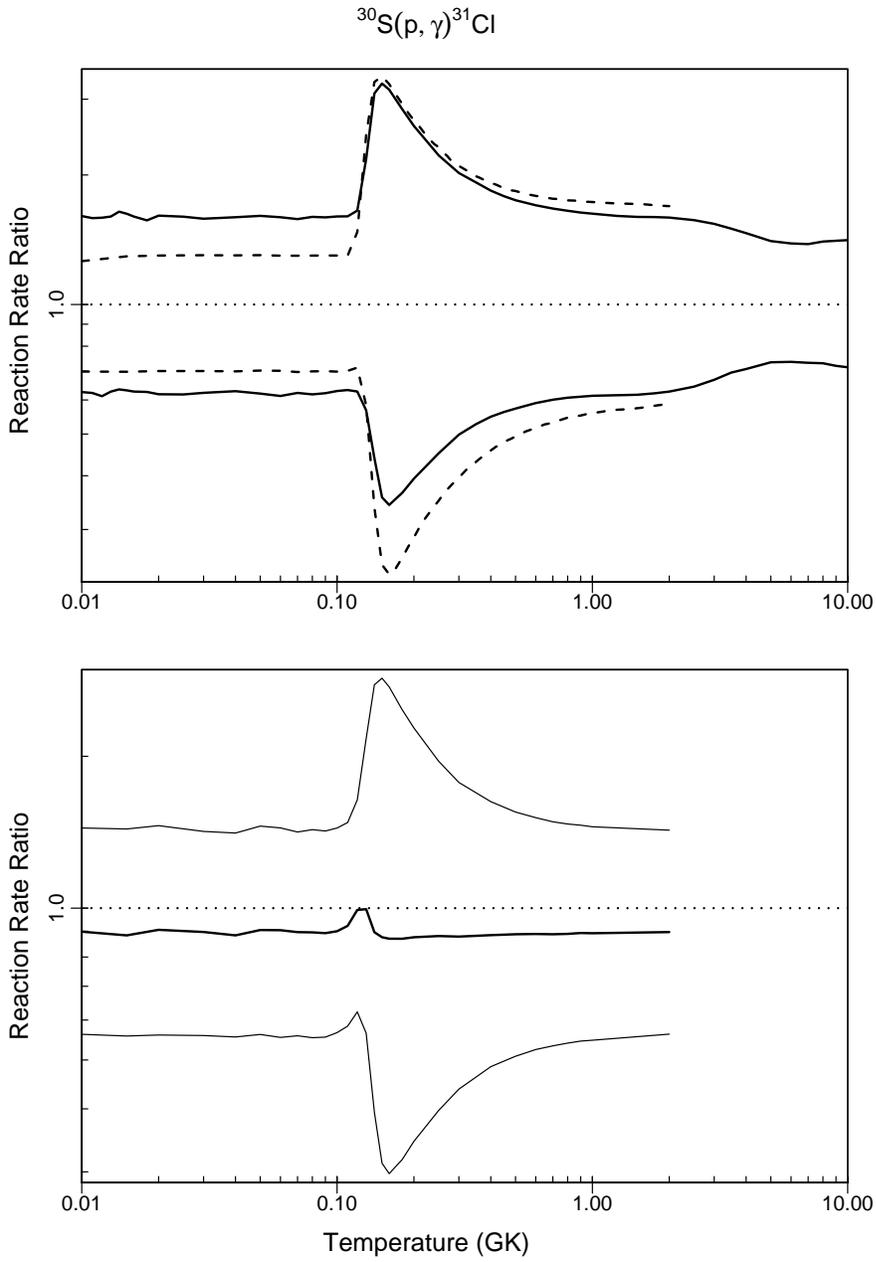}
\caption{\label{} 
Previous reaction rates: Ref. \cite{Wre09}; two values given in their Tab. IV, at T=0.04 GK (``high" value) and T=0.25 GK (``low" value), are erroneous.}
\end{figure}
\clearpage
\begin{figure}[b]
\includegraphics[height=17.05cm]{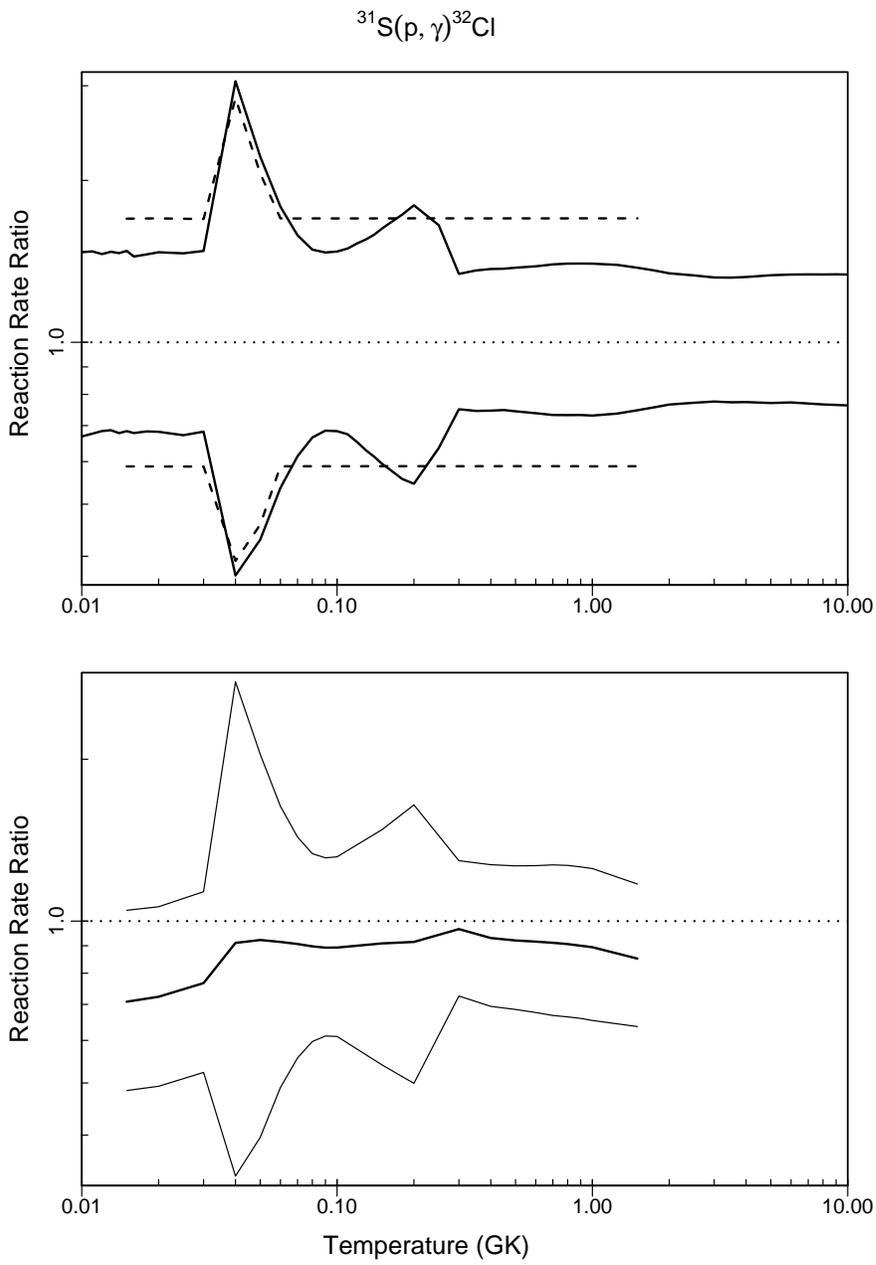}
\caption{\label{} 
Previous reaction rates: Ref. \cite{Ili99}.}
\end{figure}
\clearpage
\begin{figure}[b]
\includegraphics[height=17.05cm]{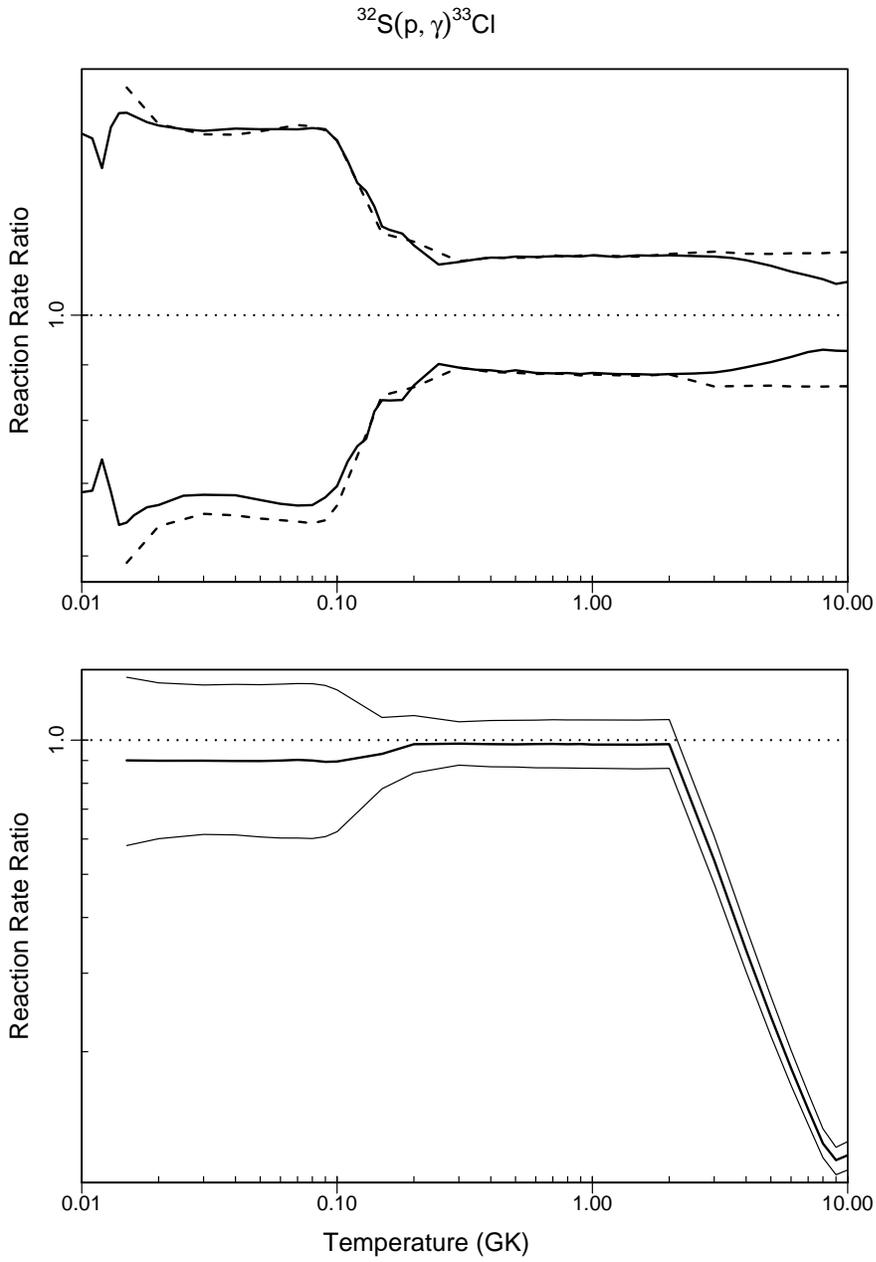}
\caption{\label{} 
Previous reaction rates: Ref. \cite{Ili01}.}
\end{figure}
\clearpage
\begin{figure}[b]
\includegraphics[height=17.05cm]{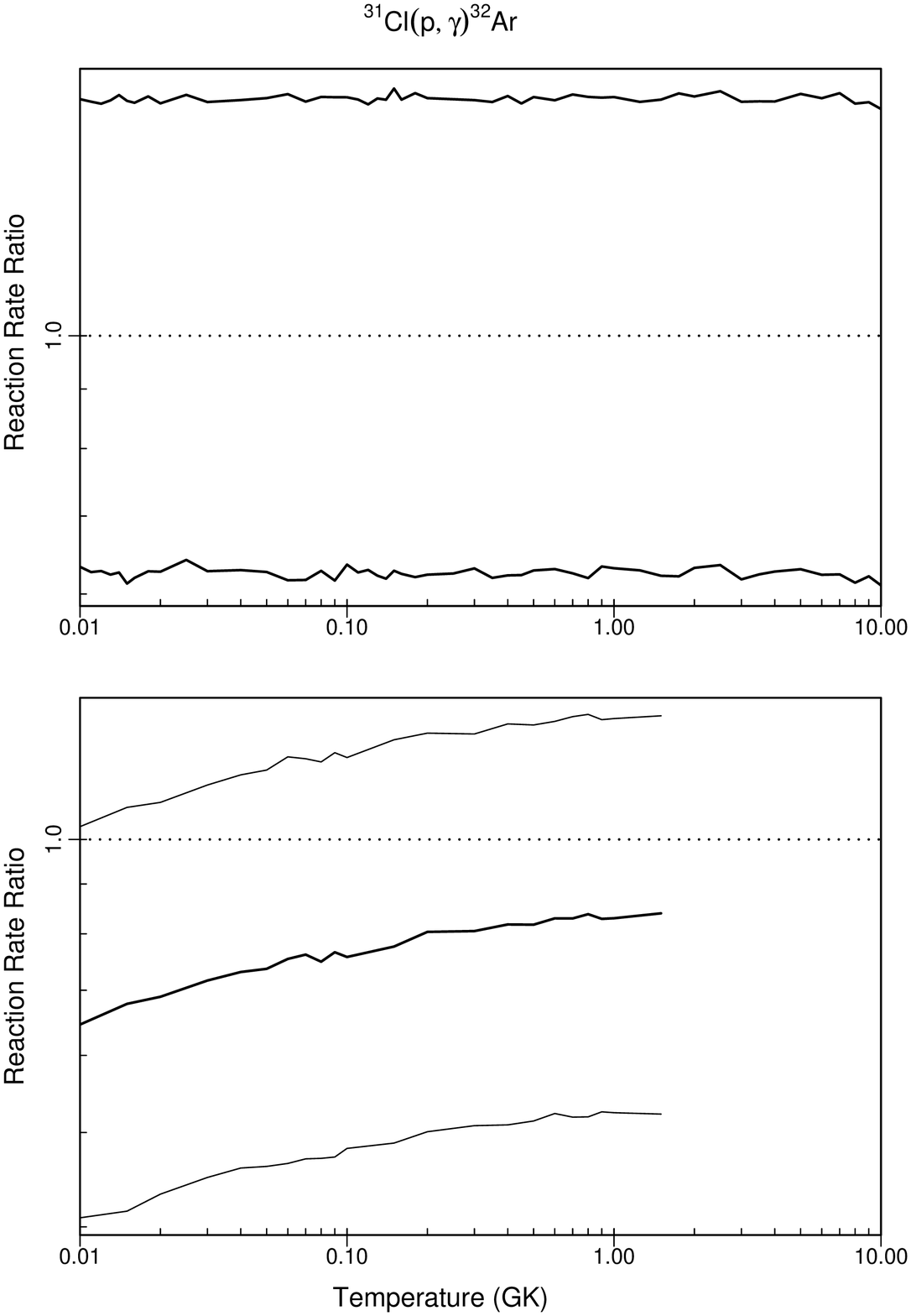}
\caption{\label{} 
Previous reaction rates: Ref. \cite{Her95}. Rate uncertainties have not been determined previously.}
\end{figure}
\clearpage
\begin{figure}[b]
\includegraphics[height=17.05cm]{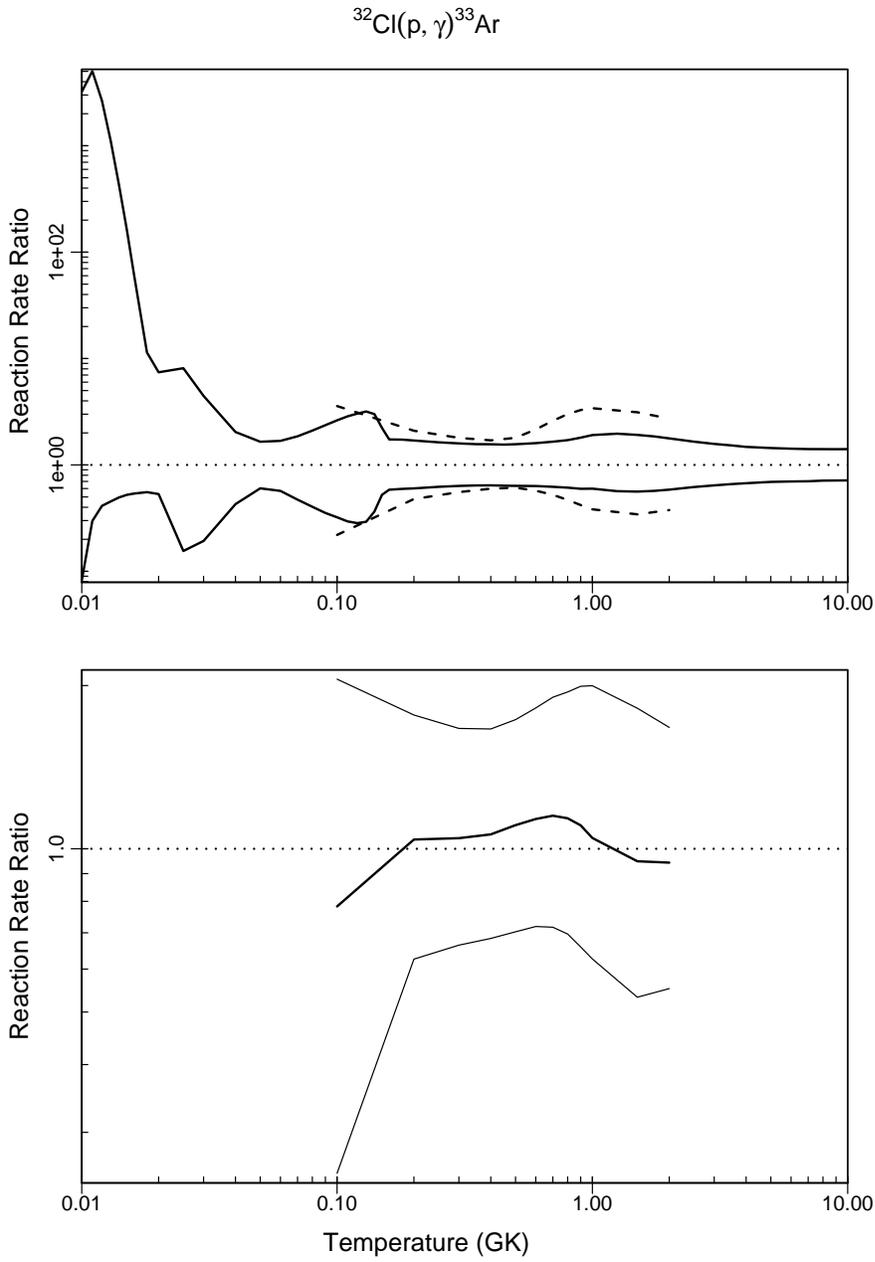}
\caption{\label{} 
Previous reaction rates: Ref. \cite{Sch05}. Note that the {\it stellar} rate is displayed in the original work, but we compare our results to the previous {\it laboratory} rate.}
\end{figure}
\clearpage
\begin{figure}[b]
\includegraphics[height=17.05cm]{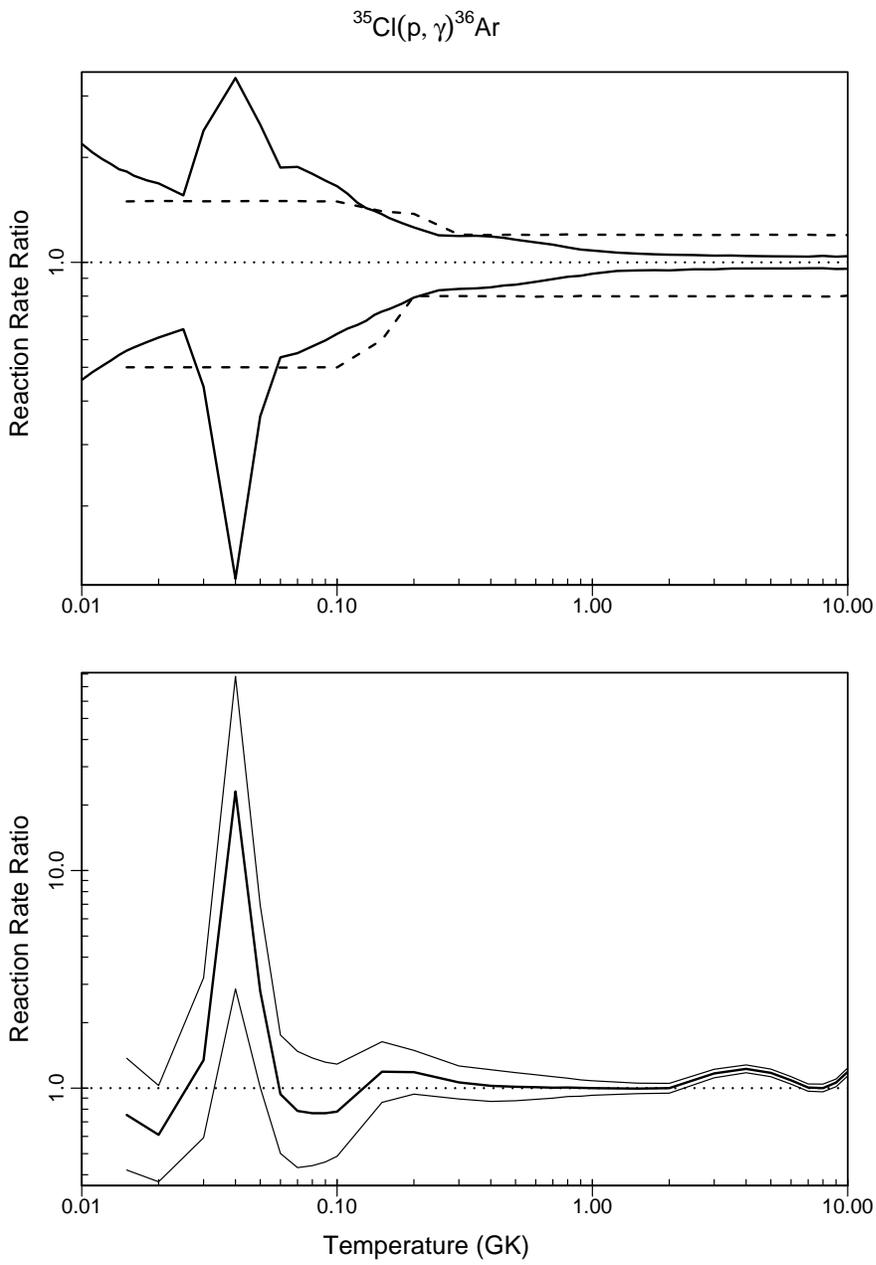}
\caption{\label{} 
Previous reaction rates: Ref. \cite{Ili01}.}
\end{figure}
\clearpage
\begin{figure}[b]
\includegraphics[height=17.05cm]{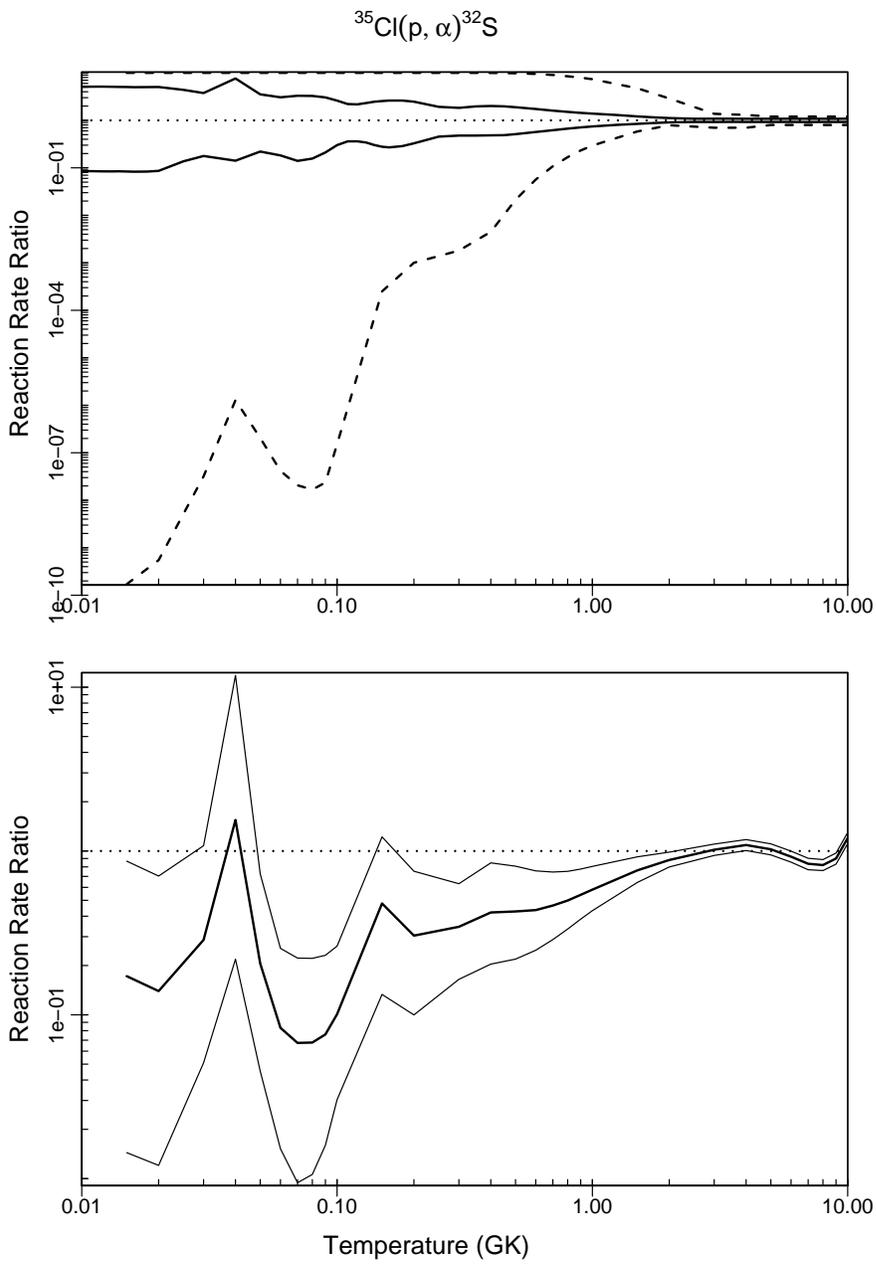}
\caption{\label{} 
Previous reaction rates: Ref. \cite{Ili01}.}
\end{figure}
\clearpage
\begin{figure}[b]
\includegraphics[height=17.05cm]{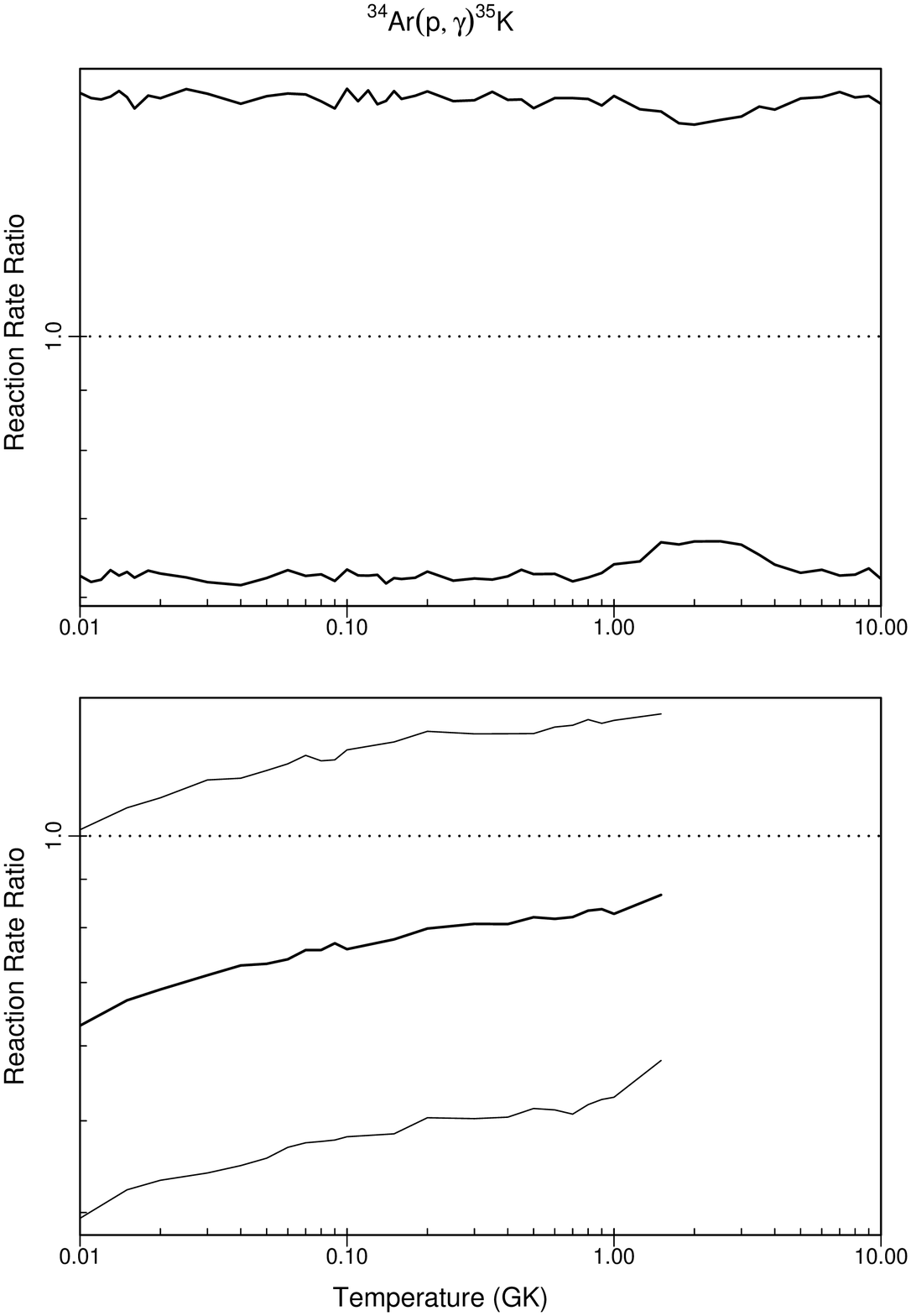}
\caption{\label{} 
Previous reaction rates: Ref. \cite{Her95}. Rate uncertainties have not been determined previously.}
\end{figure}
\clearpage
\begin{figure}[b]
\includegraphics[height=17.05cm]{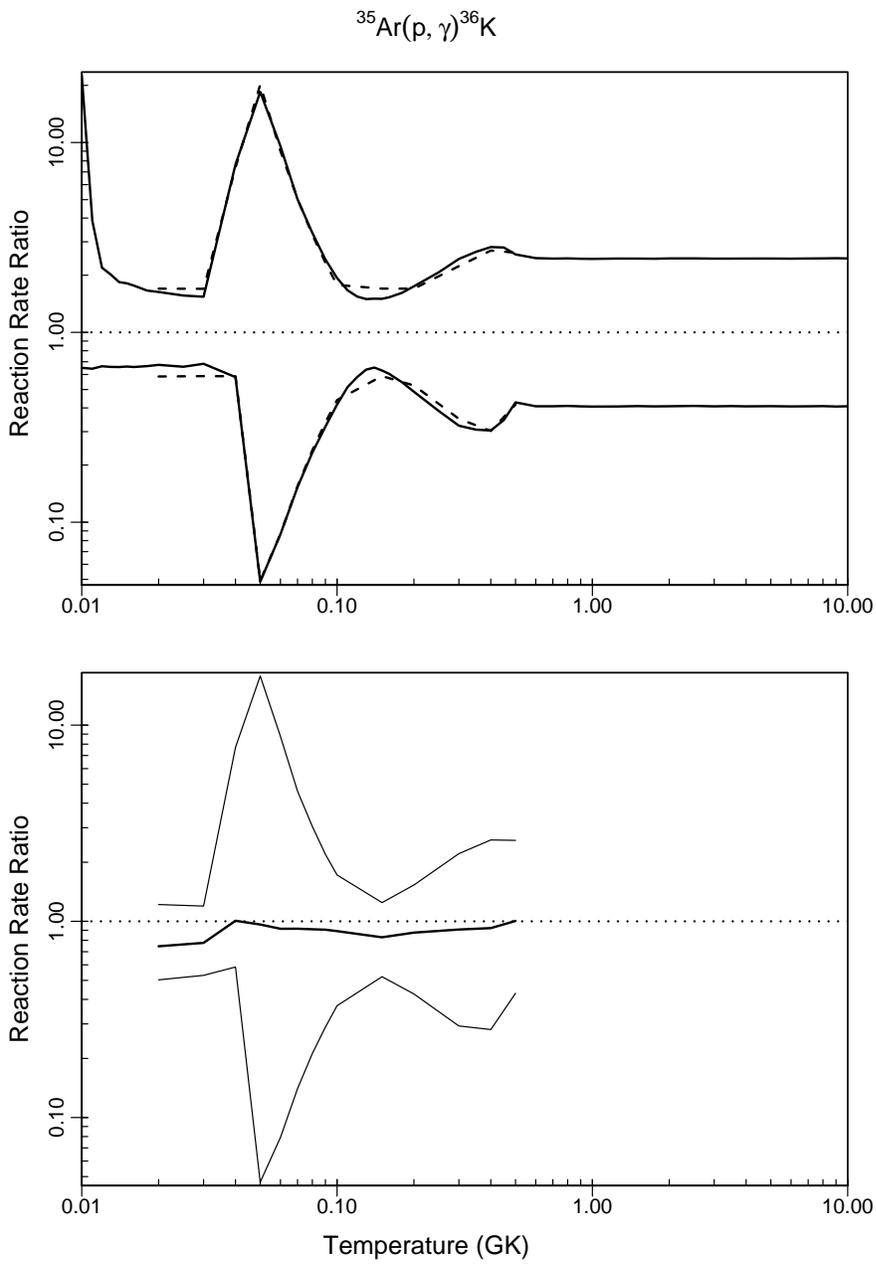}
\caption{\label{} 
Previous reaction rates: Ref. \cite{Ili99}.}
\end{figure}
\clearpage
\begin{figure}[b]
\includegraphics[height=17.05cm]{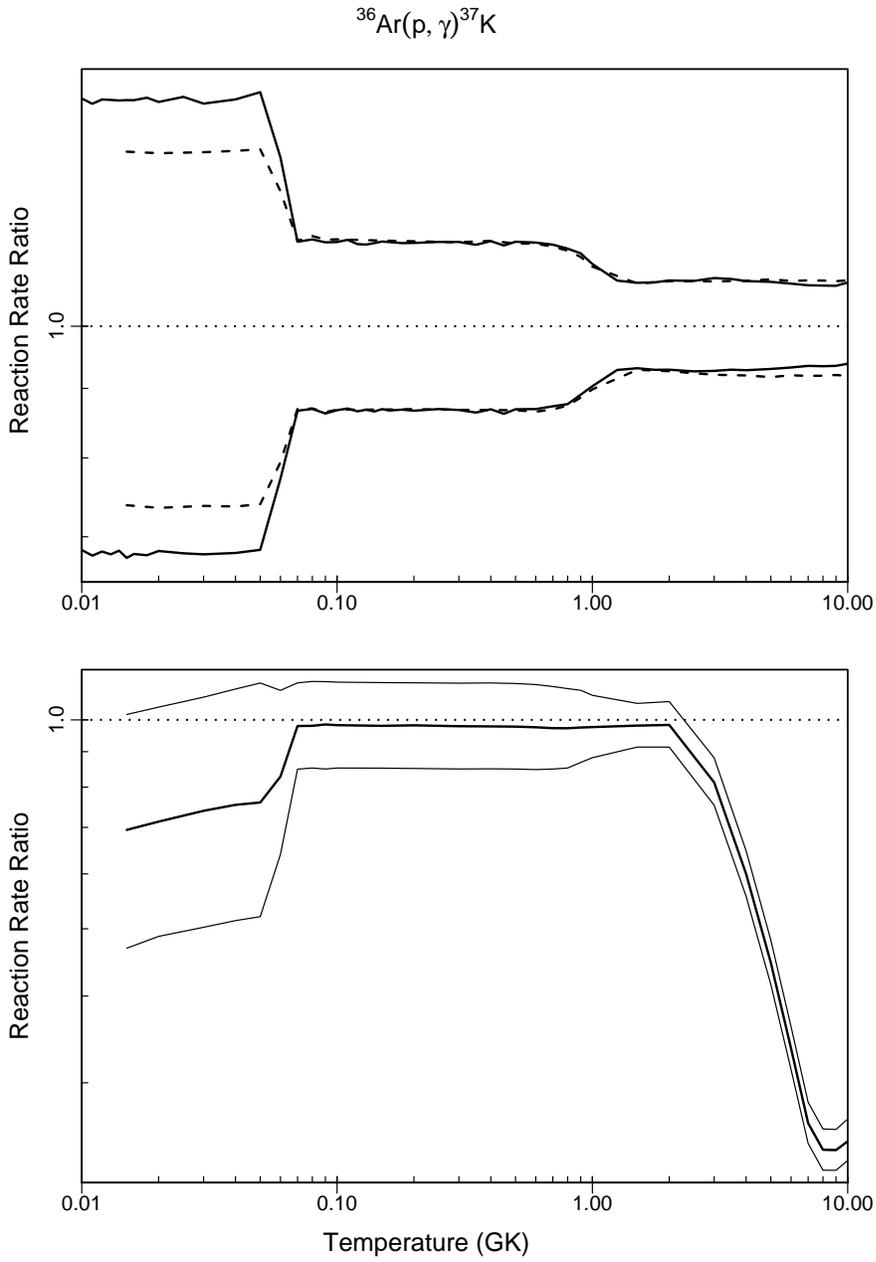}
\caption{\label{} 
Previous reaction rates: Ref. \cite{Ili01}.}
\end{figure}
\clearpage
\begin{figure}[b]
\includegraphics[height=17.05cm]{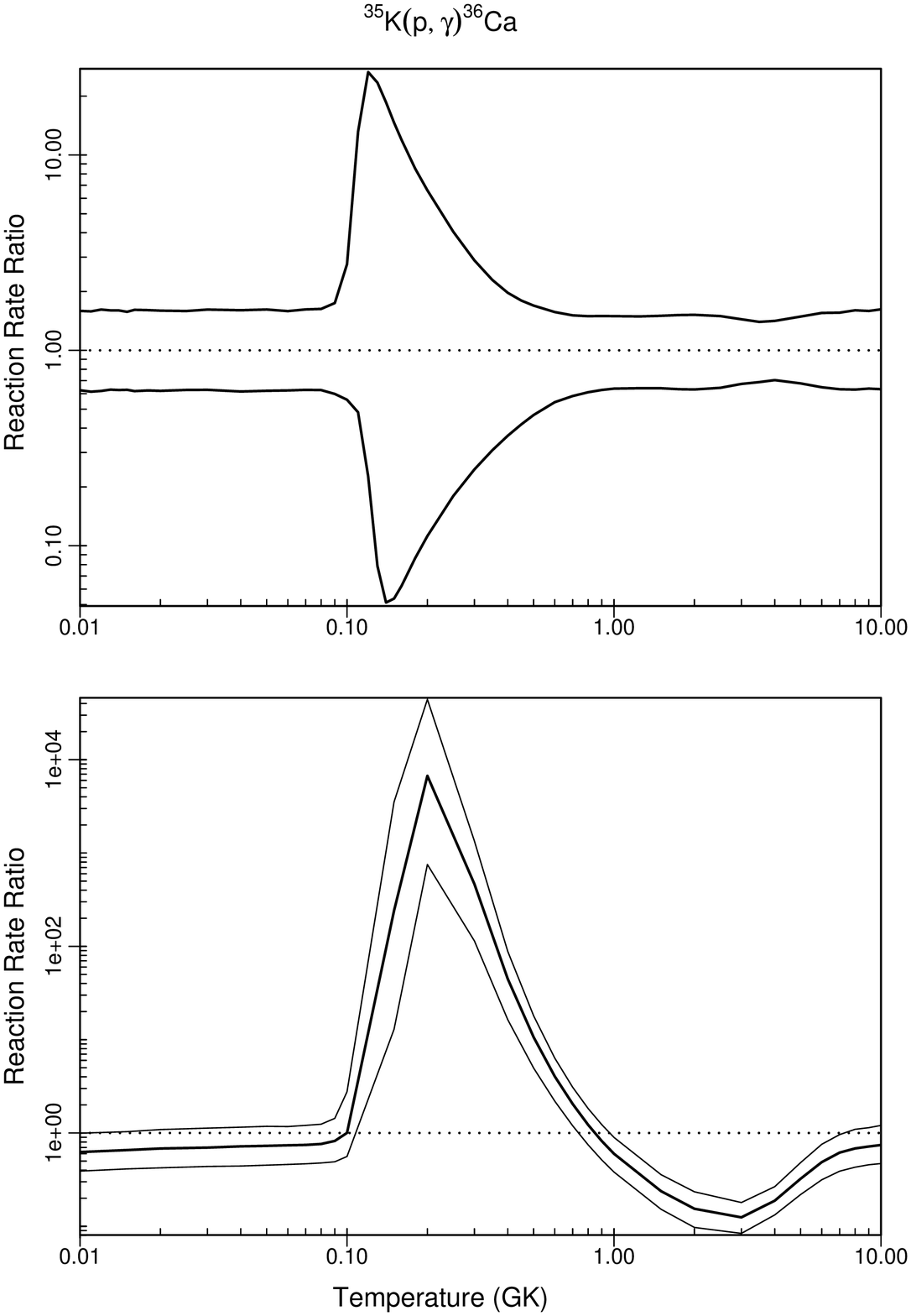}
\caption{\label{} 
Previous reaction rates: Ref. \cite{Ili01}. Rate uncertainties have not been determined previously.}
\end{figure}
\clearpage
\begin{figure}[b]
\includegraphics[height=17.05cm]{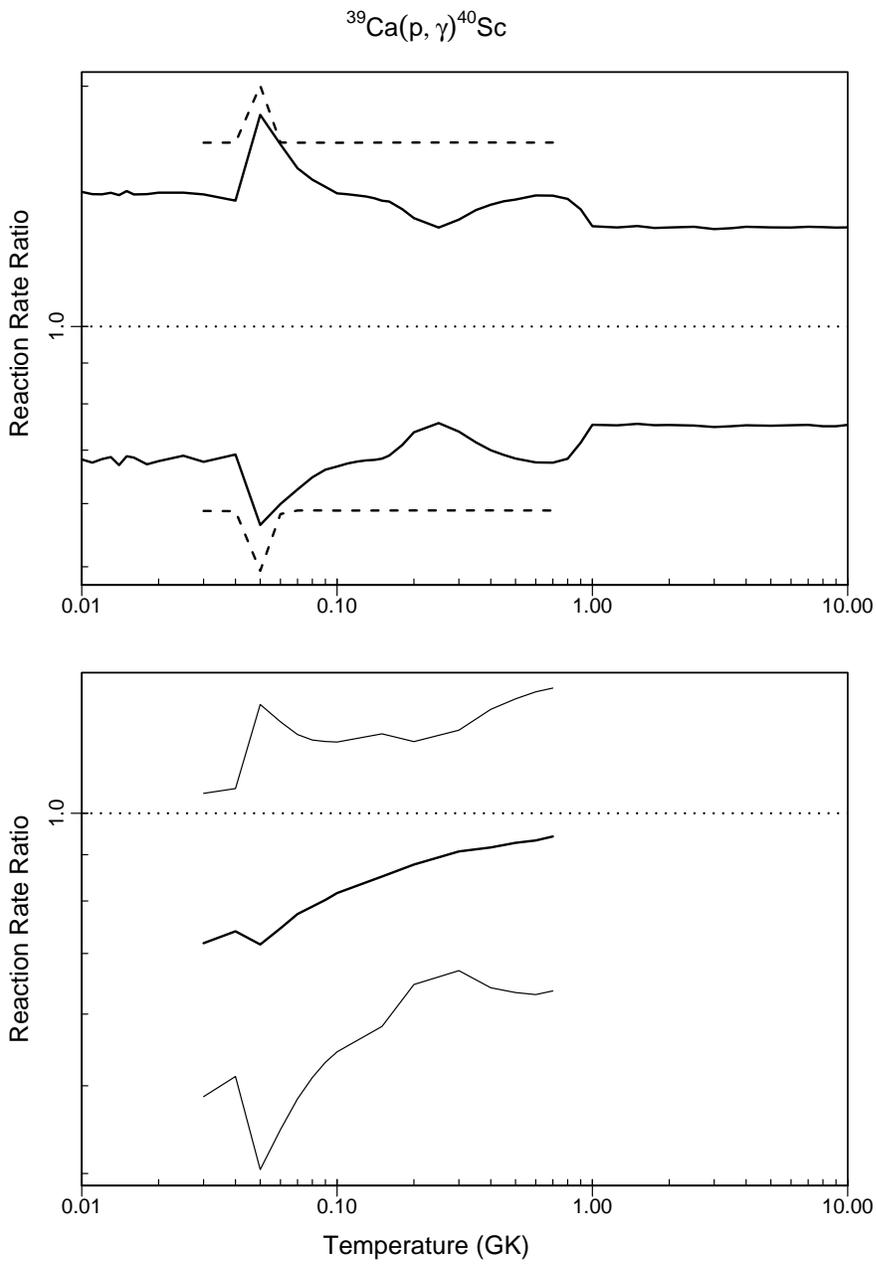}
\caption{\label{} 
Previous reaction rates: Ref. \cite{Ili99}.}
\end{figure}
\clearpage
\begin{figure}[b]
\includegraphics[height=17.05cm]{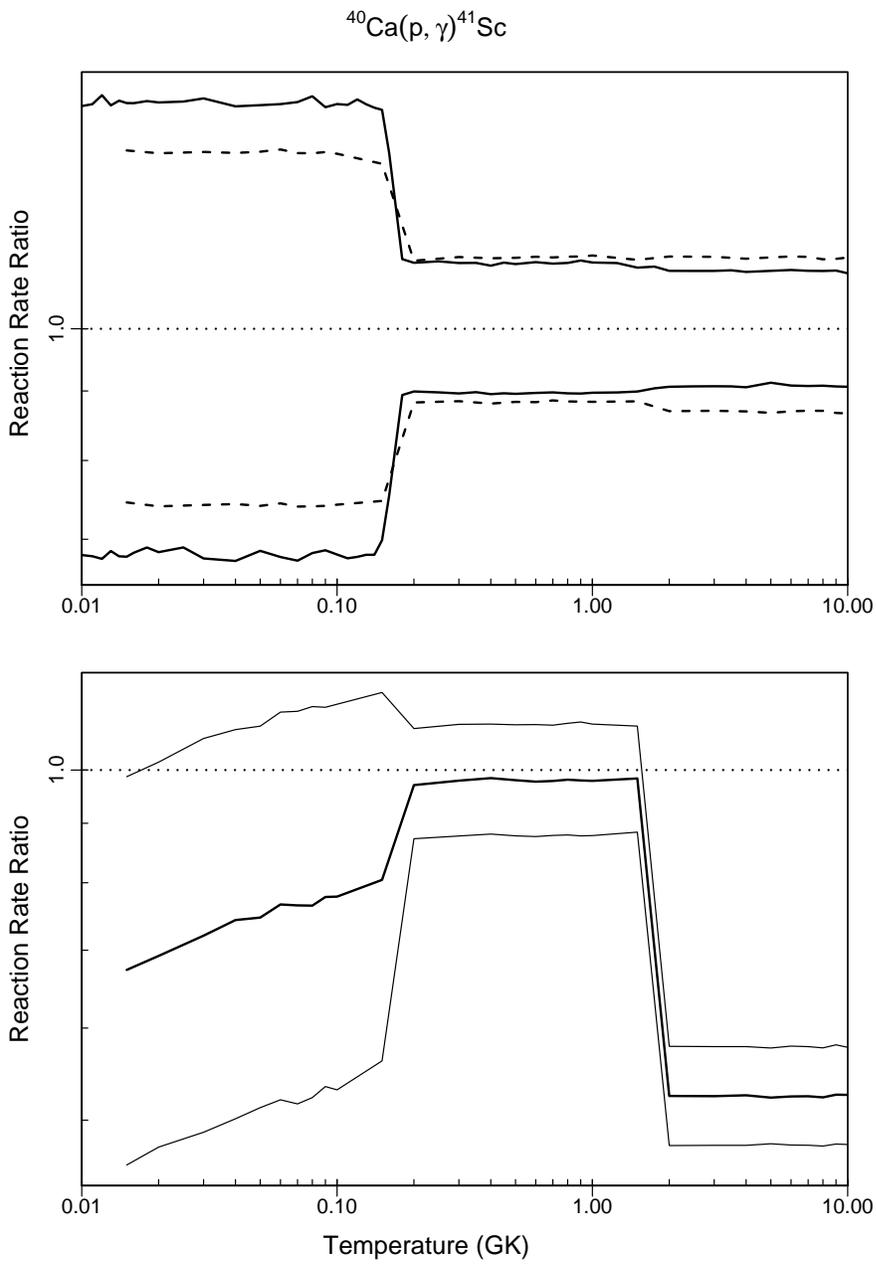}
\caption{\label{} 
Previous reaction rates: Ref. \cite{Ili01}.}
\end{figure}
\clearpage

\end{document}